\documentclass[aps,prd,nofootinbib,twocolumn,superscriptaddress,preprintnumbers,balancelastpage,longbibliography]{revtex4-1}

\usepackage{placeins}
\usepackage{amsmath,amssymb,mathtools,bm}
\usepackage{graphicx, color, hepunits}
\usepackage[dvipsnames]{xcolor}
\usepackage{float}
\usepackage{filecontents}
\usepackage{multirow}
\usepackage{fontawesome5}
\usepackage[normalem]{ulem}
 \usepackage{hyperref}
\hypersetup{
    colorlinks=true,
    linkcolor=blue,
    citecolor=blue,
    filecolor=magenta,
    urlcolor=blue
}
\usepackage[utf8]{inputenc}
\usepackage[english]{babel}

\newcommand{\gagg}{g_{a \gamma \gamma}}

\newcommand{\es}[2] {\begin{equation} \label{#1} \begin{split} #2 \end{split} \end{equation}}

\begin{document}

\title{
The Price of Discovery: Cost-Optimized Detectors for the GUT-Scale QCD Axion
}

\author{Joshua N. Benabou}
\affiliation{Center for Cosmology and Particle Physics, Department of Physics, New York University, New York, NY 10003, USA}

\author{Joshua W. Foster}
\affiliation{Department of Physics, University of Wisconsin-Madison, Madison, WI 53706, U.S.A.}

\author{Benjamin R. Safdi}
\affiliation{Leinweber Institute for Theoretical Physics, University of California, Berkeley, CA 94720, USA}
\affiliation{Theoretical Physics Group, Lawrence Berkeley National Laboratory, Berkeley, CA 94720, U.S.A.}

\date{\today}

\begin{abstract}
Grand unification and string theory point to a quantum chromodynamics (QCD) axion with mass between $10^{-11}$ and $10^{-8}$~eV, while heterotic string theory suggests a mass of 0.5--0.8~neV.
We study which laboratory experiment could discover or exclude DFSZ-strength QCD axion dark matter
at and above 0.5~neV at the lowest cost, which we estimate from scaling laws calibrated to historical data.
We find that a lumped-element detector, building on the ABRACADABRA concept, can do so with demonstrated readout technology; 
an example is a shielded toroid about 14~m across with a 6~T NbTi magnet, whose magnet and cryostat we estimate at around 120 million Swiss francs (MCHF).
If projected readout targets are met, a 5~m toroid at 25~MCHF suffices.
We find these designs by optimizing the toroid, solenoid, and core lumped-element topologies under a common cost metric.  The toroid emerges as the most promising and cost-effective topology.
The optimal field of the toroid detector is far below the 16~T of the proposed DMRadio-GUT design, which would cost nearly twice as much as our optimum.
Heterodyne searches with superconducting radio-frequency cavities, with cavity shapes that we optimize for price, cost between about the same as and twice as much as the lumped-element detectors, but require cavities of volumes $\sim$50--2000~m$^3$.
We find that nuclear-magnetic-resonance searches (CASPEr) need orders-of-magnitude gains in spin polarization and coherence.
We release \texttt{axionMultiPhysics}~\href{https://github.com/bsafdi/axionMultiPhysics}{\faGithub}, an open-source finite-element and multiphysics package for, {\it e.g.}, magnetoquasistatic and full-wave electromagnetics, superconductors, readout circuits, and mechanics, from which most of our numerical results were obtained.

\end{abstract}
\maketitle

\tableofcontents

\section{Introduction}
The quantum chromodynamics (QCD) axion with a decay constant \mbox{$f_a \sim 10^{16}$~GeV}, corresponding to a mass \mbox{$m_a \sim 0.5$~neV}, is one of the most promising new physics candidates.
Such an axion explains the strong {\it CP} problem (the absence of a neutron electric dipole moment, or EDM)~\cite{Peccei:1977hh,Peccei:1977ur,Weinberg:1977ma,Wilczek:1977pj}, can naturally accommodate the observed dark matter (DM)~\cite{Preskill:1982cy,Abbott:1982af,Dine:1982ah}, and emerges in string theory compactifications~\cite{Witten:1984dg,Svrcek:2006yi,Arvanitaki:2009fg,Halverson:2019cmy}.
It has been shown that axions arising from the dimensional reduction of higher-dimensional gauge fields in theories that accommodate grand unification have masses $10^{-11} \, \, {\rm eV} \lesssim m_a \lesssim 10^{-8} \, \, {\rm eV}$~\cite{Benabou:2025kgx}.
Even more specifically, QCD axions arising in heterotic string theory are predicted to have mass $m_a \in (0.5,0.8)$~neV, which is the mass of the so-called model-independent (MI) axion~\cite{Witten:1984dg,Choi:1985bz,Svrcek:2006yi,Benabou:2026jtv,Benabou:2026inprep}.
The MI axion mass is a natural target for laboratory axion experiments.
As we show in this work, the lumped-element approach building on the LC resonator and ABRACADABRA concepts~\cite{Sikivie:2013laa,Kahn:2016aff} appears to be the most promising approach to reach sensitivity to the MI QCD axion.
Heterodyne detectors offer an alternative, though potentially more costly, detection scheme~\cite{Berlin:2019ahk,Berlin:2020vrk,Lasenby:2019prg}, while nuclear magnetic resonance (NMR)-based approaches like CASPEr~\cite{Graham:2013gfa,Budker:2013hfa,JacksonKimball:2017elr,Aybas:2021nvn} appear furthest away at present, as we discuss.
In this work we compare these different detection concepts for the MI axion and optimize the lumped-element and superconducting radio-frequency (SRF) detector topologies and geometries with respect to a cost proxy.

Lumped-element axion DM detection is applicable when the axion Compton wavelength is much larger than the size of the experiment~\cite{Benabou:2022qpv,DMRadio:2023igr}.
In this limit, the magnetoquasistatic (MQS) approximation holds; a region containing a strong laboratory magnetic field ${\bm B}_{\rm lab}$ induces an effective current ${\bm J}_{\rm eff} = \gagg \sqrt{2 \rho_{\rm DM}} {\bm B}_{\rm lab} \cos m_a t$~\cite{Sikivie:2013laa,Kahn:2016aff}, where $\rho_{\rm DM}$ is the local DM density and where we neglect the small energy dispersion arising from the DM kinetic motion.
The effective current induces a real, oscillating, secondary magnetic field, which can induce a current in a pickup apparatus elsewhere in the detector.
That current may be resonantly enhanced through the use of circuit elements, namely capacitors and inductors, that are tuned to induce a resonant response at a frequency corresponding to $m_a$.
The induced signal is read out by inductively coupling to a superconducting quantum interference device (SQUID) circuit or similar, as we discuss more below.

\begin{figure*}[t]
    \centering
    \includegraphics[width=\textwidth]{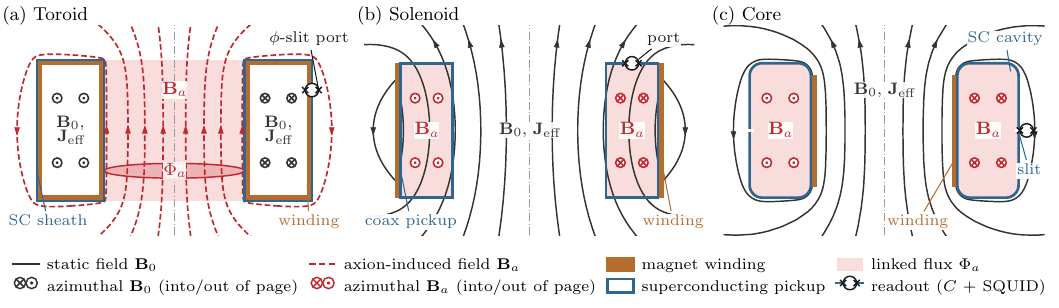}
    \caption{
Vertical cross-sections of the three detector topologies considered in this work: (a) the toroid, (b) the solenoid, and (c) the core design.
The field $\bm{B}_0$ denotes the static, laboratory magnetic field.
Axion DM induces an effective, oscillating current density $\bm{J}_{\rm eff} \propto \bm{B}_0$, which creates a real, secondary magnetic field $\bm{B}_a$.
The toroid and core detectors are dual to each other: the toroid has $\bm{B}_0$ within the superconducting toroid, with $\bm{B}_a$ creating oscillating flux through the donut hole in the middle; the core has a primary magnetic field in the middle of the donut hole and creates an axion-induced $\bm{B}_a \propto \bm{\hat \phi}$ in the superconducting toroidal box.
The solenoid detector is like the core but has $\bm{B}_0$ piercing the coaxial pickup ({\it i.e.}, the toroidal box).
The core design has the added challenge of redirecting magnetic fields away from the pickup box to keep it superconducting, while the solenoid has the challenge that the magnetic field must pierce the pickup, which prevents the use of superconducting material and thus degrades the quality factor.
}
    \label{fig:topologies}
\end{figure*}

The three lumped-element detector topologies that we consider in this work are the toroid, solenoid, and core concepts (see Fig.~\ref{fig:topologies}).
The successful ABRACADABRA-10~cm geometry was a toroid~\cite{Ouellet:2019tlz,Ouellet:2018beu,Salemi:2021gck}, as is the ongoing DMRadio-50L experiment's geometry~\cite{Rapidis:2022gti,Ankel:2026cef}, while the proposed DMRadio-m$^3$ experiment uses a solenoid topology~\cite{DMRadio:2022pkf,DMRadio:2023igr}, as does the ongoing CAL-Pathfinder experiment~\cite{Puranam:2026cpad}.
An early planning study of the DMRadio-GUT experiment~\cite{Brouwer:2022bwo}, which would nominally reach sensitivity to the MI QCD axion, uses a toroid.
A core geometry, with the field confined to the bore of a surrounding superconducting pickup, was studied in Ref.~\cite{Benabou:2022qpv} and has since been proposed as DMRadio-Core~\cite{Ankel:2026zrv}.
These experimental variants differ in the topology of the laboratory magnetic field and readout elements; the toroid geometry uses a toroidal laboratory field with a surrounding pickup sheath, the core concept has a solenoidal laboratory field with a surrounding pickup toroid, and the solenoidal design has a solenoidal magnet with a coaxial pickup structure immersed within the magnetic field region.
In this work we compare these topologies and individually optimize their geometries to make a comparison of the most cost-effective path toward GUT-scale QCD axion detection.
We account for the dominant cost drivers, which are the primary laboratory magnet and the cryogenic system, through approximate scaling laws based on historical data from the past few decades.
We show that under optimistic and conservative scenarios for the readout and the cryogenic systems, the magnetic and cryogenic costs of a lumped-element detector with sensitivity to the MI axion would be around 25 to 120~MCHF (million 2026 Swiss francs), depending on the scenario.
The total experimental cost may be on the order of twice this number~\cite{Brouwer:2022bwo}.

We compare the lumped-element detectors to the heterodyne and NMR approaches.
The heterodyne approach works by hosting a high-intensity radio-frequency (RF) mode inside a high-quality-factor SRF cavity.
Axion DM shifts power to nearly degenerate modes that differ from the primary mode in frequency by precisely the axion mass.
Signal power is read out in these adjacent modes.
The NMR approach, {\it e.g.}, the CASPEr experiment, on the other hand, uses the coupling of the axion to gluons and the induced oscillating EDM.
Spins are polarized along a laboratory magnetic field inside a ferroelectric crystal, whose internal electric field is orthogonal to it.
The axion-induced oscillating EDM drives the spins at a frequency set by the axion mass; when this matches their Larmor frequency, the spins tilt away from the magnetic field and a transverse magnetization grows linearly in time until coherence is lost.

We show that the SRF approach, with a simple cost metric that accounts for the cost of the niobium cavity, its cryostat, and the helium refrigerator, is about 30\% more expensive than the lumped-element approach under optimistic assumptions and about twice as expensive under conservative ones.
We caution, however, that our cost estimates come from historically calibrated cost data and are not robust when comparing between different types of experiments at the factor-of-two level; the cost functionals do, however, allow for optimization of topologies and geometries within a detector class.
We present optimized SRF cavities that save $\sim$30\% in cost relative to the best existing designs.
The SRF cavities would also have to change length by up to several percent to scan the full mass range, which, as far as we are aware, is unprecedented, and they must be built precisely enough that wall vibrations do not scatter pump photons into the signal mode.
On the other hand, with demonstrated nuclear spin polarization the NMR approach would require an impossibly large sample, and it becomes competitive only if the polarization and coherence times improve by many orders of magnitude.

We also prove that the toroid and core lumped-element designs are dual: under ideal conditions they give identical scan rates at the same cost.
Non-ideal effects break this tie in favor of the toroid with our accounting.
The toroid needs a superconducting shield, while the core needs the field on its superconducting pickup box held below $ \sim$100~mT, which takes $\sim$1{,}000 (10{,}000) tons of iron in the optimistic (conservative) scenario.
We then estimate that the core designs are at least $\sim$10\% more costly and require more total volume and infrastructure.
Both options are less costly than the solenoid, whose pickup sits in the static field and so must be made of copper and cooled to mK temperatures; by contrast, the pickup inductors for the toroid and core designs can likely be held closer to 4~K.
Finally, we model a realistic toroid with discrete racetrack coils, capacitor modules, and its readout beyond the MQS approximation, using the finite-element code \texttt{axionMultiPhysics}~\href{https://github.com/bsafdi/axionMultiPhysics}{\faGithub} that we release with this work, and find that under the conservative benchmark, which is based on demonstrated technology, it reaches the MI axion target for a magnetic and cryogenic cost of about 125~MCHF.

\texttt{axionMultiPhysics}~\href{https://github.com/bsafdi/axionMultiPhysics}{\faGithub}~\cite{axionMultiPhysics} is an open-source finite-element package, described in App.~\ref{app:axionmultiphysics}, that has many of the capabilities of commercial multiphysics codes such as COMSOL~\cite{multiphysics1998introduction}, including magnetostatics and magnetoquasistatics with nonlinear iron and superconductors, full-wave electromagnetics and eigenmode analysis, and structural mechanics and heat conduction, with a C++ core built on the MFEM library~\cite{mfem} that solves 3D problems in parallel with MPI.
It also has native functions for axion physics (the signal coupling, pickup inductance and scan figure of merit of lumped-element detectors, cavity form factors, SRF heterodyne  overlap factors and SQUID readout noise, with adjoint shape optimization), visualization through an interactive 3D viewer that either runs in a Jupyter notebook or exports to a standalone web page, and an AI-native interface through a Model Context Protocol (MCP) server that lets AI assistants such as Claude build, solve and optimize detector designs.

\section{Lumped-element scanning rate and cost functions}
\label{sec:scan_cost}

In this section we briefly review the physical concepts behind lumped-element axion detectors and their optimal scanning rate before discussing our choices of cost functional and benchmark parameters.
Throughout this section we use the solenoid with height $H$ much larger than its radius $R$, which we refer to as the stretched solenoid, as a simple, analytically tractable example.

\subsection{Axion-induced EMF}

In the MQS limit~\cite{Benabou:2022qpv}, the axion-induced effective current $J_{\rm eff}$ oscillates at the frequency $m_a$ and creates a magnetic flux $\Phi_a$ through the pickup~\cite{Sikivie:2013laa,Kahn:2016aff}.
For a single pickup loop, we may write $\Phi_a = \int_S {\bm dS}\cdot \bm{B}_a = \oint {\bm d \ell} \cdot {\bm A}_a$, where the first integral is over the surface the pickup loop encloses, with surface area element ${\bm dS}$, and the second is around the pickup loop itself, with vector line element ${\bm d \ell}$, and with ${\bm A}_a$ the axion-induced vector potential.
Note that the axion-induced pickup current flows in the direction ${\bm d \ell}$ by construction.
Generalizing away from a single pickup loop, we may write $\Phi_a = (1/I) \int d^3x {\bm J}_I \cdot {\bm A}_a$, where ${\bm J}_I$ is the current density through the 3D pickup that arises from injecting a current $I$ through the readout leads into the pickup.

It follows from the expressions above that we may write the axion-induced flux through the pickup as
\es{eq:flux}{
\Phi_a = \kappa_a \, \Psi \,, \qquad \Psi \equiv \frac{1}{I} \int d^3x \, {\bm A}_I \cdot {\bm B}_{\rm lab}({\bm x}) \,,
}
with $\kappa_a \equiv \gagg \sqrt{2 \rho_{\rm DM}}$ and ${\bm A}_I$ the vector potential that the current $I$ described above induces.

As a simple example, let us consider the solenoidal coax detector topology, illustrated in Fig.~\ref{fig:topologies} in the long-solenoid limit, where the solenoid height $H$ is much larger than its outer radius $R$.
The coax's inner radius is denoted by $x R$, with $0 < x < 1$.
Within the solenoid the magnetic field is taken to be ${\bm B}_{\rm lab} = B_{\rm lab} {\bm \hat z}$ for radii $0 < r < R$ from the bore center, with ${\bm \hat z}$ pointing along the solenoid's axis.
Note, importantly, that the magnetic field is assumed to penetrate the inner coax region as well as the region between the two cylinders.
Thus, the axion-induced effective current ${\bm J}_{\rm eff} \propto \bm{\hat z}$.
The pickup topology consists of a gapped connection between the inner coax cylinder and the outer coax cylinder.
The gap is connected by the LC readout circuit; axion-induced currents flow in the ${\bm \hat z}$ direction along the inner and outer walls of the coax.
Consider a current $I$ flowing in this direction.
The vector potential ${\bm A}_I$ is equal to ${\bm A}_I = - {I \over 2 \pi} \log(r/R)\, {\bm \hat z}$ for $x R < r < R$ and ${\bm A}_I = {I \over 2 \pi} \log(1/x)\, {\bm \hat z}$ for $0 < r < x R$.
Substituting this expression into~\eqref{eq:flux}, we compute
\es{}{
\Psi_{\rm solenoid} \approx {H R^2 \over 4} B_{\rm lab} (1 - x^2)\,.
}

Resonant lumped-element experiments in the MQS regime place a tunable capacitor, of capacitance $C$, in series with the pickup inductance $L_p$, and tune $C$ to achieve resonance at the frequency $\omega_0 = 1 / \sqrt{L_p C}$; here we assume that $L_p$ dominates the total inductance of the circuit.
The pickup inductance may be computed through the equivalent expressions
\es{eq:Lp}{
L_p = \frac{1}{ I^2}\int d^3x\,|\bm{B}_I|^2 = \frac{1}{I^2}\int d^3x\,\bm{J}_I\cdot\bm{A}_I \,,
}
with $\bm{B}_I = \nabla \times \bm{A}_I$ the magnetic field induced by the current $I$.
For the solenoidal coax example, this evaluates to $L_p \approx (H / 2 \pi) \log (1/x)$ in the limit $H \gg R$.
Dissipation in the circuit ({\it e.g.}, resistance of the pickup and loss in the capacitor) is described by a series resistance $R_{\rm c}$, which sets the intrinsic quality factor of the resonator, $Q = \omega_0 L_p / R_{\rm c}$.
Note that generically $R_{\rm c}$ is frequency dependent.

\subsection{The readout circuit and scanning rate}

In the nominal configuration, the current in the circuit is read out by a dc SQUID, coupled to the circuit through a mutual inductance $M$; we neglect SQUID input loading and correlations between imprecision and backaction noise.
The axion-induced EMF, with magnitude $m_a \Phi_a$, and the thermal (Johnson--Nyquist) voltage noise of the resistor at temperature $T$ drive current through the same series impedance and thus receive the same resonant enhancement.
The SQUID contributes imprecision noise, with flux power spectral density (PSD) $S_\Phi$, which enters as a current noise $S_\Phi / M^2$ that is not resonantly enhanced, and backaction noise, with current PSD $S_J$, which enters as a voltage noise $\omega^2 M^2 S_J$ in series with the resonator and is enhanced in the same way as the thermal noise.\footnote{Note that we normalize all PSDs as in Ref.~\cite{Foster:2017hbq}, such that a stationary signal $X(t)$ has $\langle X^2 \rangle = \pi^{-1} \int_0^\infty d\omega \, S_X(\omega)$.}
Thus, the sensitivity bandwidth of the experiment is set by the range of frequencies about $\omega_0$ over which thermal noise dominates over the SQUID noise.
Increasing $M$ suppresses the imprecision noise at the cost of larger backaction noise; following Ref.~\cite{Chaudhuri:2018rqn}, we choose the $M$ that optimizes the frequency-integrated sensitivity.

It is convenient to define $\eta_A \equiv 2 \sqrt{S_\Phi S_J}$, which parametrizes the SQUID noise relative to the standard quantum limit (SQL).
At the SQL, $\eta_A = 1$, while $\eta_A < 1$ is possible but requires techniques such as backaction evasion.
The resulting sensitivity bandwidth, defined as the full width over which thermal noise exceeds the total SQUID noise, exceeds the resonator linewidth $\omega_0/Q$ by a factor $\sim2T/(\eta_A \omega_0) \gg 1$.

We quantify sensitivity using the likelihood framework of Ref.~\cite{Foster:2017hbq}: the measured PSD of the circuit current in each frequency bin is exponentially distributed with mean $\lambda_a(\omega) + \lambda_B(\omega)$, with $\lambda_a$ ($\lambda_B$) the mean PSD of the axion signal (thermal and SQUID noise).
For a signal that is weak in each bin, the expected (Asimov) test statistic (TS)~\cite{Cowan:2010js} accumulated over the scan is $\Theta \simeq \sum_i (t_i/2\pi) \int d\omega \, (\lambda_{a,i}/\lambda_{B,i})^2$, where $t_i$ is the data collection time at the $i^{\rm th}$ tuning of $\omega_0$, and the median expected 95\% one-sided upper limit on $\gagg$ under the null hypothesis is the coupling for which $\Theta = \Theta_{95} \simeq 2.71$~\cite{Foster:2017hbq}.\footnote{In this asymptotic limit, $\Theta$ equals the squared matched-filter signal-to-noise ratio used in Refs.~\cite{Chaudhuri:2018rqn,DMRadio:2022pkf}.}
Note that for a $5\sigma$ discovery search the requirement $\Theta \simeq \Theta_{95}$ still suffices up to a modest overhead, since the $5\sigma$ threshold need only be reached in rescans of the small fraction of mass points that pass a pre-set significance threshold in the first scan.
Accounting for the optimal coupling, the scanning rate, which is defined as the rate $d\omega_0/dt$ at which the resonant frequency of the LC circuit may be adjusted to scan over axion masses $m_a = \omega_0$ while reaching the targeted sensitivity, is given by~\cite{Chaudhuri:2018rqn,Chaudhuri:2019ntz,DMRadio:2022pkf,Ankel:2026zrv}
\es{eq:scanrate}{
\frac{d\omega_0}{dt} = \frac{\pi^2}{8} \, \frac{Q_{\rm axion}}{\Theta_{95}} \, \frac{\Phi_a^4}{L_p^2} \, Q \, \bar G(\omega_0) \,.
}
The SQUID coupling enters through
\es{eq:Gbar}{
\bar G(\omega_0) &= \frac{\xi}{\left[\xi^2 + 2(2n+1)\xi + \eta_A^2\right]^{3/2}} \,, \\
\xi &= \frac{2\eta_A^2}{2n+1+\sqrt{(2n+1)^2 + 8\eta_A^2}} \,,
}
where $\xi \equiv 2 R_{\rm c} \, S_\Phi / (\omega_0 M^2) \propto M^{-2}$ measures the SQUID imprecision noise referred to the circuit.
In the second line of~\eqref{eq:Gbar} we tune $M$, and thus $\xi$, to  its optimal value~\cite{Chaudhuri:2018rqn,DMRadio:2022pkf}.

The quantity $n$ in~\eqref{eq:Gbar} is the thermal occupation number of the resonator, $n = (e^{\omega_0/T} - 1)^{-1}$, and at the frequencies and temperatures of interest $n \gg 1$.
Assuming $n \gg 1$ and $n \gg \eta_A$, the expression for the optimal  $\bar G$ simplifies to $\bar G \approx \omega_0/(6\sqrt{3}\, \eta_A T)$.
The DM velocity distribution enters through the axion quality factor
\es{eq:Qaxion}{
Q_{\rm axion} \equiv \int_0^\infty dv \, \frac{f(v)^2}{v} \,,
}
with $f(v)$ the local DM speed distribution in the laboratory frame, normalized such that $\int_0^\infty dv \, f(v) = 1$; an axion with laboratory-frame speed $v$ oscillates at the frequency $\omega \approx m_a (1 + v^2/2)$~\cite{Foster:2017hbq}.
For the Standard Halo Model (SHM), with local circular speed $v_0 \approx 220$~km/s and solar speed in the halo rest frame $v_{\rm obs} \approx 232$~km/s~\cite{Foster:2017hbq}, $Q_{\rm axion} \approx 6.8 \times 10^5$, which is the benchmark value we take in our projections.

\subsection{The cost functional}

Returning to the stretched solenoidal example, from~\eqref{eq:scanrate} we can compute the dependence of the scan rate on geometric parameters in the MQS limit:
\es{eq:scan_solenoid}{
{d \omega_0 \over dt} &\propto \left[ {B_{\rm lab}^2 H R^4 (1 - x^2)^2 \over \log(1/x)} \right]^2 \\
&\propto \left[ {{ W}_{\rm mag} R^2 (1 -x^2)^2 \over\log (1/x) }\right]^2\,.
}
In the second line above we substitute a proxy for the magnet cost, the stored magnetic energy, which we describe below:
\es{}{
{W}_{\rm mag} \equiv {1\over 2}\int d^3x B_{\rm lab}^2({\bm x}) \,.
}
In the limit $H \gg R$, the stored energy is dominated by that within the solenoid, with the contribution of the return field subdominant, so that ${W}_{\rm mag} \approx {1 \over 2} B_{\rm lab}^2 \pi R^2 H$.

The scan rate in~\eqref{eq:scan_solenoid} is maximized for a specific choice of $x$, corresponding to $x_{\rm max} \approx 0.534$.
The optimization of the scan rate with respect to $B_{\rm lab}$, $H$, and $R$, in the limit $H \gg R$ is more nuanced.
First, note that at fixed cost the scan rate is optimized in the limit $R \to \infty$.
In this limit, we must maintain $H \gg R$, which implies that $B_{\rm lab} \to 0$.
This is a feature of all of the detector topologies: accounting for magnetic cost alone, the optimal scan rate at fixed cost is achieved in the limit of increasingly spatially extended but weak laboratory magnetic field.
Of course, this runaway behavior is regulated by the fact that in this limit the detector becomes increasingly large and thus requires an increasingly large room and increasing cryogenic costs.

\begin{figure}[t]
\centering
\includegraphics[width=\columnwidth]{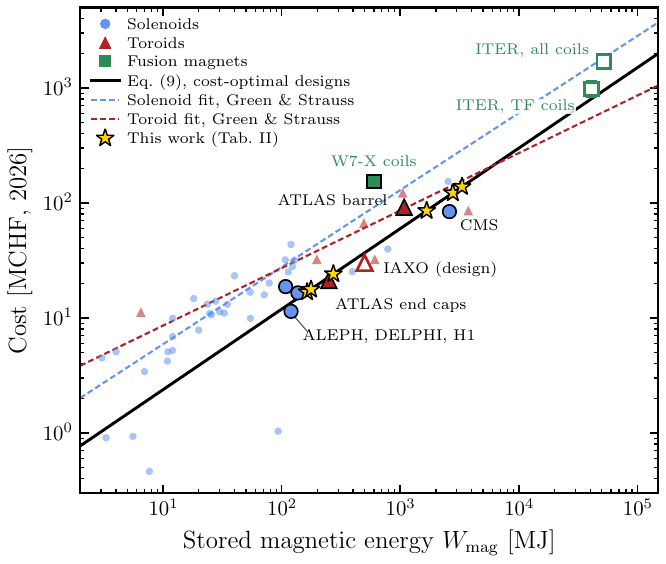}
\caption{Cost ${\mathcal C}$ as a function of stored magnetic energy of historical superconducting solenoids (blue circles) and toroids (red triangles), in 2026 MCHF.
For reference, we show the historical magnets of Ref.~\cite{Green:2008cost} (faint points), and power-law fits to these data (dashed) for solenoids and toroids independently.
We also include the CMS coil~\cite{Herve:2004cms}, the ATLAS barrel and end-cap toroids~\cite{tenKate:2008lhec}, and pre-construction estimates for the ALEPH, DELPHI, and H1 solenoids~\cite{ALEPH:1983bkx,Bartl:1983zq,H1:1986ab} in the figure (outlined points), as well as the design estimate for the IAXO toroid~\cite{Armengaud:2014gea} (open triangle).
We additionally include two fusion magnets (squares), which use Nb$_3$Sn or NbTi cable-in-conduit conductor: the full ITER magnet system~\cite{ITER:FDR2001,Sauthoff:2003fesac}, the part of it that we attribute to the ITER toroidal-field coils (open squares), and the  Wendelstein 7-X coils~\cite{Klinger:W7Xiter}.
We show our cost-function proxy~\eqref{eq:cost_HB}, including both the magnetic energy term and the cryogenic term, evaluated at the scan-rate cost optimum (see text for details).
We overlay the costs of the designs in Tab.~\ref{tab:numerics} (stars).
}
\label{fig:magnet_costs}
\end{figure}

To regulate this runaway we add the cost of the cryostat that keeps the magnet superconducting at $\sim$4~K, using the two-term form of Refs.~\cite{ISSDetectorWorkingGroup:2007uvu,nuSTORM:2013cqr},
\es{eq:cost_HB}{
{\mathcal C} = a \left[ \left( \frac{W_{\rm mag}}{1\,{\rm MJ}} \right)^{0.7} + \kappa \left( \frac{A_{\rm cold}}{1\,{\rm m}^2} \right)^{0.8} \right] ,
}
where $A_{\rm cold}$ is the area of the magnet's cold region, which we approximate by the area of the closed envelope of the magnet.
The first term roughly accounts for the magnet and associated costs, such as the conductor costs, quench protection, and the structure that holds the magnetic forces.
The exponent of the stored-energy term in~\eqref{eq:cost_HB} is empirical, taken from Refs.~\cite{ISSDetectorWorkingGroup:2007uvu,nuSTORM:2013cqr}, and agrees well with fits of the cost of historical detector magnets to their stored energy, which give $W_{\rm mag}^{0.69}$~\cite{Green:2008cost} (see Fig.~\ref{fig:magnet_costs}).
The second term in~\eqref{eq:cost_HB} accounts for the cryostat, including the vacuum vessel, thermal shields, and insulation built over $A_{\rm cold}$; the cryostat exponent in~\eqref{eq:cost_HB} is also taken from Refs.~\cite{ISSDetectorWorkingGroup:2007uvu,nuSTORM:2013cqr}.

We keep the exponents in~\eqref{eq:cost_HB} but refit the two coefficients ourselves to the projected costs of three magnets from which Refs.~\cite{ISSDetectorWorkingGroup:2007uvu,nuSTORM:2013cqr} derive their scaling law: the ALEPH and CMS detector magnets and a design estimate for the magnet of the Gammas, Electrons and Muons (GEM) detector, which was planned for the Superconducting Super Collider, together spanning 0.14--3~GJ of stored energy.
We use their quoted radii, lengths, stored energies, and prices, and take $A_{\rm cold}$ to be the closed-envelope area as described above.
A least-squares fit in log price then gives $\kappa \approx 1.2$.
In 2026 Swiss francs, we find $a \approx 0.25$~MCHF.

We emphasize that~\eqref{eq:cost_HB} is simply used as a metric by which we can optimize designs and roughly estimate cost.
This formula should not be considered accurate for realistic cost estimates.
In particular, it omits several sources of cost, such as the pickup, which we discuss later, and oversimplifies others.
Our optimized designs store 70~MJ--3.3~GJ, close to the cost calibration range of 0.14--3~GJ, but in some cases extending beyond this range, so our results rely on a mild extrapolation (we find the cold areas are 60--830~m$^2$, close to the three magnets discussed above).
In addition, the cost~\eqref{eq:cost_HB} accounts for magnetic energy but not field strength, while in reality the cost may rise near the upper critical field of the superconductor.
Our optimized toroids reach peak fields of up to $\sim$8~T, close to the limit of NbTi at 4.5~K, where its critical current density falls steeply and more conductor is needed.
We discuss in Sec.~\ref{sec:add_cost} how we deal with larger field strengths in our cost estimates.
Further, while the exponent of the stored-energy term in~\eqref{eq:cost_HB} agrees with fits to solenoid magnets and to an ensemble of (mostly solenoidal) detector magnets, Ref.~\cite{Green:2008cost} showed that historical toroidal magnets alone give a cost scaling of $W_{\rm mag}^{0.50}$.
If the latter scaling is more appropriate, the weaker energy penalty may favor smaller toroids operating at higher field.
Throughout this work we follow European (CERN-style) accounting and include hardware and industrial labor in our costs but not, {\it e.g.}, laboratory personnel.
Lastly, note the calibration prices are converted from 1992 Swiss francs (CHF) to 2026 CHF accounting for inflation.  

In the MQS limit the scan rate goes as $(W_{\rm mag} R^2)^2$.
This implies that if we increase all magnet dimensions by $s$, at fixed scan rate, $W_{\rm mag} \propto 1/s^2$ while $A_{\rm cold} \propto s^2$.
We may then analytically optimize~\eqref{eq:cost_HB} with respect to $s$; at the cost minimum we find that the cryostat term is $\sim$$0.875$ times the magnetic energy term, so that ${\mathcal C} \approx 1.875 \times a\, (W_{\rm mag} / 1 \, {\rm MJ})^{0.7}$.
This is the cost-optimized function, for iron-free designs, that we illustrate in solid black in Fig.~\ref{fig:magnet_costs}.

We compare the predicted costs from~\eqref{eq:cost_HB} to historical and future detector magnet cost projections in Fig.~\ref{fig:magnet_costs}.
The fit of Ref.~\cite{Green:2008cost} to historical toroids alone gives costs $1.2$--$1.8$ times higher than~\eqref{eq:cost_HB} for the toroids in Tab.~\ref{tab:numerics}.
On the other hand,~\eqref{eq:cost_HB} overestimates the costs of, {\it e.g.}, the ATLAS barrel toroid and two end-cap toroids, which are $\sim$130~MCHF together in 2026 CHF ($\sim$90~MCHF for the barrel and $\sim$20~MCHF for each end cap)~\cite{ATLAS:1997magnetTDR,ATLAS:RRB2002062,ATLAS:RRB2006069,tenKate:2008lhec}.
Evaluated with their actual stored energies and cold areas,~\eqref{eq:cost_HB} predicts $\sim$260~MCHF.\footnote{Note that these low-field toroids have a large cold area for their stored energy and so are not  cost-optimal, while in Fig.~\ref{fig:magnet_costs} we show their costs against the cost-optimal line.}
In  Fig.~\ref{fig:magnet_costs} we also include the estimated cost of the ITER magnet system, which is approximately captured by~\eqref{eq:cost_HB}.
We emphasize that the optimized lumped-element designs of Tab.~\ref{tab:numerics} are modest compared to the ITER magnet.
They reach peak fields of at most 8.3~T and store less than 3.3~GJ, while the ITER toroidal-field coils reach 11.8~T and store $\sim$40~GJ.
Additionally, the ITER magnets sit in a cryostat 28~m in diameter and 24~m high~\cite{ITER:FDR2001}, larger than all our designs.

\subsubsection{Quality factor considerations for the pickup inductor temperature}
\label{sec:4K_pickup}

Note that~\eqref{eq:cost_HB} also completely neglects the $\sim$10~mK cooling stage.
Previous lumped-element experimental designs have assumed that the full resonator circuit, including the pickup inductor, is held near 10~mK.
This presents an immediate experimental challenge, since the pickup inductor is very large ($\sim$10~m) in our optimized designs.
On the other hand, as we point out here, it is likely not necessary for the pickup inductor to be so cold, except for the solenoidal topology.
Let us define $Q_i \equiv \omega_0 L_p / R_i$ for each loss channel in the resonator circuit, with $R_i$ the effective series resistance of that circuit element, which is assumed to be at temperature $T_i$.
The contribution that enters the scan rate in~\eqref{eq:scanrate} in the limit $n \gg 1$ is $(T / Q)^{-1} \equiv \left( \sum_i T_i / Q_i \right)^{-1} = \left( \sum_i T_i R_i / (\omega_0 L_p) \right)^{-1}$.
A central question is then whether, for the target $T  / Q = 10 \, \, {\rm mK} / 2 \times 10^7$ ($T  / Q = 30 \, \, {\rm mK} /  10^6$) in the optimistic and conservative benchmarks, discussed more below, the pickup inductor must be at 10~mK (30~mK) or can be warmer, such as at 4~K.

Let us consider the optimistic benchmark scenario first, since it is more stringent.
At 4~K the surface resistance of the pickup inductor, for the optimal toroid and core designs discussed later, needs to be less than roughly tens of p$\Omega$ to not degrade $(T/Q)^{-1}$ at hundreds of kHz; as discussed more below, this is theoretically possible for the toroid and core designs.
(In the conservative scenario, the resistance needs to be less than a few n$\Omega$, which is a much less stringent requirement.)
If the pickup needs to be at the lowest cooling stage, that could substantially modify the cooling cost estimate used in this work, which only accounts for the 4~K magnet.
This is likely the case for the solenoid design, since in that case the pickup is likely copper, or a more expensive material, and thus not superconducting.

It is worth considering more carefully the possible sources of loss in the superconducting pickup inductor for the toroid and core designs.
The effective series resistance, $R_p$, of the pickup is related to the surface resistance, $R_s$, by $R_p = R_s L_p / \ell_p$, where $\ell_p \equiv \int d^3x |{\bm B}_I|^2 / \oint dS\,|{\bm B}_{I}|^2 $.
Here, ${\bm B}_{I}$ is the magnetic field generated by a current $I$ flowing back through the pickup leads, and the surface integral is taken along all superconducting surfaces.
(Note that $\ell_p$ is proportional to $G$ as defined in Sec.~\ref{sec:heterodyne} in the context of SRF cavities.)
A first point to note is that $L_p / \ell_p$ is scale invariant, which means that if we increase the size of the detector, the series resistance of the pickup inductor $R_p$ stays fixed; for our optimal toroid and core geometries, $L_p / \ell_p \sim 2$ and $L_p / \ell_p \sim 0.5$, respectively.
On the other hand, since $Q_p = \omega_0 L_p / R_p$ and $L_p$ increases linearly with the overall size of the detector, bigger detectors increase the pickup quality factor with the linear scale size of the apparatus.

The surface resistance has, in principle, multiple contributions.
The  Bardeen--Cooper--Schrieffer (BCS) superconductor surface resistance of niobium is commonly modeled from the fit~\cite{Dhakal:2012iq}
 \es{}{
 R_{\rm BCS} \approx 0.2\, {\rm m}\Omega\,\left({1~{\rm K}\over T }\right)\,\left({f \over 1.5~{\rm GHz}}\right)^2 e^{- {17.67~{\rm K} \over T}} \,,
 }
with an additional logarithmic enhancement: the Mattis--Bardeen conductivity carries a factor $\log(4T/\omega)$, valid for $\omega \ll T$, as is the case here, which the fit holds at its 1.5~GHz value.
Restoring its frequency dependence raises $R_{\rm BCS}$ by a factor of 3 (2) at 120~kHz (1~MHz).
The physical intuition behind this formula is to think of the current as carried by two fluids: the Cooper pairs, which have no resistivity, and then thermally excited quasiparticles, which are not superconducting and contribute in the normal way to conductivity.
The exponential above is the Boltzmann suppression for exciting the quasiparticles.
The surface resistance thus evaluates to $\sim$5~f$\Omega$ (0.3~p$\Omega$) at 120~kHz (1~MHz) and 4.2~K.
We should approximately double these values to obtain the series resistance, since $L_p / \ell_p \sim 2$ for the toroid.
Since these resistances are less than the p$\Omega$ level, they are completely negligible even for our optimistic benchmarks.

What is left to worry about, then, are resistance contributions that do not go to zero as $T \to 0$.
These include non-idealized contributions from, {\it e.g.}, impurities, hydrides, lossy oxide, grain boundaries, and surface contamination, among other possible sources of loss.
There is no first-principles theory for these loss sources, and as far as we are aware their effects have never been measured at the target frequencies $\sim$100~kHz--1~MHz.
Measurements at $\sim$$80$~MHz and $\sim$2~K in SRF cavities show residual losses up to 5~n$\Omega$~\cite{Brown:2026cqz}, which is around 100 times higher than needed to not degrade the temperature-weighted quality factors in the optimistic scenario.
In the conservative scenario, however, this level of residual loss would be marginally acceptable.
Note, however, that the resistance is expected to drop substantially over the two orders of magnitude in frequency in going from 80~MHz to the MHz scale for the following reason.
Assume that the resistance comes from normal Ohmic loss within the penetration depth, with a frequency-independent conductivity and penetration depth.
The frequency dependence then arises from the induced electric field within the conductor, which scales as $E \propto \omega_0 B$.
The dissipated power goes as $E^2 \propto \omega_0^2$, giving the expectation $R_s \propto \omega_0^2$.
If this scaling holds, then going from 80~MHz to 1~MHz leads to a negligible contribution to the surface resistance from, {\it e.g.}, impurities, though measurements are needed to definitively resolve this question.

Trapped magnetic vortices could also, in principle, add sources of dissipation.
These vortices can enter if the pickup becomes superconducting when the magnetic field is already turned on and nonzero at the sheath surface, and they may also enter if the lower critical field $B_{c1}$ is exceeded anywhere after the sheath is superconducting.
On the other hand, we assume in all detectors below that the pickup inductors become superconducting {\it before} the magnetic field is ramped up and that the pickup inductors have no non-contractible curves, so that persistent currents are not induced when the magnet turns on.
(If persistent currents were induced, they would in fact cancel the signal.)
Note that for the toroid, this means that in addition to the azimuthal gap, which serves as the readout, there also needs to be a poloidal gap to stop currents from being generated during the magnetic ramp-up.

\subsubsection{Additional cost considerations}
\label{sec:add_cost}

Almost all of our optimal configurations have peak magnetic fields below $\sim$9~T, which is roughly the practical upper limit on field strength for magnets made from NbTi conductors at $\sim$4~K.
Our pricing model is based on such magnets.
Magnets made from Nb$_3$Sn can in principle go to $\sim$16~T, while higher fields require rare-earth barium copper oxide (REBCO) tape.
We do not consider REBCO tape magnets because such high fields are never preferred under our cost function, even discounting the extra cost associated with REBCO.
We also do not consider  Nb$_3$Sn magnets above $\sim$9~T with one exception.
The DMRadio-GUT proposal~\cite{Brouwer:2022bwo} used a toroidal topology with a benchmark peak field strength of around 16~T.

As a point of comparison, we contrast our optimal detectors, determined in this work, to the DMRadio-GUT benchmark with a 16~T peak field strength.
We assume that field strength is achieved with Nb$_3$Sn  conductors.
We use a very approximate cost increase for Nb$_3$Sn magnets relative to those at lower fields made from NbTi according to the following procedure.
First, we compute the required $I \times L_{\rm mag}$ to source the magnetic field, with $I$ the current flowing through the coils and $L_{\rm mag}$ the total length of conductor material in the coils.
Nb$_3$Sn costs approximately 30~CHF per kA$\cdot$m of critical current at 16~T and 4.2~K~\cite{Cooley:2005ize}.
We account for a possible factor-of-two increase in the amount of conductor to lower the current density from the critical current density for a safety margin; we treat this factor of two as a source of uncertainty in the price.
We also subtract at most 20\% of the magnet cost from the formula in~\eqref{eq:cost_HB}~\cite{Green:2008cost}, as that is associated with NbTi, which is replaced.

An additional consideration, emphasized already and which we return to later in this work, is that the resonant pickup for the solenoid is necessarily immersed in a large magnetic field, which makes it challenging to have a low-loss superconducting pickup without the use of, {\it e.g.}, REBCO, which we do not consider here.
The DMRadio-m$^3$ experiment, which plans to use a solenoid design, will have an Oxygen-Free High Conductivity (OFHC) copper pickup, which we estimate has a surface resistance $R_s \sim 100~\mu\Omega$ at tens of MHz, thus limiting the quality factor to $\sim 10^5$ and requiring the pickup to be ultracold ($\sim$20~mK) to not substantially degrade the noise~\cite{DMRadio:2022pkf}.

The core designs of Ref.~\cite{Ankel:2026zrv} use soft-iron segments, at room temperature in the GUT-scale design, to redirect part of the return flux of the magnet, so that the field on the niobium part of the pickup stays below $\approx 170$~mT, the lower critical field of niobium.
For such designs we add the cost of the iron, $c_{\rm Fe} M_{\rm Fe}$, to~\eqref{eq:cost_HB}, where $M_{\rm Fe}$ is the iron mass.
We take $c_{\rm Fe} \approx 5$~CHF/kg in 2026 Swiss francs~\cite{ILDConceptGroup:2020sfq}.
Because the iron is warm, it does not add to $A_{\rm cold}$.
When iron is present, $W_{\rm mag}$ in~\eqref{eq:cost_HB} is the stored energy $\int d^3x \int_0^{B} {\bm H}\cdot d{\bm B}$, which reduces to $\frac{1}{2}\int d^3x\, B^2$ outside the iron.

\subsection{Benchmark parameters}

\begin{table}[t]
\caption{Benchmark parameters.
The two benchmarks differ only in the quality factor $Q$, the loss-weighted temperature $T$, and the amplifier noise $\eta_A$.}
\label{tab:benchmarks}
\begin{ruledtabular}
\begin{tabular}{lccc}
Quantity & Optimistic & Conservative & Ref. \\
\hline
$Q$ (toroid, core) & $2\times10^7$ & $10^6$ & \cite{Brouwer:2022bwo,Ankel:2026zrv} \\
$Q$ (solenoid) & $2\times10^7$ & $10^5$ & \cite{Brouwer:2022bwo,DMRadio:2022pkf} \\
$T$ (loss-weighted) & 10~mK & 30~mK & \cite{Brouwer:2022bwo,DMRadio:2022pkf,Ankel:2026zrv} \\
$\eta_A$ & 0.1 & 10 & \cite{Brouwer:2022bwo,Ankel:2025nld} \\
\hline
$\rho_{\rm DM}$ & \multicolumn{2}{c}{0.3~GeV/cm$^3$} & \cite{Baxter:2021pqo} \\
$Q_{\rm axion}$ (SHM) & \multicolumn{2}{c}{$6.8\times10^5$} & \cite{Foster:2017hbq} \\
$\Theta_{95}$ & \multicolumn{2}{c}{2.71} & \cite{Foster:2017hbq} \\
$\gagg\, 2\pi f_a/\alpha$ (DFSZ) & \multicolumn{2}{c}{$8/3 - 1.77$} & \cite{Brandt:2026wir} \\
$m_a f_a$ & \multicolumn{2}{c}{$5.70\times10^{-3}$~GeV$^2$} & \cite{diCortona:2015ldu} \\
Scan range $\omega_0/2\pi$ & \multicolumn{2}{c}{120.9~kHz--1~MHz} & \\
Live time & \multicolumn{2}{c}{3~yr} & \\
\end{tabular}
\end{ruledtabular}
\end{table}

We consider the two benchmarks of Tab.~\ref{tab:benchmarks}, which differ only in $Q$, $T$, and $\eta_A$.
The optimistic benchmark takes the resonator and amplifier targets of the DMRadio-GUT baseline~\cite{Brouwer:2022bwo}, $Q = 2\times10^7$, $T = 10$~mK, and $\eta_A = 0.1$, for all topologies.
This $Q$ is an order of magnitude above the best demonstrated lumped-element superconducting resonator~\cite{Kolevatov:2025lzx}, and for the solenoid, whose pickup sits in the static field, it is far beyond what could reasonably be reached with copper; we use this benchmark only to compare the topologies at equal noise.
Reaching $\eta_A = 0.1$ requires going beyond the SQL, {\it e.g.}, with RF quantum upconverters, which can implement backaction-evading and squeezing protocols~\cite{Kuenstner:2022gyc}.
Following Ref.~\cite{Brouwer:2022bwo}, we use $\eta_A = 0.1$ in~\eqref{eq:Gbar} as a phenomenological target.
The conservative benchmark values, on the other hand, match or are closer to demonstrated technology.
These values are meant to indicate what could be accomplished today, with minimal R\&D.
We take $T = 30$~mK and $\eta_A = 10$, which is the noise predicted for dc SQUID amplifier chains built from demonstrated components below $\sim$50~MHz~\cite{Ankel:2025nld}.
The sheath of the toroid lies outside the confined static field and can be superconducting throughout.
In our core topology design the pickup box lies outside the coil and is superconducting, with the field on its surface held below $\sim$100~mT, below the lower critical field of niobium at 4~K (in the core design of Ref.~\cite{Ankel:2026zrv} the section of the pickup crossing the magnet bore is instead a normal metal).
For both we take $Q = 10^6$, which is comparable to the $Q \approx 2\times10^6$ demonstrated by a fixed-frequency superconducting resonator with a $\sim$1-liter inductor at 250~kHz~\cite{Kolevatov:2025lzx}.
The pickup of the solenoid is immersed in a large, static field and likely requires a non-superconducting copper pickup, and so for this topology we take $Q = 10^5$, which matches the DMRadio-m$^3$ target at 5--200~MHz~\cite{DMRadio:2022pkf}.

The remaining parameters are common to both benchmarks.
We assume $\rho_{\rm DM} = 0.3$~GeV/cm$^3$~\cite{Baxter:2021pqo} for the local DM energy density, since determinations of this quantity span roughly 0.3--0.6~GeV/cm$^3$~\cite{deSalas:2020hbh} and taking a conservative value is important to make sure we do not miss a potential signal.
For the  QCD contribution to $\gagg$ we take the recent lattice value 1.77(8)~\cite{Brandt:2026wir}, which is important because it cancels less of the DFSZ value $E/N = 8/3$ than the chiral perturbation theory value 1.92(4)~\cite{diCortona:2015ldu}.\footnote{Note, however, that an independent lattice calculation finds 1.965(35)~\cite{Lin:2026hbw}.}
We require that the scan cover $m_a = 0.5$--$4.1$~neV at the DFSZ-strength coupling in 3~years of live time.
We assume DFSZ-strength couplings because this is the coupling predicted in theories that undergo grand unification~\cite{Agrawal:2022lsp}, including in the string theory constructions that motivate this search.

\section{Analytic lumped-element examples}
\label{sec:long}

We now work out the three topologies of Fig.~\ref{fig:topologies} analytically in the long limit, $H \gg R$, where the ends of the detectors can be neglected.
First, note that recalling the definition of $\Psi$ in~\eqref{eq:flux}, we may integrate the scan rate formula between a lower frequency $\omega_1$ and an upper frequency $\omega_2$, assuming that we maintain QCD axion sensitivity, for which $g_{a\gamma\gamma} \propto m_a$, to arrive at the requirement on $\Psi$:
\es{}{
{\Psi^4 \over L_p^2} > {\mathcal F}_{\rm req} \,,
}
where we define
\es{eq:Freq_req}{
{\mathcal F}_{\rm req} \equiv \frac{3\sqrt{3}}{\pi^2}\,\frac{\Theta_{95}\,\eta_A T}{Q_{\rm axion}\, Q\, \rho_{\rm DM}^2\, \gagg^4(m_1)\, t}\left[1 - \frac{\omega_1^4}{\omega_2^4}\right] ,
}
with $t$ the total observing time.
Throughout this work we assume the scan runs from $\omega_1 = m_1 = 0.5$~neV to $\omega_2 = 4.1$~neV, with $\gagg(m_1) = 9.1\times10^{-20}$~GeV$^{-1}$ the DFSZ coupling at the lowest mass, which dominates the scan time.
For the benchmarks of Tab.~\ref{tab:benchmarks},
\es{eq:Freq_num}{
{\mathcal F}_{\rm req} \approx
\begin{cases}
1.04\times10^{4}~{\rm T^4\,m^{10}} & \text{optimistic}, \\
6.26\times10^{7}~{\rm T^4\,m^{10}} & \text{cons., toroid/core}, \\
6.26\times10^{8}~{\rm T^4\,m^{10}} & \text{cons., solenoid} \,.
\end{cases}
}
As we show below, for each of the three long detectors we find
\es{eq:g_def}{
\frac{\Psi^2}{L_p} = g\, W_{\rm mag} R^2 \,,
}
where $g$ is a dimensionless geometric factor.
The requirement therefore fixes $W_{\rm mag} = \sqrt{{\mathcal F}_{\rm req}}/(g R^2)$.

\subsection{Long solenoid}
\label{sec:long_solenoid}

For the long solenoid with a coaxial pickup of inner radius $xR$, discussed above, $\Psi_{\rm solenoid} = HR^2 B_{\rm lab}(1-x^2)/4$, $L_p = (H/2\pi)\log(1/x)$, and $W_{\rm mag} = \frac{1}{2}\pi R^2 H B_{\rm lab}^2$, so that~\eqref{eq:g_def} holds with
\es{eq:g_solenoid}{
g_{\rm sol} = \frac{(1-x^2)^2}{4\log(1/x)} \,.
}
The quantity $g_{\rm sol}$ is maximized at $x_{\rm max} \approx 0.534$, with the maximum value $g_{\rm sol} \approx 0.204$.

Let us consider first the magnetic cost alone, which is given by the first term in~\eqref{eq:cost_HB}, which we define as ${\mathcal C}_{\rm mag} = a\,(W_{\rm mag}/1\,{\rm MJ})^{0.7}$.
This evaluates to ${\mathcal C}_{\rm mag} \approx 17~{\rm MCHF}\,(1\,{\rm m}/R)^{1.4}$ ($\approx 790~{\rm MCHF}\,(1\,{\rm m}/R)^{1.4}$) for the optimistic (conservative) benchmark, independent of $H$.
The magnetic field within the solenoid, on the other hand, is given by $B_{\rm lab} \approx 10 \, \, {\rm T} (1 \, \, {\rm m} / R)^2 \sqrt{3 \, \, {\rm m} / H}$ ($B_{\rm lab} \approx 10 \, \, {\rm T} (3 \, \, {\rm m} / R)^2 \sqrt{9 \, \, {\rm m} / H}$) for the optimistic (conservative) case.

Adding in the cryogenic costs, we assume that the cold area, in the large $H/R$ limit, is dominated by the outer coax wall at radius $R$, with $A_{\rm cold} \approx 2 \pi R (H + R)$, also including the two end caps.
(Note that we assume the inner bore is cold so that no additional cost is associated with cooling the inner coax wall at radius $x R$.)
At fixed $H/R$, the cost necessary to reach ${\mathcal F}_{\rm req}$ is obtained by fixing $W_{\rm mag} = \sqrt{{\mathcal F}_{\rm req}} / (g R^2)$, with $g\approx 0.204$ fixed, and minimizing ${\mathcal C}$ with respect to $R$.
This yields the optimal radius
\es{eq:R_star}{
R_\star^3 = \frac{7}{8}\, \frac{\left[W_{\rm mag}R^2/(1\,{\rm MJ\,m^2})\right]^{0.7}}{\kappa\,[2\pi(1+H/R)]^{0.8}}~{\rm m}^3 \,,
}
where $W_{\rm mag}R^2 = \sqrt{{\mathcal F}_{\rm req}}/g$ is 399~MJ\,m$^2$ ($9.8\times10^4$~MJ\,m$^2$) for the optimistic (conservative) benchmark. %
For the optimistic benchmark, $H/R = 10$ gives $R_\star \approx 1.2$~m, $B_{\rm lab} \approx 3.7$~T, and ${\mathcal C} \approx 25$~MCHF, and $H/R = 5$ gives $R_\star \approx 1.4$~m and ${\mathcal C} \approx 20$~MCHF.
For the conservative benchmark, $H/R = 10$ (5) gives $R_\star \approx 4.3$ (5.0)~m and ${\mathcal C} \approx 190$ (150)~MCHF.
The cost keeps falling as $H/R$ decreases, which pushes the optimum toward $H \sim R$, where the long-solenoid expressions fail.
In the next section we numerically optimize the cost without assuming the limit $H \gg R$.

\subsection{Long toroid}
\label{sec:long_toroid}

The toroid detector is assumed to be described by the following geometry, in cylindrical coordinates with the $\bm{\hat z}$ direction aligned with the toroid's symmetry axis, $r = 0$ at the center (the center of the donut hole), and $z = 0$ the midplane, so that the toroid extends in the $z$-direction between $z = -H/2$ and $z = + H/2$.
The toroid cross-section is rectangular, extending in radius between $y R < r < R$ and $-H/2 < z < H/2$.
The magnetic field is assumed to be ${\bm B}_{\rm lab} = B_{\rm in}\,(yR/r)\,\hat{\bm \phi}$, where $B_{\rm in}$ is the field at the inner radius, within the toroid's volume, and it is zero everywhere else.
The axion-induced effective current ${\bm J}_{\rm eff} = \kappa_a {\bm B}_{\rm lab}\cos m_a t$ is thus also in the $\bm{\hat \phi}$ direction.
In this section we again work in the limit $H \gg R$.
In this limit, and assuming for a moment no shielding around the magnet, the strength of the axion-induced magnetic field in the center of the toroid is given by $B_a = \kappa_a B_{\rm in}\, yR \log(1/y)$.
The field points in the $\bm{\hat z}$ direction and oscillates at frequency $m_a$.

The pickup sheath is assumed to be a hermetic enclosure of the magnet that tightly surrounds the magnet on all sides.
The sheath is gapped in the azimuthal ($\phi$) direction at a fixed $\phi$ value, corresponding to slicing the sheath in the $z$ direction at a fixed $\phi$ value.
(In reality, the two sides of the sheath may overlap in $\phi$, but the gap must be maintained.)
The gap is assumed to be bridged by a wire that runs to a capacitor and is then inductively coupled to the readout circuit (see Sec.~\ref{sec:realistic_toroid}).
In the limit $H \gg R$ the signal strength coupling $\Psi_{\rm toroid}$, inductance $L_p$, and magnetic energy $W_{\rm mag}$ are analytically approximated by
\es{eq:toroid_long}{
\Psi_{\rm toroid} &= \pi (yR)^3 B_{\rm in} \log(1/y) \,, \qquad L_p = \frac{\pi (yR)^2}{H} \,, \\
W_{\rm mag} &= \pi H (yR)^2 B_{\rm in}^2 \log(1/y) \,.
}
It follows that~\eqref{eq:g_def} holds with
\es{eq:g_toroid}{
g_{\rm tor} = y^2 \log(1/y) \,.
}
The quantity $g_{\rm tor}$ is maximized at $y = e^{-1/2} \approx 0.61$, where $g_{\rm tor} = 1/(2e) \approx 0.184$.
Then
\es{eq:W_toroid}{
W_{\rm mag} = \frac{2e\sqrt{{\mathcal F}_{\rm req}}}{R^2} \,, \qquad B_{\rm in}^2 = \frac{4 e^2 \sqrt{{\mathcal F}_{\rm req}}}{\pi H R^4} \,.
}
These expressions evaluate to $W_{\rm mag} = 442~{\rm MJ}\,(1\,{\rm m}/R)^2$ and $B_{\rm in} \approx 30~{\rm T\,m^{5/2}}/\sqrt{H R^4}$ for the optimistic benchmark, and $W_{\rm mag} = 3.4\times10^4~{\rm MJ}\,(1\,{\rm m}/R)^2$ and $B_{\rm in} \approx 272~{\rm T\,m^{5/2}}/\sqrt{ HR^4}$ for the conservative case.

The magnetic cost is \mbox{${\mathcal C}_{\rm mag} \approx 18~{\rm MCHF}\,(1\,{\rm m}/R)^{1.4}$} for the optimistic benchmark and $\approx$$380~{\rm MCHF}\,(1\,{\rm m}/R)^{1.4}$ for the conservative scenario.
Just as for the solenoid, these expressions become smaller at large $R$.
To regulate this runaway, we include the cryogenic cost function using~\eqref{eq:cost_HB}.
We assume the cold surface is $A_{\rm cold} = 2\pi R(R+H)$, which is dominated by the outer cylinder in the limit $H \gg R$.
Minimizing the full cost functional yields the optimal radius $R_\star \approx 1.2$~m ($R_\star \approx 1.4$~m) for $H/R = 10$ ($H/R = 5$), with $B_{\rm in} \approx 6.0$~T ($B_{\rm in} \approx 5.7$~T) for the optimistic benchmark.
In the conservative benchmark scenario, the costs rise to ${\mathcal C} \approx 130$ (100)~MCHF for $H/R = 10$ ($H/R = 5$).
As for the solenoid, the smaller the ratio $H/R$, the cheaper the design.
This again pushes us to consider, numerically, the small $H/R$ regime, which we do in the next section.

As a side note, the hermetic sheath pickup, with a slit at fixed $\phi$, is the optimal pickup for which the axion-induced magnetic flux is completely non-overlapping with the laboratory magnetic volume itself.
This pickup does not, however, achieve the optimal value of $g$.
For example, consider the pickup that consists of an outer cylinder at radius $R$ with height $H$ that has an azimuthal slit.
This pickup achieves $g \approx 0.204$, which is slightly better than the hermetic sheath.
The issue, however, is that such a pickup would be more susceptible to noise.
By the fluctuation-dissipation theorem, an object $X$ adds noise to a readout circuit proportional to $\int_X d^3x B_I^2$, where $B_I$ is the magnetic field generated by a current $I$ flowing through the readout circuit and $\int_X$ is taken over the volume of the object $X$ in question.
In the case of the hermetic sheath, $\int_X d^3x B_I^2 = 0$ identically, since the current $I$ flows across the inner wall of the pickup sheath.
In the case where the pickup is simply a cylinder at $r = R$, however, the quantity  $\int_X d^3x B_I^2 > 0$ since the region interior to the toroid contributes, which means that, {\it e.g.}, thermal and other noise from the magnet itself would contribute to the pickup circuit.
Note also that in the large $H$ limit, the hermetic sheath pickup performs identically to the pickup cylinder located at $r = y R$.

\subsection{Long core}
\label{sec:long_core}

The idealized core detector is illustrated in the rightmost panel of Fig.~\ref{fig:topologies}.
The detector consists of a solenoidal magnetic field through the central bore, with radius $y R$ and height $H$.
The magnetic field region is surrounded by a superconducting cylindrical box that has azimuthal symmetry and extends in radius from $y R$ to $R$ and has height $H$.
The axion-induced oscillating magnetic field, which is azimuthal, induces a poloidal current on the surface of the box.
For the moment, we ignore the fact that the box is immersed in a magnetic field, which in reality would prevent the box from becoming superconducting at the large field values preferred by the optimal geometry.
In Sec.~\ref{sec:numerics}, we include this constraint and use iron to route the field around the box.
Ignoring this point for the moment, the quantities of interest evaluate to
\es{eq:core_long}{
\Psi_{\rm core} &= \frac{H}{2}\,(yR)^2 B_{\rm lab} \log(1/y) \,, \qquad L_p = \frac{H}{2\pi}\log(1/y) \,, \\
W_{\rm mag} &= \frac{\pi}{2}\, H (yR)^2 B_{\rm lab}^2 \,,
}
where $B_{\rm lab}$ is the magnitude of the field strength through the central solenoid.
Interestingly, this implies that~\eqref{eq:g_def} holds with $g_{\rm core} = y^2\log(1/y)$, which is the same quantity found for the toroid.
This is not a coincidence, as the toroid and core designs are dual to each other in a precise sense described in the next section.

\section{Lumped-element topology dualities}
\label{sec:duality}

We saw in Sec.~\ref{sec:long_core} that in the large $H/R$ limit the toroid and core topologies give identical scan rates at fixed $W_{\rm mag}$.
This duality between the toroid and core detectors actually holds at any $H/R$, which we proceed to demonstrate here.
In this treatment, we neglect practical considerations related to the magnetic field requirement at the surface of the superconducting box in the case of the core detector, which we deal with in the next section.

\subsection{Toroid--core duality}

First, note that the toroid and core detectors are topologically and geometrically identical, except that the laboratory and axion-induced fields are interchanged.
Let us consider this geometry.
There are two distinct inductances that may be defined.
We call $L_{\rm box}$ the inductance corresponding to magnetic fields going through the interior of the box, with $\bm{B} \propto \bm{\hat \phi} /r$ identically due to the boundary conditions of the box.
We call $L_{\rm out}$ the inductance corresponding to magnetic fields that pierce the center of the interior hole in the $\bm{\hat z}$ direction.
Due to the boundary conditions within the superconducting box, $L_{\rm box} = (H/2 \pi) \log(1/y)$ identically for any $H/R$.
In the limit $H \gg R$, we may calculate $L_{\rm out} = \pi (y R)^2/H$; at finite $H/R$, it follows by Cauchy--Schwarz and the energy of the return flux that $L_{\rm out} < \pi (y R)^2 /H$ (see App.~\ref{sec:optimality}).
For the toroid, $L_p = L_{\rm out}$ while $L_p = L_{\rm box}$ for the core.
The key point, which we show below, is that for both the toroid and core the final scan rate only depends on $L_{\rm box} \times L_{\rm out}$.

Let us consider first the toroid.
The field $B_I$ that is created by the current $I$ flowing through the pickup circuit lies entirely outside of the box interior.
It thus follows that the vector potential generated by the current $I$ inside of the box is $\bm{A}_I = (L_p I)/(2 \pi r) \bm{\hat \phi}$.
Then, using the expression $\Psi = (1/I)\int d^3x\, \bm{A}_I \cdot \bm{B}_{\rm lab}$, it follows that $\Psi = L_p \Phi_{\rm lab}$, where $\Phi_{\rm lab}$ is the flux from the laboratory magnetic field through a vertical slice of the box interior.
A similar calculation for the core geometry leads to the same result, $\Psi = L_p \Phi_{\rm lab}$, with $\Phi_{\rm lab}$ in this case the flux through the center of the hole due to the laboratory magnetic field in the $\bm{\hat z}$ direction.

Let us define $L_{\rm static}$ to be the inductance in the current-free region of static magnetic field: $L_{\rm static} = L_{\rm box}$ for the toroid detector and $L_{\rm static} = L_{\rm out}$ for the core.
It follows from the Cauchy--Schwarz inequality that
\es{}{
W_{\rm mag} \geq {\Phi_{\rm lab}^2 \over 2 L_{\rm static}} \,,
}
with the equality holding if the region of laboratory magnetic field is current free; see App.~\ref{sec:optimality} for details.
This assumption holds true in all of the idealized detectors, since the source currents are located on the boundaries.
The scan rate scales as $(\Psi^2 / L_p)^2$, and substituting the relations above we find $\Psi^2 / L_p \leq 2 L_p L_{\rm static} W_{\rm mag}$, with the inequality saturated for both idealized detectors.
This tells us that for both the toroid and core detectors $d \omega_0 / dt \propto W_{\rm mag}^2 L_{\rm box}^2 L_{\rm out}^2$.
Note that the cold area is equivalent for both detectors, since they share the same geometry, so the total cost is equivalent.
Thus, the idealized toroid and core detectors are formally dual to each other at finite $H/R$.
In the next section, however, we show that the requirement that the magnetic field be small enough at the surface of the superconducting box is a severe restriction for the core at small sizes, which breaks the duality, while the toroid requires an additional superconducting shield.

In App.~\ref{sec:optimality}, we show that the idealized toroid and core are the optimal detectors among long detectors whose laboratory and pickup fields do not overlap (and do not vary along the $\hat{\bm{z}}$ direction), including geometries beyond the box-and-hole one considered above.
We also find that allowing the fields to overlap can reduce the cost by at most a few tens of percent, at the price of placing the pickup in the magnet's static field; in this case, the optimal laboratory magnetic field is a force-free helical configuration.

\subsection{Toroid--solenoid duality}

We may derive a similar duality between the toroid and solenoid detector topologies, but only in the case where the toroid pickup is an outer cylinder at $r = R$.
As discussed in Sec.~\ref{sec:long_toroid}, this pickup topology has a slightly larger coupling factor $g$ in the large $H/R$ limit than the hermetic sheath pickup, but it is more susceptible to noise.

To see this duality, note that integrating by parts gives $\int d^3x \, {\bm A}_I \cdot {\bm B}_{\rm lab} = \int d^3x \, {\bm A}_{\rm lab} \cdot {\bm B}_I$, so that
    \es{eq:g_symmetric}{
    {\Psi^2 \over L_p W_{\rm mag}} = {2 \left( \int d^3x \, {\bm A}_I \cdot {\bm B}_{\rm lab} \right)^2 \over \int d^3x \, |{\bm
    B}_I|^2 \, \int d^3x \, |{\bm B}_{\rm lab}|^2} \,.
    }
This expression is symmetric under the interchange of the laboratory magnetic field and the magnetic field ${\bm B}_I$ generated by the current $I$ flowing through the pickup.
Any pair of detectors that are obtained by swapping field profiles at fixed $W_{\rm mag}$ thus have the same scan rate.
In fact, the toroid--core duality above is a special case of this symmetry, since, {\it e.g.}, the core's laboratory field has the same shape as the toroid's pickup field.
For the solenoid, the laboratory field generated by the winding at $r = R$ has, in the limit $H \gg R$, the same shape as the field generated by a current flowing azimuthally around the toroid's outer cylinder pickup, while the solenoid's coax pickup field, ${\bm B}_I = I/(2\pi r) \, \bm{\hat \phi}$ for $xR < r < R$, has the same shape as the toroid's laboratory field.
It follows that the solenoid and the toroid with the outer cylinder pickup have identical $g$ in the limit $H \gg R$, given by~\eqref{eq:g_solenoid} (with $x \to y$).
At finite $H/R$ the two differ, because the current on a superconducting cylinder is not uniform along its height, while that in a solenoid winding is.

Note, however, that this duality does not extend to the noise.
In the toroid with the outer cylinder pickup, ${\bm B}_I$ fills the box and thus overlaps the magnet winding, so that $\int_X d^3x \, B_I^2 > 0$ for the winding.
In the solenoid case, on the other hand, ${\bm B}_I$ is confined between the coax conductors and the winding lies outside, so $\int_X d^3x \, B_I^2 = 0$ for the winding, just as for the hermetic sheath.
In the solenoid case the laboratory field fills the pickup region.
This does not add noise, but it means that the coax walls sit in the full static field and cannot be conventional flux-excluding superconductors, which limits the pickup quality factor $Q$ for more realistic designs and requires the pickup to be held at a low temperature.

\section{Solenoid, toroid, and core at finite size}
\label{sec:numerics}

We now optimize the three detector topologies at finite $H/R$.
We require the scanning coverage described in Sec.~\ref{sec:scan_cost} for the optimistic and conservative benchmarks, and for this scanning coverage we find the geometry, at fixed topology, that minimizes the cost function in~\eqref{eq:cost_HB}.
We compare the solenoid, the core, and three versions of the toroid: unshielded, inside a superconducting shield, and with an iron core inside the shield.
Note that for the core detector we add the additional constraint that the surface field along the superconducting box be below 100~mT.
The pickup box sits on the 4~K stage, where the lower critical field of niobium is about 140~mT.
We use 100~mT here and in the realistic toroid in Sec.~\ref{sec:realistic_toroid} to keep a safety margin.
We allow for iron surrounding the core pickup box to help achieve this requirement, with the cost of warm iron added to the cost function as described in Sec.~\ref{sec:scan_cost}.
Readers interested mainly in the results can skip to Sec.~\ref{sec:numerics_results}.

\subsection{Numerical methods}
\label{sec:numerics_method}

At fixed $H/R$ and for a fixed topology, we solve for the scanning rate using the code package \texttt{axionMultiPhysics}~\href{https://github.com/bsafdi/axionMultiPhysics}{\faGithub}~\cite{axionMultiPhysics}, which is an open-source finite-element code for lumped-element axion detectors that we release with this work.
The code is written in C++ and built on the MFEM library~\cite{mfem}, with a Python interface.
The code allows for both axisymmetric and fully 3D detector simulations, though axisymmetric simulations suffice for this section.
\texttt{axionMultiPhysics} solves the MQS equations of motion (see App.~\ref{app:axionmultiphysics}) in terms of the magnetic vector potential; for the axisymmetric calculations used here, the poloidal magnetic field is described by the flux function $\psi = rA_\phi$.
The code also has a full-wave solver, but the optimization here uses the MQS approximation.
For each design we compute the laboratory field $\bm B_{\rm lab}$ and the pickup field $\bm B_I$, which is generated by a current $I$ injected across the readout port.
We assume that $\psi$ is constant on each superconducting surface.
Iron follows the nonlinear $B$--$H$ law described more below.
On a sphere enclosing the detector, the solution is matched to the vacuum field extending to infinity, so that no artificial outer boundary is needed.
The code outputs give $\Psi$ in~\eqref{eq:flux}, $L_p$ in~\eqref{eq:Lp}, and $W_{\rm mag}$, which together with the cold area $A_{\rm cold}$ determine the scan rate and the cost~\eqref{eq:cost_HB}.
We also record the peak laboratory field and, for the core, the maximum field on the superconducting box.

\subsection{Toroid and solenoid scans}
\label{sec:numerics_toroid}

Our scans for the toroid and solenoid differ distinctly from those for the core.
We begin by describing our toroid and solenoid analyses.
We scan over geometries using the global optimization differential evolution algorithm implemented in \texttt{SciPy}~\cite{Storn:1997,Virtanen:2019joe}.
The loss function for the minimizer is taken to be the cost of the design, with the scan rate requirement ${\mathcal F} = {\mathcal F}_{\rm req}$.
We scan over the overall size, the ratio $H/R$, and the inner-to-outer radius ratio $y$; for any choice of geometry, we may then solve for the peak $B$ field that minimally achieves the scan-rate requirement.
(For the solenoid we denote the inner-to-outer radius ratio of the coaxial pickup by $x$, as in Sec.~\ref{sec:long_solenoid}.)
We also perform scans in which the three gaps between the toroid winding and its superconducting sheath (on the inner and outer faces and at the top and bottom) and, for the solenoid, the height of the pickup relative to that of the winding are free parameters.
However, the optimizer drives the gaps to their lower bounds, so that the sheath essentially lies on the winding, and makes the pickup and the winding equally tall.
We therefore fix the gaps at these bounds and the pickup height to the winding height.

In the toroid case, the one numerical computation needed at a given geometry is the computation of the pickup inductance $L_p$.
We compute the pickup inductance with \texttt{axionMultiPhysics} (see App.~\ref{app:axionmultiphysics}).
We assume axisymmetry, to simplify the computation of $L_p$; in practice, this means assuming the vertical readout slit is much smaller than the radius $R$.
To compute $L_p$, we inject a current along the readout ports, and we solve for the unique surface-current and magnetic field profile that obeys the superconducting boundary conditions on the sheath.
\texttt{axionMultiPhysics} then meshes space in the $r$-$z$ directions and solves the MQS equations for the magnetic field (in practice, for the flux function $\psi = rA_\phi$), subject to the superconducting boundary condition that $\psi$ is constant along the sheath.

We also consider two variants of the toroid.
In the first, the toroid is placed inside a closed superconducting cylinder (a shield), which confines the return flux of the pickup mode to the apparatus, as is automatically the case for the solenoid and core pickups.
Without the shield, the pickup mode of the toroid is a magnetic dipole that couples to, and radiates into, the environment.
The shield is then the cold envelope and we add it to the cryogenic cost.
We additionally scan over the radial and vertical gaps between the sheath and the shield, whose optimal values are 20--30\% of the sheath radius.
We account for the shield in computing $L_p$ as well; $\psi$ is constant on both the sheath and the shield.
In the second variant, motivated by the SHAFT experiment~\cite{Gramolin:2020ict}, the region enclosed by the winding of the shielded toroid is filled with a magnetic core with the $B$--$H$ law of soft iron (see below).
For an azimuthally symmetric winding, Amp\`ere's law fixes $H = NI/(2\pi r)$ inside the core regardless of its permeability.
The core therefore adds roughly its saturation magnetization, about 2~T, to the field that sources the axion signal, while the stored magnetic energy grows only slightly, since $H$ is fixed by the winding current.
At fixed geometry, $L_p$ is unaffected by the core, since the sheath keeps the pickup field out of the region enclosed by the winding.
We idealize the core as electrically insulating, though this may be difficult to achieve in practice.
The core should be cold, and though we add its cost $c_{\rm Fe}M_{\rm Fe}$, we do not add an additional cost term for its cooling.

For the solenoid geometry, on the other hand, the numerical difficulty is reversed.
In this case, the pickup inductance $L_p$ is analytically computed, but the static laboratory magnetic field must be computed numerically at finite $H/R$.
We compute this again using \texttt{axionMultiPhysics}, which again solves for the flux function $\psi$ on a finite-element mesh in the $r$-$z$ plane with an open boundary condition.
From this field profile, we may compute the stored energy, the peak magnetic field, and the coupling $\Psi$.

Two additional choices are made for the toroid and solenoid computations, though neither choice strongly affects the physical results.
First, we assume that the primary magnetic field is generated by a current sheet of thickness $10^{-3}$ times the inner winding radius for the toroid, with or without a shield, and times the bore radius for the solenoid.
In practice, this thickness will be set by current-density requirements, but we find it has a negligible effect.
Second, the corners of the toroidal winding are rounded with a radius of at least $10^{-3}$ times the smaller of the winding's radial width and height (about 1~mm at the optimistic optimum, where this bound is reached); the sheath and shield and the solenoid are modeled with sharp corners, whose associated field singularities are integrable, so $L_p$, $W_{\rm mag}$, and $\Psi$ remain finite even in the sharp-corner limit.
Note that we perform scans treating the rounding radii as free parameters but find they prefer their smallest allowed values.

\subsection{Core scans}
\label{sec:numerics_core}

The core scans require more numerical computations because of (i) the field requirement at the superconducting box, which must be checked numerically, and (ii) the use of iron, which causes the computation of the magnetic field profile to become nonlinear.
The solenoidal magnetic field in the central bore is generated by a coil, which has a total height $H_{\rm coil}$ and radius $R_{\rm coil}$.
(The coil is approximated to be a thin current sheet of thickness $10^{-3}R_{\rm coil}$, as for the toroid and solenoid, for simplicity, but accounting for realistic thicknesses only has a percent-level effect on our results.)
The superconducting, toroidal box that serves as the pickup is taken to have a gap of width $c$ relative to the coil outer edge.
The box is centered in the $z$-direction on the coil and extends symmetrically in height to $H_{\rm box}$ and has width $w_{\rm box}$.
The box has a 5~mm niobium wall and is read out through an ideal, thin slit running around its outer wall.
The edges of the box are allowed to be rounded with curvature radius $t_{\rm edge}$, with $t_{\rm edge}$ between 0.02 and 0.45 times the minimum of $w_{\rm box}$ and $H_{\rm box}$.
The edges are allowed to be rounded to help keep the local field values at the superconducting box from crossing the critical field strength.
Warm iron is allowed to surround the toroidal box but has a fixed 10~cm offset, both radially and vertically, from the cold components; doubling this offset to 20~cm raises the cost by 10\% (5\%) in the optimistic (conservative) case.

The iron thickness varies with radius like $1/r$, and the overall iron thickness is treated as a nuisance parameter.
The iron consists of a cylindrical shell and two end caps; the end-cap thickness is $qR_{\rm coil}^2/r$ (held constant for $r < 0.3R_{\rm coil}$), and the shell thickness is $qR_{\rm coil}^2/R_s$, with $R_s$ the inner radius of the shell and $q$ the overall iron-thickness parameter.
We allow for a central hole in the iron at the two end caps of radius $R_{\rm hole}$.
We take the iron thickness to fall as $1/r$ so that, approximately, the magnetic field entering the iron is constant in magnitude as a function of $r$.
We scan over all of these geometric and iron parameters: six without iron, including the overall size, and eight with iron.
Without iron and without the field constraint on the box, the core reduces to the idealized detector of Sec.~\ref{sec:long_core}, which by the toroid--core duality of Sec.~\ref{sec:duality} has the same optimal cost as the unshielded toroid.
The coil current is not scanned.
Instead, at each geometry we solve for the current at which $\mathcal{F} = \mathcal{F}_{\rm req}$, which for designs without iron is a simple rescaling, since then $\mathcal{F} \propto I^4$.

In \texttt{axionMultiPhysics} we treat the surface of the box as a perfect superconductor ($\psi = {\rm const.}$), and only after solving for the field profile do we check if the field at the surface is low enough to be superconducting.
The magnetic field is solved through the iron regions using a $B$--$H$ law, which is the relation between the magnetic field strength and magnetic flux density; this relation is nonlinear for iron and is modeled with a ``knee'' profile, with a relative permeability of 1000 up to $B = 2$~T and the differential permeability of vacuum at higher field values [\eqref{eq:axionmultiphysics_knee}].
This makes the full MQS equations of motion nonlinear because, {\it e.g.}, two solutions to the equations of motion no longer superimpose due to the nonlinear $B$--$H$ law.
\texttt{axionMultiPhysics} solves for $\psi$ by Newton iteration on the magnetostatic energy functional.

\subsection{Optimal designs}
\label{sec:numerics_results}

  \begin{figure*}[t]
    \centering
    \includegraphics[width=\textwidth]{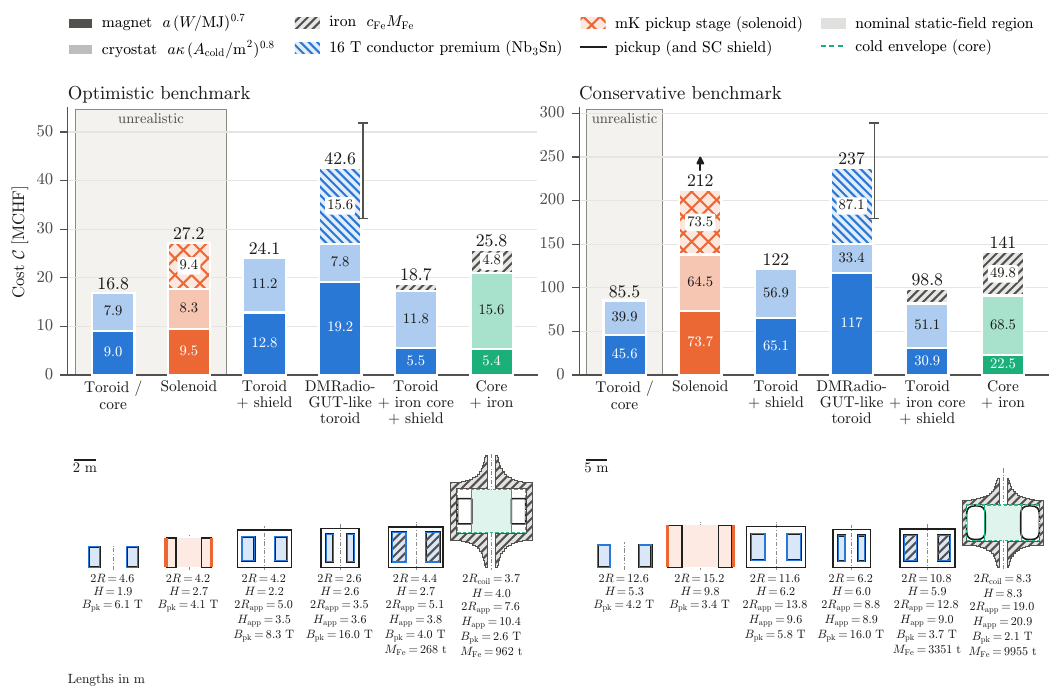}
    \caption{(Top panels) The minimum costs ${\mathcal C}$ for each detector topology, varying the geometry, magnetic field strength, and, where applicable, the amount of iron.
We show this for the optimistic benchmark scenario (left) and the conservative benchmark scenario (right).
For each of these optima, we show the dimensions of the magnet and the peak laboratory magnetic field values in the bottom panels.
The designs in the gray boxes are not realistic: the unshielded toroid pickup couples to surrounding conductors and external magnetic noise, while the idealized core omits the critical-field constraint on its superconducting pickup.
The two designs have identical costs; only the toroid is drawn.
As the pickup of the solenoid sits in the static field, it likely requires a non-superconducting copper pickup, which cannot realize the optimistic quality factor.
We indicate a rough lower bound on the cost of cooling the copper pickup to millikelvin temperatures in the cross-hatched segments, with the arrow emphasizing that the conservative solenoid cost is a lower bound.
We also show the DMRadio-GUT-like toroid (16~T peak field, with the shape of Ref.~\cite{Brouwer:2022bwo}) with a shield added, scaled to the optimal size (see text); the hatched segment and associated error bar are the approximate additional cost of the Nb$_3$Sn winding needed at 16~T.
(Note that $R_{\rm app}$ and $H_{\rm app}$ refer to the outer radius and height of the apparatus, including the shield or the iron envelope.)}
    \label{fig:numerics_optima}
\end{figure*}

\begin{table*}[t]
\caption{Numerical values for the geometric and magnetic parameters for the cost-optimal detectors subject to the constraints of achieving the required QCD axion sensitivity for the optimistic and conservative benchmarks (see Fig.~\ref{fig:numerics_optima} for a graphical illustration).
For the core design, we impose the additional constraint that the field on the niobium box not exceed 100~mT.}
\label{tab:numerics}
\begin{ruledtabular}
\begin{tabular}{lcccccccccccc}
& $2R$ & $H$ & $r_{\rm in}/r_{\rm out}$ & $B_0$ & $B_{\rm pk}$ & $W_{\rm mag}$ & $A_{\rm cold}$ & $M_{\rm Fe}$ & \multicolumn{4}{c}{$\mathcal{C}$ [MCHF]} \\
& [m] & [m] & & [T] & [T] & [MJ] & [m$^2$] & [t] & magnet & cryostat & iron & total \\
\hline
\multicolumn{13}{l}{\it Optimistic benchmark} \\
Toroid & 4.6 & 1.9 & 0.54 & 6.1 & 6.1 & 164 & 60 & 0 & 9.0 & 7.9 & 0.0 & 16.8 \\
Toroid + shield & 5.0 & 3.5 & 0.45 & 8.3 & 8.3 & 273 & 94 & 0 & 12.8 & 11.2 & 0.0 & 24.1 \\
Toroid + iron core + shield & 5.1 & 3.8 & 0.42 & 6.0 & 4.0 & 81 & 100 & 268 & 5.5 & 11.8 & 1.3 & 18.7 \\
Solenoid & 4.2 & 2.7 & 0.56 & 2.4 & 4.1 & 177 & 64 & 0 & 9.5 & 8.3 & 0.0 & 17.7 \\
Core & 7.6 & 10.4 & 0.58 & 2.6 & 2.6 & 79 & 142 & 962 & 5.4 & 15.6 & 4.8 & 25.8 \\
\multicolumn{13}{l}{\it Conservative benchmark} \\
Toroid & 12.6 & 5.3 & 0.54 & 4.2 & 4.2 & 1{,}670 & 456 & 0 & 45.6 & 39.9 & 0.0 & 85.5 \\
Toroid + shield & 13.8 & 9.6 & 0.45 & 5.8 & 5.8 & 2{,}777 & 712 & 0 & 65.1 & 56.9 & 0.0 & 122.0 \\
Toroid + iron core + shield & 12.8 & 9.0 & 0.46 & 5.7 & 3.7 & 958 & 623 & 3{,}351 & 30.9 & 51.1 & 16.8 & 98.8 \\
Solenoid & 15.2 & 9.8 & 0.56 & 1.5 & 3.4 & 3{,}319 & 833 & 0 & 73.7 & 64.5 & 0.0 & 138.2 \\
Core & 19.0 & 20.9 & 0.49 & 2.5 & 2.1 & 610 & 898 & 9{,}955 & 22.5 & 68.5 & 49.8 & 140.8 \\
\end{tabular}
\end{ruledtabular}
\end{table*}

The results of the scan over geometries for each topology considered are shown in Fig.~\ref{fig:numerics_optima}.
(See Tab.~\ref{tab:numerics} for the numerical values.)
We show the minimal cost for each design that covers the required QCD axion parameter space under the optimistic benchmark scenario (left panel) and the conservative benchmark scenario (right panel).
The conservative benchmark scenario is roughly a factor of five more expensive for each detector topology.
For the toroid topology, we illustrate results for the unshielded toroid, the toroid with a shield, and the toroid plus iron core plus shield.
The unshielded toroid is not realistic for the reasons discussed above; its noise profile and quality factor would be too degraded.
The optimized core with iron costs 7\% (15\%) more than the optimized toroid with a shield in the optimistic (conservative) case.
The solenoid result for the optimistic benchmark is comparable to the toroid alone, though we caution that the quality factor assumed for the optimistic solenoid may not be realistic, due to the fact that the magnetic field must pierce the pickup; at the least, achieving the required quality factor for the optimistic benchmark for the solenoid would likely vastly increase the cost relative to the estimate here, as this would require superconductors beyond the conventional low-temperature superconductors.

On the other hand, for the conservative benchmark, the copper pickup must be held below $\sim$40~mK: below $\sim$10~K its resistivity, and hence its quality factor, no longer improves, so the noise budget $T/Q=30~{\rm mK}/10^5$ can only be met by lowering the pickup temperature.
As far as we are aware, no cost data exists for millikelvin systems of this size.
We thus conservatively estimate the cost of this stage by applying the $\sim$4~K cryostat term of~\eqref{eq:cost_HB} to the copper surface of the pickup, without reoptimizing the design.
This gives an additional cost of 9 (74)~MCHF in the optimistic (conservative) case, such that the total cost for the solenoid is at least 212~MCHF.
Since a millikelvin enclosure requires at least the same shielding and insulation as a 4~K cryostat, and we do not include the price of dilution refrigerators, this likely underestimates the true cost.
Further, a millikelvin stage of this size, spanning a $\sim$10$^3$~m$^3$ bore, would be unprecedented as far as we can tell.
(The largest existing millikelvin stage cools $\sim$1~m$^3$~\cite{Alduino:2019xia}.)

We also compare the optimized costs of the solenoid, toroid, and core designs with that of a detector similar to the proposed DMRadio-GUT design~\cite{Brouwer:2022bwo}, which is a toroid with a 16~T peak field and $r_{\rm in}/r_{\rm out} = 0.5$.
To estimate the latter cost, we adopt the height-to-radius ratio of the DMRadio-50L magnet~\cite{Droster:2025utl} (the GUT detector would share approximately the same ratio~\cite{Brouwer:2022bwo}), and use a hermetic sheath pickup and a superconducting shield.
We let the total volume vary with these geometric ratios, along with the peak field, held fixed.
Using the cost functional in~\eqref{eq:cost_HB}, we find that this DMRadio-GUT-like toroid costs 27 (150)~MCHF in the optimistic (conservative) scenario, which is 12\% (23\%) more than our optimized shielded toroid.
On the other hand, the actual cost is likely much higher as fields above $\sim$9~T cannot be produced from NbTi conductors.

As discussed in Sec.~\ref{sec:add_cost}, magnets between $\sim$9~T and $\sim$16~T require Nb$_3$Sn or REBCO.
We apply the additional, though approximate, cost increase for Nb$_3$Sn conductor magnets discussed in that section, which increases the cost of the DMRadio-GUT-like toroid to an expected value of 43 (237)~MCHF, which is 1.8 (1.9) times that of our optimized shielded toroid in the optimistic (conservative) scenario.
We show this cost increase in Fig.~\ref{fig:numerics_optima}, along with uncertainty bands associated with our Nb$_3$Sn cost estimate, following the procedure described in Sec.~\ref{sec:add_cost}.

In the bottom panels of Fig.~\ref{fig:numerics_optima} we show the dimensions of the optimal detector designs.
Notably, the optimal core detectors are much larger than the optimal toroid and solenoid detectors.
The reason for this is that the core design is fighting the requirement that the magnetic field at the surface of the superconducting box stay below 100~mT, and the way to satisfy this constraint is to make the primary magnetic field smaller in amplitude, which requires a bigger detector to keep the necessary sensitivity; the larger detector then drives up the cryogenic cost.

  \begin{figure*}[t]
    \centering
    \includegraphics[width=\textwidth]{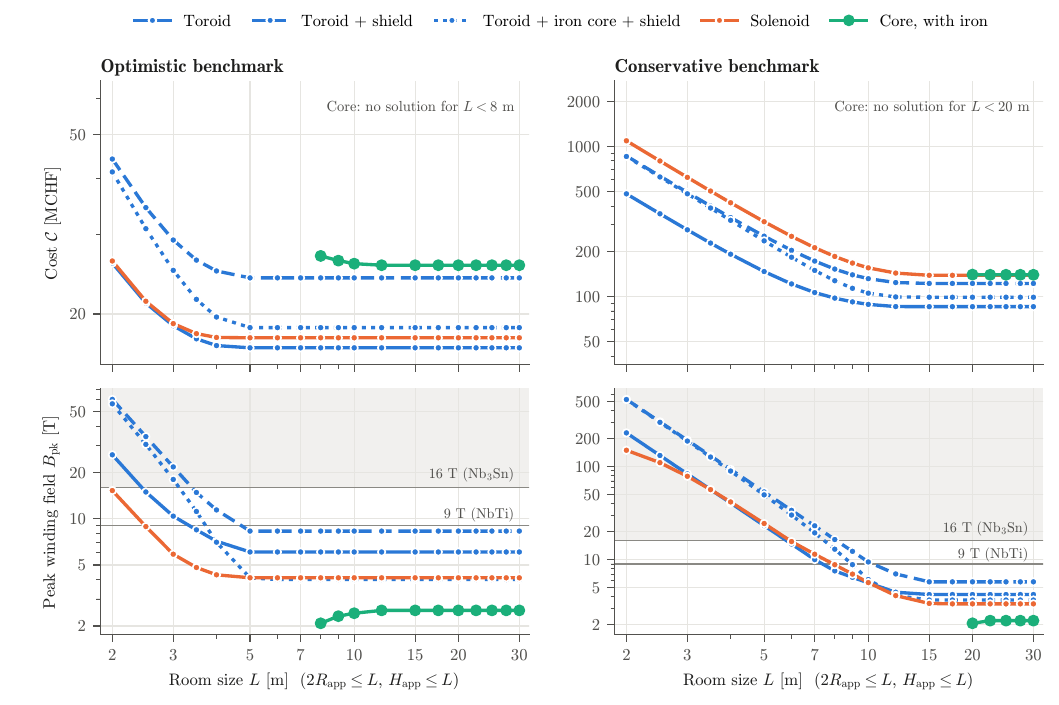}
    \caption{(Top panels) The cost ${\mathcal C}$ as a function of the room size $L$ for the optimistic benchmark scenario (left panel) and the conservative benchmark (right panel).
We assume the room is a cube with side length $L$, and we require the experiment to fit within the room.
No solution is found for the core experiment with $L < 8$~m ($L < 20$~m) for the optimistic (conservative) benchmark.
(Note that $R_{\rm app}$ and $H_{\rm app}$ refer to the radius and height of the core detector accounting for the iron envelope.)
(Bottom panels) The peak laboratory magnetic field associated with each design as a function of $L$.
The magnetic fields must be below the indicated values to use NbTi and Nb$_3$Sn conductors to generate the laboratory field.}
    \label{fig:numerics_rooms}
\end{figure*}

As a further illustration of how the cost and peak magnetic field scale with the size of the detector, in Fig.~\ref{fig:numerics_rooms} we show how the cost of the detectors (top panels) and the peak laboratory field strength (bottom panels) scale with the size of the room, $L$.
The room is taken to be cubic, with volume $L^3$, and the detector is required to fit within the room.
In the optimistic benchmark scenario, the core design requires a room of at least $L \approx 8$~m, while the room must be larger than $\sim$20~m in the conservative benchmark scenario.
Otherwise, the magnetic field cannot be kept small enough at the surface of the pickup box.
The toroid plus shield scenario requires a room of at least $\sim$5~m ($\sim10$~m) for the optimistic (conservative) benchmark to stay within the magnetic fields allowed by NbTi superconductors; going to Nb$_3$Sn would allow for slightly more compact, though more expensive, designs.
We also note that for all designs the cost is relatively flat as a function of $L$ past a critical $L$ value, which differs per design and per benchmark.

\subsection{Cost versus target axion mass}

The benchmarks of this paper target the axion mass range from roughly 0.5~neV to 5~neV.
The bottom end of this mass range is motivated by the prediction in heterotic string theory, in particular~\cite{Benabou:2026jtv}.
On the other hand, the mass of the QCD axion in string theory could potentially be lower, to near $10^{-11}$~eV~\cite{Benabou:2025kgx,DiUbaldo:2026rly}.
From a data-driven perspective as well, axion masses at and  below around $10^{-11}$~eV will likely be covered by black hole superradiance searches ({\it e.g.},~\cite{Arvanitaki:2010sy,Arvanitaki:2014wva,Baryakhtar:2020gao,Hoof:2024quk,Ning:2026ebu}).
Thus, the lowest mass that a laboratory detector would likely need to probe is around $10^{-11}$~eV.
Here, we briefly describe how to estimate the cost of such a detector and, more generally, how cost scales with the minimum QCD axion mass probed.

As an example, suppose we design a lumped-element axion experiment to achieve sensitivity to QCD axions for masses $m_1 < m_a < 2 m_1$.
It is straightforward to show that the cost of the optimal detector scales with $m_1$ as ${\mathcal C} \propto m_1^{-3/4}$, while the volume of the apparatus scales roughly as $m_1^{-1.4}$.
(Note that these cost scalings apply to detectors without iron.)
Very roughly, this implies that a toroidal experiment sensitive to DFSZ axions down to $m_a \sim 10^{-11}$~eV would cost around 450--2300~MCHF depending on the optimistic or conservative scenario.
In the conservative scenario this detector would store more than four times the 41~GJ of the ITER toroidal-field coils, though at a peak field of only $\sim$3~T, compared to 11.8~T in ITER, so it could be wound from NbTi rather than Nb$_3$Sn~\cite{ITER:magnets}.
While these would clearly be expensive experiments, it is reassuring that the cost needed to cover the full axion mass range down to the superradiance regime is well within the costs of previous particle physics and astro-particle physics experiments and observatories.

Alternatively, an intermediate-scale project might target a higher mass band at a lower cost.
As an example, if we set $m_1 = 10^{-8}$~eV, then the cost of the experiment is approximately 2.5--13~MCHF, depending on the optimistic or conservative benchmark.
Note that the MQS approximation continues to apply well to these detectors at this higher mass range.

\section{A realistic toroid beyond MQS}
\label{sec:realistic_toroid}

As a specific case study, we work within the conservative benchmark scenario and consider the associated cost-optimal shielded toroid identified in Sec.~\ref{sec:numerics}, whose magnetic plus cryogenic cost is $\approx 122$~MCHF.
The conservative benchmark parameters match or are close to demonstrated technology (Tab.~\ref{tab:benchmarks}), so such a detector could be built today with minimal R\&D.
The purpose of this section is to show that no hidden obstacles challenge its performance once the idealizations of Sec.~\ref{sec:numerics} are removed, and that this performance holds without the low-frequency MQS approximation.

\begin{figure*}[t]
\centering
\includegraphics[width=\textwidth]{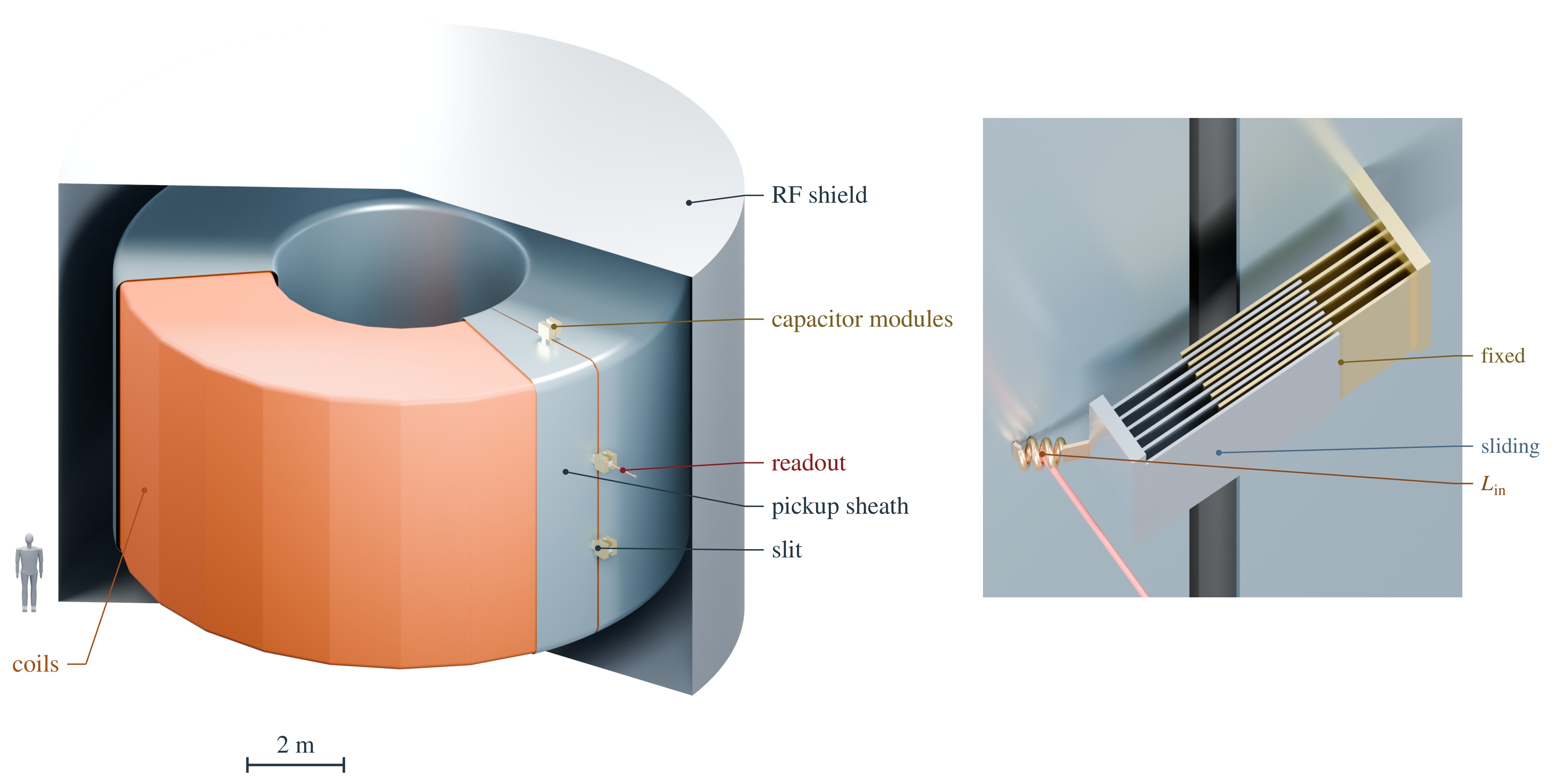}%
\caption{\textit{Left}: Cutaway of the 25-racetrack toroid (Sec.~\ref{sec:realistic_toroid}), showing the coil cases, slit pickup and readout, RF shield, and eight capacitor modules (three are visible).
We show a $1.75$~m human figure for reference.
\textit{Right}: the example capacitor module connected to the SQUID input coil, drawn as two interleaved combs of flat plates (fewer plates than in the design).
The fixed comb is bonded to one side of the slit; the sliding comb connects to the opposite side of the slit through a superconducting strap and the input coil.
Windings, sheath and RF shield operate at $4$~K, with the capacitor plate stacks at $30$~mK.
The readout line and the capacitor module are illustrative and are not meshed in the \texttt{axionMultiPhysics} model.
The design is sized to meet our conservative benchmark.}
\label{fig:realistic_3d}
\end{figure*}

The realistic design is shown in Fig.~\ref{fig:realistic_3d}.
Its field is generated by $N = 25$ discrete racetrack coils with winding packs of finite thickness, whose fringe fields set how close the superconducting pickup sheath can sit to the windings (Sec.~\ref{sec:realistic_magnet}).
The sheath is broken by a $0.5^\circ$ readout slit, which is bridged by eight tunable capacitor modules, and an input coil in series with one module couples the circuit to the SQUID (Sec.~\ref{sec:realistic_readout}).
We then estimate the noise from mechanical vibration and environmental electromagnetic pickup (Sec.~\ref{sec:realistic_noise}), and simulate the full detector beyond the MQS limit (Sec.~\ref{sec:realistic_fullwave}).
The finite winding packs, the gaps that keep the static field on the sheath below 100~mT, and the capacitor layout together require dilating the detector by a factor $s \approx 1.02$ to reach ${\mathcal F}_{\rm req}$, which raises its cost by $\sim$4\%, to $\approx 127$~MCHF.

\subsection{Racetrack magnet and pickup sheath}
\label{sec:realistic_magnet}

One primary consideration in going to a more physical toroidal design is how the laboratory magnetic field is generated.
The experiments ABRACADABRA-10~cm~\cite{Ouellet:2019tlz} and DMRadio-50L~\cite{Rapidis:2022gti} used continuously wound magnets, but going to the large sizes required to reach GUT-scale QCD axions means that racetrack-style magnets are likely more realistic.
An example of such a racetrack design is shown in Fig.~\ref{fig:realistic_3d}, with $N = 25$ total windings; this design produces a magnet comparable to that given by the toroid plus shield benchmark in Tab.~\ref{tab:numerics}.
In particular, we require the peak field strength to be 5.8~T.
The winding packs themselves are assumed to be wide (effectively connecting with their neighbors), to form an $N$-gon, with finite thickness needed to keep the current density through the packs less than an assumed engineering current density of $228$~A/mm$^2$~\cite{Bottura:1999vz}.
This requirement sets the thickness of the winding packs to $\sim$20~mm, which is much smaller than the necessary gap between the winding packs and the pickup sheath and thus has minimal effect on the final result.

For a perfect toroid, the magnetic field is completely confined to the toroid interior.
For a racetrack toroid generated by $N$ windings, the magnetic field outside of the toroid drops as $e^{-2\pi\,\delta r / d}$, where $d$ is the distance between windings and $\delta r$ is the distance from the toroid in the radial direction.
Given that $d \propto r / N$ at a radius $r$, the magnetic field drops off more quickly at the inner boundary of the toroid than at the outer boundary.
This is important because we want to place the superconducting sheath as close as possible to the toroid, but we need the magnetic field to be less than the lower critical field of niobium, $\sim$140~mT at 4~K; we adopt a more conservative threshold that the field strength be below 100~mT everywhere along the sheath, as we impose for the core design, and we also account for the finite thickness of the winding packs.

For the $N= 25$ design, keeping the field less than 100~mT at the sheath requires a gap between the magnet winding packs and the sheath.
The innermost gap is $\sim$110~mm, measured from the innermost point of the winding packs, while the outer gap is $\sim$50~mm, measured from the outermost point of the winding packs.
The top and bottom gaps need to be 50~mm as well.
To accommodate the gaps and the finite winding packs, the detector is dilated by $s \approx 1.02$ to maintain the target sensitivity.
The optimal dilation is not uniform, however; the inner radius, outer radius, and height of the windings are re-optimized including the field constraint at the sheath surface.
Relative to Tab.~\ref{tab:numerics}, the inner radius of the windings grows by 130~mm, the outer radius by 80--120~mm (the windings are flat), and the half-height by around 70~mm.
This results in a negligible few percent increase in total magnetic and cooling cost relative to the ideal design.

The more realistic design shown here, with the same costing, must be scaled by a factor $s \approx 1.02$ to achieve ${\mathcal F}_{\rm req}$, which increases the cost by $\sim$3\%, to $\approx 126$~MCHF, for the $N=25$ benchmark.
This arises primarily from the finite winding packs and the gaps that keep the static field on the sheath below 100~mT.
Note that a design with $N = 15$ ($N = 50$) racetrack coils would require $s \approx 1.041$ ($s \approx 1.012$), illustrating that the toroid design works even for a modest number of magnetic windings.

\subsection{Readout slit, tuning capacitors, and SQUID coupling}
\label{sec:realistic_readout}

In going beyond the ideal detector, we also account for the locations of the tunable capacitor modules.
We do not present the fully cost-optimal configuration but rather show that for a marginal increase in cost an effective design is achievable.
We take the pickup sheath to be broken by a $0.5^\circ$ readout gap (see Fig.~\ref{fig:realistic_3d}).
Note that this corresponds to a $\sim$22~mm gap at the inner boundary of the sheath and a $\sim$52~mm gap at the outer boundary. 
As shown in Fig.~\ref{fig:realistic_3d}, we take the gap to be bridged by capacitor modules, described more below.
Note that we do not optimize the gap; smaller gaps may be advantageous for environmental noise purposes, as we discuss later. 

Each tunable capacitor module is small compared with the height and width of the pickup sheath.
The modules bridge the slit, so that current crosses it only as displacement current.
For the $N=25$ benchmark, each module must tune from about 130~nF at 120~kHz down to about 1.5~nF at 1~MHz.
A simple realization is two interleaved sets of flat plates separated by vacuum gaps, one set attached to each side of the slit.
Neglecting fringing fields, $C =  N_{\rm gaps} w x / d$, where $N_{\rm gaps}$ is the number of gaps between adjacent plates, $w$ is the plate width along the slit, $x$ is the overlap of the plates across the slit, and $d$ is the gap, normal to the sheath.
The module is tuned by sliding one set of plates to change $x$.
For example, $d = 0.25$~mm, $N_{\rm gaps} = 128$ and $w = 24$~cm give 130~nF at $x = 12$~cm and 1.5~nF at $x \approx 1.4$~mm.
With plates as thick as the gaps, the stack is about 6~cm thick.
We emphasize that this is simply an example and other capacitor module designs are possible.

Given the capacitor modules, the next question is where they should be located and how many we should have.
In the ideal limit the capacitor module is continuously distributed across the slit so that current crosses the sheath evenly.
Having finite, localized capacitor modules forces the current to cross the pickup gap at local points, which increases the pickup inductance and thus decreases ${\mathcal F} \propto L_p^{-2}$.
Thus the more tunable capacitor modules, the better.
For example, with a single tunable capacitor module the pickup inductance rises by $\sim$20\% to 40\% depending on whether the capacitor module is on the inner or outer ring of the pickup sheath, respectively.
With two capacitor modules the pickup inductance can be brought within $\sim$7\% of its optimal value.
In our fiducial design we use eight modules, which are naturally thought of as four pairs.
The eight-module design requires a $\sim$0.5~MCHF increase in cost, by increasing the size of the detector, to account for the 1\% increase in $L_p$ and maintain the target sensitivity.
Two capacitor modules are at $r \approx 2.57$~m (inner ring of sheath) at $z \approx \pm 0.51$~m and two more at $z \approx \pm 2.04$~m,  while there are two more modules at the top and bottom of the sheath at $ r \approx 4.52$~m and $z \approx \pm 3.25$~m, and two more modules along the outer ring of the sheath at $r \approx 5.98$~m and $z \approx \pm 1.02$~m.
This is the 8-capacitor design that gives the lowest $L_p$, with $L_p$ only 1\% above the optimal value.
In this design each of the capacitor modules is assumed to be tunable by its own actuator, which must enter through the RF shield.

It is worth noting that the capacitor modules can be kept at lower temperatures than the pickup sheath.
Recall that each loss channel enters the scan rate through $T_i/Q_i$.
For a capacitor at resonance, $Q_C = 1/\tan\delta$, so with the entire conservative benchmark noise budget the capacitors need $\tan\delta \lesssim 10^{-8}$ at 4~K but only $\tan\delta \lesssim 10^{-6}$ at 30~mK.
Because the modules are small, they can be held on thermally isolated millikelvin stages, connected to the 4~K slit lips by superconducting straps, with a heat load of order $10~\mu$W, as in the DMRadio designs, which separate a millikelvin resonator from a 4~K magnet~\cite{DMRadio:2022pkf,Brouwer:2022bwo,Ankel:2026cef}.
Cooling the capacitors does not, however, relax the requirement on the 4~K sheath and RF shield, whose surface resistance must remain below $\approx 5$~n$\Omega$ at 120~kHz.

Lastly, we must consider the component that couples the detector to the SQUID or other readout device.
(Note that in the conservative benchmark scenario the readout uses a SQUID.)
The inductive coupling to the SQUID circuit may be performed, {\it e.g.}, through a small mutual inductance between the SQUID and an input coil in series with one of the capacitor modules, as illustrated in Fig.~\ref{fig:realistic_3d}.
At 120~kHz (1~MHz) the optimal mutual inductance between the SQUID and the pickup circuit is $M \approx 1.0$~nH ($M \approx 0.4$~nH).
However, since the input coil carries only the current through its own module, about 12\% of the total, it must couple more strongly to the SQUID, with $M_{\rm in} \approx 8$~nH (3~nH), to make up for the smaller current; with the increased coupling, this circuit performs equivalently to the simpler circuit diagram of a single capacitor, whose value is the total of the eight modules, in series with $L_p$ and coupled inductively to the SQUID.
For a SQUID of loop inductance $L_{\rm SQ} = 200$~pH and coupling efficiency $k^2 \approx 0.5$, this corresponds to $L_{\rm in} = M_{\rm in}^2/(k^2 L_{\rm SQ}) \approx 0.7~\mu$H ($0.1~\mu$H), {\it i.e.}, a centimeter-scale superconducting coil of $\sim$15 ($\sim$5) turns in series with the module.

\subsection{Beyond the MQS limit}
\label{sec:realistic_fullwave}

All of the calculations to this point have assumed the MQS limit, where the axion Compton wavelength is assumed to be much larger than the size of the pickup inductor~\cite{Benabou:2022qpv}.
This approximation should break down when the axion Compton wavelength becomes comparable to the circumference of the pickup toroid, which is at a frequency of around 3~MHz ($m_a \sim 10$~neV).
At higher frequencies we pass the first self-resonance, where the reactance of the pickup switches from inductive to capacitive.
Searches above the first resonance are possible, but would generally require inductive tuning until the pickup reactance turns inductive again.

\begin{figure*}[t]
\centering
\includegraphics[width=\textwidth]{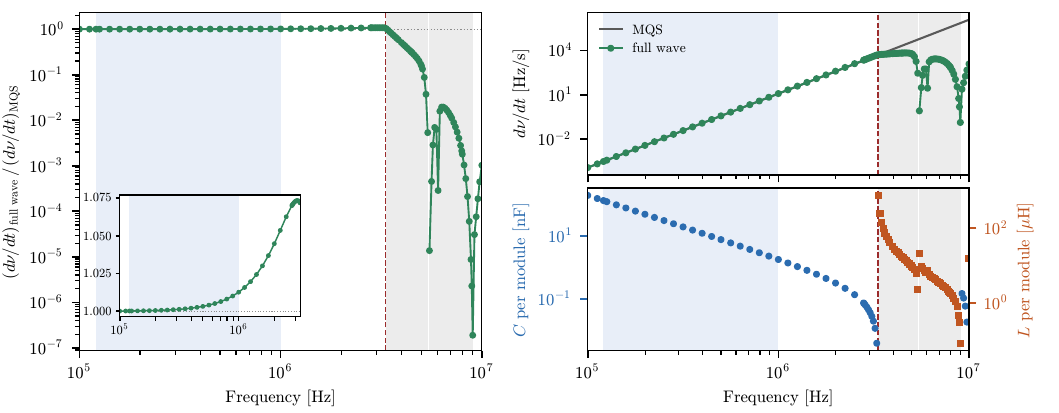}
\caption{
Full-wave scan of the 25-racetrack toroid design with \texttt{axionMultiPhysics}.
\textit{Left}: scan rate $d\nu/dt$ relative to the MQS estimate.
The inset zooms in below the first self-resonance (dashed).
\textit{Right, top}: scan rate for the DFSZ axion, in the full-wave calculation and in the MQS approximation.
\textit{Right, bottom}: the tuning element needed in each capacitor module (we show the largest of the eight).
Below the first self-resonance the pickup behaves as an inductor, so it is tuned with a capacitor ($C$, left axis); above it the pickup behaves as a capacitor, so it must be tuned with an inductor ($L$, right axis).
We assume ideal tuning elements with $Q = 10^6$ independent of frequency.
We indicate the target band (blue) and the band in which inductive tuning is required (gray).
}
\label{fig:scan_rate_MQS}
\end{figure*}

We use \texttt{axionMultiPhysics} to perform a full-wave 3D detector simulation, solving~\eqref{eq:axionmultiphysics_fullwave}, of the fiducial $N = 25$ geometry described above, in a model that includes the conducting coil cases and the eight capacitor modules, each of which is a port connected to its tuning capacitor (or, above the first self-resonance, a tuning inductor).
We set the SQUID input coil at the single-mode optimal coupling~\cite{Ankel:2025nld}, and the resonator's intrinsic quality factor to $Q = 10^6$, as in the conservative benchmark.
In Fig.~\ref{fig:scan_rate_MQS} we show the numerically computed scan rate relative to that estimated from the MQS limit for the same hardware (left panel), while in the right panel we show the absolute scan rate itself from the numerical scan along with the tuning needed to achieve resonance.
Since the scan time is dominated by the lowest mass, where the MQS limit is a good approximation, going beyond the MQS limit does not appreciably change the total estimated scan time to cover the target frequency range (the total change in needed scan time is less than $0.002$ years).

We emphasize that racetrack-style toroids are not unprecedented at the scale proposed here, as illustrated in Fig.~\ref{fig:magnet_costs}.
Examples include the ATLAS barrel toroid, which is roughly 20~m in diameter and $25$~m long and has a peak field strength of $3.9$~T in its windings; it is generated by 8 racetrack coils~\cite{ATLAS:2008xda}.
The IAXO axion experiment magnet is planned to also be a toroidal magnet with 8 racetrack coils, reaching a peak field of $5.4$~T in its windings, with an overall diameter of $5.2$~m and length of $25$~m~\cite{Armengaud:2014gea}.
There are a number of comparable superconducting toroidal-field systems in fusion devices, although their coils are typically D-shaped rather than racetracks~\cite{ITER:magnets,Wan:2002fec}, including ITER, which has 18 Nb$_3$Sn coils, each 17~m high and 9~m wide, with a peak field of 11.8~T~\cite{ITER:facts,ITER:magnets}.

To our knowledge, all superconducting toroidal-field magnets at the few-meter scale are assembled from discrete, separately manufactured coils rather than wound continuously.
(Note that the ABRACADABRA-10~cm~\cite{Ouellet:2019tlz,Ouellet:2018beu,Salemi:2021gck} and DMRadio-50L~\cite{Rapidis:2022gti,Ankel:2026cef} experiments use smaller, continuously wound magnets.)
Note that typically the coils are connected in series by inter-coil joints and fed by a single power supply, as in ATLAS, whose barrel and two end-cap toroids are connected in series to one $20.5$~kA supply~\cite{ATLAS:2008xda}, and IAXO~\cite{Armengaud:2014gea}.
Larger fusion systems divide the series circuit among several pairs of current leads.
Moreover, at the few-meter scale, the racetrack magnets are not fundamentally more costly than solenoidal magnets since large detector toroids and solenoids, such as the ATLAS barrel toroid and the CMS solenoid, are both wound from the same aluminum-stabilized NbTi Rutherford-cable technology~\cite{Vedrine:2014bdk}.

\subsection{Mechanical vibration and environmental electromagnetic pickup}
\label{sec:realistic_noise}

Beyond thermal noise, two other noise sources are important to consider: mechanical vibrations and electromagnetic interference (EMI).

\subsubsection{Mechanical vibration estimate}

Mechanical vibrations can have two classes of contributions, one being external vibrations and the other thermal vibrations of detector components.
Only the latter is not reducible in principle, at a fixed detector temperature.

Let us consider one source of thermal vibrations and track through how it affects the noise budget in our toroid example with eight capacitor modules.
Let us suppose that the NbTi coil packs are wrapped with thin copper or steel encasements at $T \approx 4$~K.
These encasements can be modeled as perfect conductors.
Let us characterize the vibrations of the encasements through the displacement field ${\bm u}({\bm x},t)$, which describes the magnitude and direction of displacement of any point of the conductors at any moment of time.
We claim that a displacement ${\bm u}$ in the encasement induces a change in the pickup flux $\delta\Phi_p = \oint \bm w\cdot\bm u\, dA$ with $\bm w = (\bm K_p/I_p)\times\bm B_{\rm lab}$, where $\bm K_p$ is the surface current of the pickup mode (the mode generated by injecting $I_p$ through the pickup leads).
This relation may be proven by using reciprocity, and the integral is taken over all conducting surfaces that carry pickup-mode current; here we evaluate it on the encasements.

We note that ${\bm w} = 0$ for an ideal pickup toroid, since $\bm{K}_p$ and ${\bm B}_{\rm lab}$ are parallel.
The only nonzero contribution to ${\bm w}$ comes near the slit in the pickup, where the pickup current is forced into the poloidal directions to cross at the capacitor modules.
(In the limit of a continuous distribution of such modules, ${\bm w} \to 0$ on the sheath.)
The slit also allows the pickup field, which the sheath otherwise excludes from the magnet volume, to leak into the volume near the slit; the screening currents this drives on the neighboring encasements dominate ${\bm w}$, since there the static field is that of the winding packs rather than the much smaller fringe field at the sheath.
This contribution is suppressed by narrowing or overlapping the slit, and only in part by increasing the number of modules.
We may compute ${\bm w}$ numerically for our specific eight-module setup.

To compute $\delta \Phi_p$ we need to also have a theory describing the displacement ${\bm u}$.
Let us consider the transverse vibrations of the encasement, for which ${\bm u}$ and ${\bm w}$ are parallel.
Then, we can describe the PSD of ${\bm u}$ from thermal fluctuations by $S_u(\omega) \approx (4 T / \omega^2) \cdot {\rm Re}\, Y$, where $Y$ is the mechanical point admittance (mobility)~\cite{Callen:1951vq,Levin:1997kv}, here the out-of-plane mobility of the encasement wall.
As an example, for a $\sim$3~mm thick encasement of copper or steel, $Y \approx 10^{-3}$~m/s/N~\cite{Cremer:2005sbs}, which gives $\sqrt{S_u} \approx 10^{-18}$~m$/\sqrt{\rm Hz}$ at 120~kHz and falling with frequency as $\omega^{-1}$.
Substituting this expression into that for $\delta \Phi_p$, we compute an equivalent noise temperature of $T_{\rm vibrations} \sim 10$~mK at 120~kHz, with the equivalent noise temperature scaling with frequency as $\omega^{-1}$.\footnote{In detail, the flux noise depends on how the displacement is correlated over the whole encasement.
We compute this directly using Levin's method~\cite{Levin:1997kv}, which involves driving the encasements with a distributed force weighted by the ${\bm w}$ of our \texttt{axionMultiPhysics} solution.  The formalism for performing these computations is built into \texttt{axionMultiPhysics}.}
This is subdominant, though only slightly, compared to thermal noise.  In addition to thermal fluctuations of the encasements, we also verify through the same formalism that the thermal fluctuations of the sheath, shield, and capacitor modules are far subdominant relative to the contributions of the encasement fluctuations. 

Note, however, that the estimate above is crude, and additional sources of vibrations would almost certainly be present in a realistic design, requiring more careful estimates.
We also note that, unlike for the toroid, the solenoid has no ideal configuration in which ${\bm w} \approx 0$.
Its pickup currents are parallel to the bore field only in the limit of a long solenoid, whereas the optimal geometries have height comparable to radius, and in any case the pickup circuit must close by carrying the full current across the field at the end of the coax.
This means that solenoids are more susceptible to vibrational noise, which could be severe in certain designs.

  \subsubsection{Electromagnetic interference and screening}

The toroid's unshielded pickup mode is a magnetic dipole, which is why the design already carries a  superconducting shield.
With that shield closed, external fields are excluded entirely.
The remaining question, then, relates to EMI that arrives inside the shield through the ports needed for wiring.
The magnet may be charged and then subsequently disconnected, so in principle ports associated with charging the magnet are not a concern during data taking.
Similarly, the actuators needed to tune the capacitor modules can be disconnected and isolated during the observing period.
At the least, however, live wires are needed during data taking for the SQUID in order to both power the SQUID and bring the signal out of the shielded area.

The cables running from outside the superconducting shield to inside carry whatever EMI exists outside the shielded region along themselves to the inside; if they run within superconducting tubes welded into the shield, the superconducting surface remains closed and this conducted path is the only one.
However, this source of noise is reducible in principle, for reasons discussed below, in contrast to the thermal vibrational noise discussed in the previous subsection.
First, the room containing the experiment can itself be made radio quiet, filtering out external EMI and, {\it e.g.}, radio lines associated with radio stations, through passive shielding.
Let us refer to this shielded layer as the room-temperature shield, to be distinguished from the 4~K-stage superconducting shield surrounding the detector.
EMI will still be generated by electronics between the two shields.

The following consideration can be used to, in principle, make the detector insensitive to EMI associated with the SQUID port and cables.
First, we note that for a gapless sheath, current noise in the sheath would only be generated from the EMI if the wires that are the noise source wrapped through the center of the torus; this, practically, is easy to avoid.
On the other hand, the sheath must necessarily have a gap.
The concern with EMI from the cables is that the common-mode voltage they carry would couple capacitively to the two sides of the sheath gap, generating a potential difference across it and causing current to flow.
A natural solution to this problem is to make the geometry symmetric with respect to the gap, so that even if EMI is present, it does not generate a potential difference across the gap and thus does not generate current noise in the pickup.
This may be achieved by having the SQUID cables approach in a straight line to a point directly above the gap, symmetric between the two sides, with the SQUID located exactly in between the two sides; the required symmetry is a millimeter-scale alignment on a centimeter-scale termination, which is a machining tolerance rather than a research problem.
Note that such external symmetry does not constrain the capacitances internal to the SQUID package itself, which are instead a requirement on its layout.
How well this alignment may be achieved, and how much EMI there is in the first place, are design requirements that are reducible in principle, so we do not consider them further here.

\section{Comparison with other detection concepts}
\label{sec:other}

In the previous sections we estimated the cost of a lumped-element axion detector sensitive to the MI QCD axion at roughly 20 to 120~MCHF for the shielded toroids, depending on the optimistic or conservative benchmark parameters and the detector topology, with the geometries optimized.
In this section we compare the lumped-element experiments to other detector proposals that are in principle sensitive to the MI QCD axion, beginning with the proposal to use SRF cavities~\cite{Berlin:2019ahk,Lasenby:2019prg,Berlin:2020vrk}.

\subsection{Optimized SRF cavities for heterodyne axion detection}
\label{sec:heterodyne}

\begin{figure*}[t]
\centering
\includegraphics[width=\textwidth]{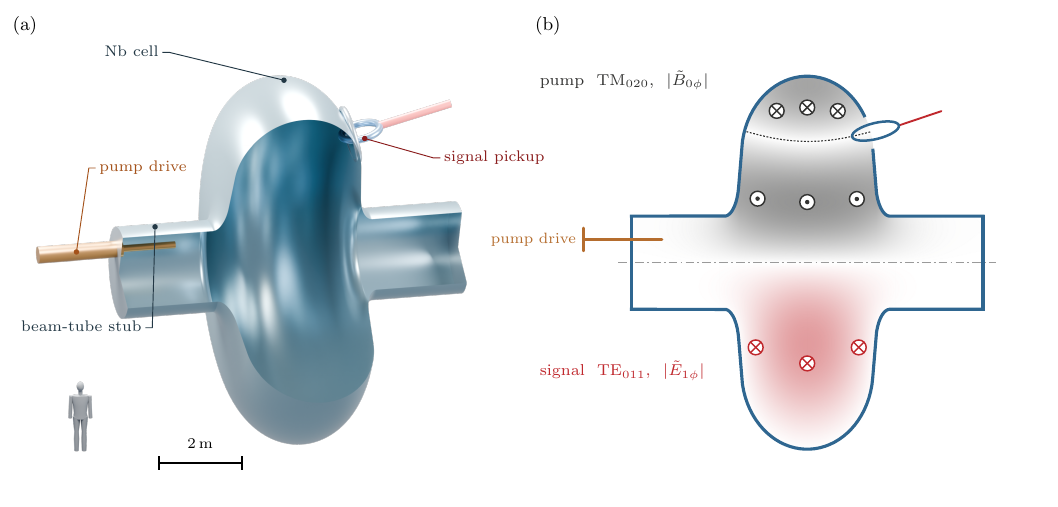}
\caption{
The optimistic SRF benchmark.
(a)~Cutaway of the scaled single-cell elliptical cavity, drawn to scale ($V \approx 180$~m$^3$), with a 1.75~m human figure for scale.
The pump drive and the signal pickup are schematic.
(b)~Meridional cross-section of the same cell, with the symmetry axis indicated by the dot-dashed straight line, showing the magnitude of the azimuthal pump TM$_{020}$ magnetic field $\tilde B_{0\phi}$ (top) and of the signal TE$_{011}$ electric field $\tilde E_{1\phi}$ (bottom), with their directions marked as in Fig.~\ref{fig:topologies}.
The pump field changes sign across the dotted line and the signal field does not, which reduces the overlap to $\eta_{10} \approx 0.45$.
The pickup sits near the node of the pump field on the wall and links the signal magnetic field $\tilde{\bm{B}}_1$.
}
\label{fig:srf_design}
\end{figure*}

The SRF cavity detection concept works as follows.
Consider a superconducting cavity pumped with a mode at frequency $\omega_0 \gg m_a$, such that in the presence of axion DM the effective current in the cavity is
\es{}{
\bm{J}_{\rm eff} = {g_{a\gamma\gamma} \sqrt{2 \rho_{\rm DM}} \over 2} \tilde{\bm{B}}_0(\bm{x}) \big[ &\sin[(\omega_0 + m_a)t] \\
&- \sin[(\omega_0 - m_a)t] \big] \,,
}
where $\tilde{\bm{B}}_0(\bm{x})$ is the spatial profile of the pump magnetic field in the cavity; that is, $\bm{B}_{\rm lab}(\bm{x},t) = \tilde{\bm{B}}_0(\bm{x}) \cos \omega_0 t$.
The key point is that the axion populates two side-band modes at frequencies $\omega_0 \pm m_a$, and we can search for power in these modes to detect or exclude axions.
In particular, we look for power at the frequency $\omega_1 \equiv \omega_0 + m_a$.

The oscillating axion-induced effective current generates a secondary electromagnetic field that obeys
\es{}{
(\partial_t^2 - \nabla^2) \bm{E}(\bm{x},t) = -\partial_t \bm{J}_{\rm eff}(\bm{x},t) \,.
}
The key advantage of the heterodyne approach is that the time derivative on $\bm{J}_{\rm eff}$ brings down a factor $\sim$$\omega_0 \gg m_a$, whereas the equivalent derivative in the MQS designs brings down a factor of $m_a$.
This means that the heterodyne signal is less suppressed than the lumped-element signal in the limit of long axion wavelength relative to the size of the experiment~\cite{Berlin:2019ahk,Lasenby:2019prg}.

The signal power transferred by the axion to the mode $\omega_1$ may be computed to be~\cite{Berlin:2019ahk}
\es{}{
P_{\rm sig} = {1 \over 4} g_{a\gamma\gamma}^2 \eta_{10}^2 B_0^2 \rho_{\rm DM} V {Q_1 \over \omega_1} \,,
}
where $V$ is the cavity volume, $Q_1$ is the loaded quality factor of the mode at frequency $\omega_1$, and $B_0^2 \equiv V^{-1} \int_V | \tilde{\bm{B}}_0|^2\,dV$ is the square of the spatial RMS of the laboratory magnetic field strength.
The dimensionless quantity $0 \leq \eta_{10} \leq 1$ is an overlap factor between the induced electric field and the laboratory magnetic field:
\es{}{
\eta_{10} \equiv \frac{\left| \int_V \tilde{\bm{E}}_1^{*} \cdot \tilde{\bm{B}}_0 \right|}
{\sqrt{\int_V |\tilde{\bm{E}}_1|^2}\;\sqrt{\int_V |\tilde{\bm{B}}_0|^2}} \,,
}
where $\tilde{\bm{E}}_1$ is such that $\bm{E}_1 = {\rm Re}\left[ \tilde{\bm{E}}_1 e^{i \omega_1 t}\right]$.

The signal mode is read out through a port with coupling $\beta \equiv Q_i/Q_{\rm cpl}$, where $Q_i$ is the intrinsic quality factor of the mode, $Q_{\rm cpl}$ is its coupling quality factor, set by the power that leaks out through the port, and $Q_1 = Q_i/(1+\beta)$ is its loaded quality factor.
The port feeds a transmission line with power transmission $\tau$, followed by a phase-preserving amplifier that adds $n_{\rm add}$ quanta of noise.
Using the same likelihood framework as for the lumped-element detectors, the rate at which the splitting $\omega_1 - \omega_0$ may be scanned while reaching the expected sensitivity is
\es{eq:SRF_scan}{
{d(\omega_1 - \omega_0) \over dt} = {\pi^2 \over 32} {Q_{\rm axion} \over \Theta_{95}} \, {g_{a\gamma\gamma}^4 \rho_{\rm DM}^2 \, \eta_{10}^4 B_0^4 V^2 \, Q_i \over m_a \, \omega_1^3} \, \bar G_{\rm SRF} \,,
}
with
\es{eq:G_SRF}{
\bar G_{\rm SRF} &= {\tau^2 \beta^2 \over (1+\beta)^3 \sqrt{A} \, (A + D)^{3/2}} \,, \\
A &= \tfrac{1}{2} + n_{\rm add} + \tau n_e + (1-\tau) n_\ell \,, \\ 
D &= {4 \tau \beta \, (n_i - n_e) \over (1+\beta)^2} \,.
}
Here $n_i$, $n_e$, and $n_\ell$ are the thermal occupation numbers of the cavity walls, of the noise incident on the cavity from the readout line, and of the loss in the line, respectively.
$A$ is the noise floor that is not resonantly enhanced (vacuum, amplifier, and line noise), while $D$ is the cavity thermal noise, which is filtered by the same Lorentzian as the signal.
For $n_i \gg 1$, $\tau n_i/A \gg 1$, and a cold line ($n_e = n_\ell = 0$), the optimal coupling is $\beta \approx 2\tau n_i/A$, and $\bar G_{\rm SRF} \approx \tau\,\omega_1/(6\sqrt{3}\,A\,T)$.
This is the analog of the lumped-element optimum $\bar G \approx \omega_0/(6\sqrt{3}\,\eta_A T)$, with $A = 1$ at the SQL.
This correspondence is exact under the replacements $2R_{\rm c}S_\Phi/(\omega_0 M^2) \to A/(\tau\beta)$, $\eta_A \to A/\tau$, and $2n+1 \to 2n_i + A/\tau$; the function $\bar G$ of~\eqref{eq:Gbar}, before optimization over $M$, becomes $\bar G_{\rm SRF}$ of~\eqref{eq:G_SRF}.
The coupling $\beta$ thus plays the role of the mutual inductance $M$.
Overcoupling broadens the response, which reduces the flat noise floor relative to the signal away from resonance, as increasing $M$ reduces the imprecision noise.
The parameters $\eta_A$ and $A/\tau$ therefore enter the scan rate in the same way, but they are related differently to the noise temperature of the amplifier.
A SQUID with uncorrelated imprecision and backaction noise adds $\eta_A/2$ quanta when noise matched, while for $\tau = 1$ and $n_e = 0$ a phase-preserving amplifier at the port adds $n_{\rm add} = A - 1/2$; the two coincide only at the SQL.
Values $A < 1$ require a phase-sensitive measurement, as is the case for $\eta_A < 1$ (see Sec.~\ref{sec:scan_cost}).

\subsubsection{Cost projections for a SHADE-like detector}

While the cavity mode frequencies are determined by the precise details of the cavity geometry, they scale inversely with its size at fixed shape, so that we can write $\omega_0 = \xi\, V^{-1/3}$ for some dimensionless $\xi$.
For the lowest mode of a compact cavity (a cube, a sphere, or a cylinder with $L = 2R$) $\xi \approx 4.4$; however, the pump and signal modes must be nearly degenerate and have a non-vanishing overlap $\eta_{10}$.
For the TM$_{0,n+1,0}$ (pump) and TE$_{0n1}$ (signal) pairs of a cylindrical cavity proposed in Ref.~\cite{Berlin:2019ahk}, $\xi \approx 7.5$ (TM$_{020}$/TE$_{011}$, $L/R \approx 0.79$) and $\xi \approx 10.8$ (TM$_{030}$/TE$_{021}$, $L/R \approx 0.62$).
For the TE$_{012}$ (pump) and TM$_{013}$ (signal) modes of a cylinder with $L/R \approx 2.35$, the pair that maximizes the signal power per volume among the low-lying cylinder modes at fixed peak wall field~\cite{Lasenby:2019prg}, $\xi \approx 9.1$.
For the TM$_{020}$ (pump) and TE$_{011}$ (signal) modes of the single-cell elliptical cavity of the SHADE experiment considered below, $\xi \approx 7.6$.
At fixed volume the scan rate scales as $\omega_0^{-2}$, which implies that low frequencies perform better than high frequencies with all else held fixed; with $\omega_0 \propto  V^{-1/3}$, the scan rate scales as $V^{8/3}$ at fixed $\xi$.

In the following estimates, we benchmark on an idealized cavity modeled on that of the SHADE experiment~\cite{Giaccone:2022pke,Castaneda:2024shade}, which uses a single-cell elliptical niobium cavity derived from an accelerator cell (the 650~MHz high-velocity cell of PIP-II), pumped in its TM$_{020}$ mode and read out in its TE$_{011}$ mode.
These modes are made nearly degenerate by adjusting the length of the cell.
In Sec.~\ref{sec:srf_opt}, we consider a more optimized detector, which minimizes the cost for the required science reach.

Since the dimensions of the SHADE cell have not been published, we construct our cell from the 650~MHz regular cell of the Project~X reference design~\cite{Holmes:2013vpa} that was designed for particles with velocity $0.9\,c$.
The cell's contour consists of an iris ellipse and an equator ellipse joined by a common tangent, with iris radius 50~mm, equator radius 200.3~mm, half-cell length 103.8~mm, and (axial, radial) semi-axes of $(18, 38)$~mm for the iris ellipse and $(82.5, 84)$~mm for the equator ellipse.
We shorten the cell along its axis by a factor of 0.86, scaling the half-cell length and the axial semi-axes of both ellipses, which makes the TM$_{020}$ and TE$_{011}$ modes degenerate at 1.42~GHz, with $V \approx 17$~L at this size.
The cell is closed by beam-tube stubs of radius 50~mm and length 100~mm.
For each design below, the cell geometry is scaled uniformly to reach the required volume.
We caution that our results below only apply to a scaled SHADE experiment; in Sec.~\ref{sec:srf_opt} below, we optimize the cavity geometry and find designs about 30\% cheaper.

We solve for the axisymmetric modes of this cell using our code package \texttt{axionMultiPhysics} (App.~\ref{app:axionmultiphysics_cavity}).
We find $\xi \approx 7.6$, a ratio of peak wall magnetic field to $B_0$ of $B_{\rm pk}/B_0 \approx 1.5$, a peak wall electric field of the pump $E_{\rm pk} \approx 1.8\, B_0$, and a surface area $A_{\rm cav} \approx 6.2\,V^{2/3}$.
For the overlap we find $\eta_{10} \approx 0.45$, close to the value 0.44 for the TM$_{020}$/TE$_{011}$ pair of a cylindrical cavity~\cite{Thomson:2023moc} and to the estimate $\eta_{10} \simeq 0.5$ of Ref.~\cite{Berlin:2019ahk} for this family of mode pairs.\footnote{We caution that conventions differ in the literature.
Some works quote the square of this overlap ({\it e.g.}, $C_{01} = \eta_{10}^2$ in Ref.~\cite{Lasenby:2019prg}).
The SHADE design value of 0.19~\cite{Castaneda:2024shade} agrees with $\eta_{10}^2$ for our cell.}
SRF cavities are commonly characterized through their geometry factors, which are defined by
\es{eq:geom_factor}{
G \equiv \omega \int |\bm{B}|^2 dV / \int |\bm{B}|^2 dA \,;
}
we find $G \approx 520$~$\Omega$ for the pump mode and 600~$\Omega$ for the signal mode.
The quantity $G$ measures how much of the magnetic energy is stored in the volume of the cavity relative to the surface, where it may dissipate.
Note that $G$ is only a function of the shape of the cavity and its mode, not of the overall size of the cavity; thus, $G$ remains constant under our rescalings to larger $V$.

\begin{table}[t]
\caption{SRF benchmark parameters that differ from those of the lumped-element benchmarks in Tab.~\ref{tab:benchmarks}.
The quality factors are computed from $Q = G/R_s$.}
\label{tab:srf_benchmarks}
\begin{ruledtabular}
\begin{tabular}{lccc}
Quantity & Optimistic & Conservative & Ref. \\
\hline
$R_s$ & 1.5~n$\Omega$ & 20~n$\Omega$ & \cite{Brown:2026cqz,Miyazaki:2023zye} \\
$B_{\rm pk}$ (wall) & 150~mT & 90~mT & \cite{Castaneda:2024shade} \\
$T$ (cavity wall) & 2~K & 4.5~K & \\
$A$ & 0.1 & 10 & \cite{Brouwer:2022bwo,Ankel:2025nld} \\
\end{tabular}
\end{ruledtabular}
\end{table}

The SRF benchmark parameters that differ from those of Tab.~\ref{tab:benchmarks} are given in Tab.~\ref{tab:srf_benchmarks}.
We compute the quality factors self-consistently from the surface resistance $R_s$ of the niobium, $Q = G/R_s$.
For the operating fields below we assume $R_s \approx 1.5$ (20)~n$\Omega$ for both the signal and pump modes under the optimistic (conservative) scenario.
The optimistic value lies at the low end of the residual resistances measured at 2~K in production quarter-wave resonators at 80.5~MHz, 1.5--5.5~n$\Omega$~\cite{Brown:2026cqz}, and in production spoke cavities at 352~MHz, 2--7~n$\Omega$~\cite{Miyazaki:2023zye}, both at much lower fields and on much smaller surfaces than considered here; the conservative value allows for the growth of $R_s$ with the RF field and for these larger surfaces.
At the frequencies of interest the BCS resistance is below 1~n$\Omega$ at both 2 and 4.5~K~\cite{Kolb:2020jab}.
More precise values, including the field dependence of $R_s$ and the loss of the weak signal mode in the presence of a strong pump, would require a dedicated design study.

We take the peak wall magnetic field to be $B_{\rm pk} = 150$~mT (90~mT) in the optimistic (conservative) case.
The conservative value is typical of continuous-wave production cavities, while the optimistic value is close to but above the SHADE design, $B_0 = 0.1$~T~\cite{Castaneda:2024shade}, and may require aggressive surface treatment.

For the readout, a common optimistic benchmark for heterodyne searches is a receiver at or near the SQL, $A = 1$~\mbox{\cite{Berlin:2019ahk,Lasenby:2019prg,Berlin:2020vrk}}.
We instead take $A = 0.1$ in our optimistic scenario and $A= 10$ in our conservative case in order to mirror our assumptions for the lumped-element approach.
Note that $A = 1$ is typically assumed at the GHz frequencies where most design studies of heterodyne experiments have taken place, but in the 10--100~MHz range, where we find the large cavities naturally sit, new detection techniques would be needed.
The natural readout at these frequencies may instead be the one used for the lumped-element designs: a superconducting pickup loop would couple the signal mode's magnetic field to a dc SQUID or RF quantum upconverter~\cite{Kuenstner:2022gyc}.
A sensor coupled in this way has $A = \eta_A$, and the two readout parameters can be compared directly, justifying our choices.

We emphasize, however, that there are a few reasons why $A = 0.1$ may be more difficult to achieve than $\eta_A = 0.1$ in the lumped-element searches, though neither sensitivity has been demonstrated to date.
In particular, the SRF target $A = 0.1$ must meet the same requirements as the lumped-element target $\eta_A = 0.1$, but it must also operate next to the pump mode.
For example, pump power that leaks into the sensor would compete with the noise floor, and thus as $A$ is lowered, so too is the allowed leakage.
Another challenge that the SRF approach faces is that it requires a link from the $\sim$2 K cavity to the quantum sensor, while for the lumped-element approach the sensor is baked into the detector itself.
This is important because loss in the link adds noise that competes with the sensor noise through $(1-\tau)n_\ell$; for example, if the connection has $10^{-3}$~dB of loss at 2 K ($n_\ell \sim 550$), this effectively adds $\sim$0.1 to $A$.
Lastly, we note that vibrations are a principal concern in the SRF approach, as vibrations at frequency $m_a$ can convert pump modes to signal modes, which adds to the noise floor; a related concern is also present for the lumped-element approaches, as we discuss in Sec.~\ref{sec:realistic_noise}, though a more detailed calculation shows that this concern may be more pronounced in the SRF case (see below).

Given these design requirements, we can then calculate the $V$ needed to achieve sensitivity to the QCD axion with three years of scan time over the mass and coupling range specified in Tab.~\ref{tab:benchmarks}.
In the optimistic (conservative) case we find $V \approx 185$~m$^3$ ($V \approx 4800$~m$^3$), with $Q_i \approx 4\times10^{11}$ ($Q_i \approx 3\times10^{10}$) and $f_0 \approx 64$~MHz ($f_0 \approx 22$~MHz).
The associated surface areas for the optimistic and conservative scenarios are $A_{\rm cav} \approx 200$~m$^2$ and $A_{\rm cav} \approx 1750$~m$^2$, respectively.
Next, we turn to how to roughly estimate the cost of such an experiment.

\begin{figure}[t]
\centering
\includegraphics[width=\columnwidth]{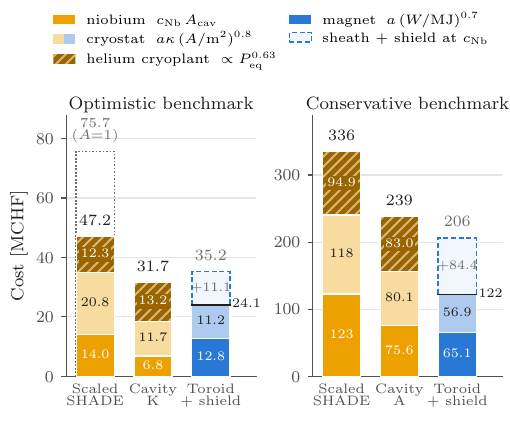}
\caption{Cost of the optimistic and conservative SRF benchmark designs broken down by the three cost categories we consider.
We compare these costs to the costs of our optimized lumped-element toroids with shields, where we add in an additional cost category for the niobium to the lumped-element detectors to make a more direct comparison to the SRF costs.
On the other hand, we caution that the niobium costs for the lumped-element detectors use the identical cost formula as for the SRF cavities, but this is likely an overestimate of the cost for the lumped-element designs since the niobium surfaces are not subject to large RF fields like in the heterodyne detectors.
The left SRF bars use scaled SHADE-like cells.
Under both benchmark scenarios the SRF cavity designs are more expensive than the lumped-element detectors (see text for details).
The optimistic SRF bars use $A = 0.1$, matching $\eta_A = 0.1$ of the lumped-element designs; the dotted outline shows the total cost of the scaled SHADE cell for $A = 1$.
The middle bars show cavities K (optimistic) and A (conservative) of Sec.~\ref{sec:srf_opt} and Fig.~\ref{fig:srf_optimum}.
The black line in each toroid bar marks the total toroid + shield cost of Tab.~\ref{tab:numerics}, printed to its right, and the dashed box above it adds the sheath and shield priced at $c_{\rm Nb}$, which is likely an overestimate (see text).}
\label{fig:srf_cost}
\end{figure}

We include three contributions to our heterodyne axion experiment cost estimate:
\es{eq:cost_SRF}{
  {C_{\rm SRF} } &\approx c_{\rm Nb} {A_{\rm cav} \over {\rm m}^2} + {a \kappa } \left( {A_{\rm cav} \over {\rm m}^2} \right)^{0.8} \\
  &\quad + 60\, {\rm MCHF} \left( {P_{\rm eq} \over 18\, {\rm kW}} \right)^{0.63} \,.
  }
The coefficient $c_{\rm Nb}$ is the cost per unit area of the niobium cavity, including its fabrication and surface preparation; we roughly estimate $c_{\rm Nb} \approx 0.07$~MCHF/m$^2$ by scaling the European XFEL cavity costs~\cite{Altarelli:2006zza} to a large cavity.
The combination $a \kappa \approx 0.3$~MCHF is the exact same combination as in~\eqref{eq:cost_HB} and accounts for the cryostat.
The last term is the cost of the helium refrigerator, scaled from the 60~MCHF cost of the LCLS-II cryoplant, which has a capacity of 18~kW at 4.5~K-equivalent cooling~\cite{Fauve:2024lcls}; the exponent $0.63$ is taken from Ref.~\cite{Green:2008refrigerators}.
Here, $P_{\rm eq}$ is the 4.5~K-equivalent heat load on the cavity.
(To convert a 2~K heat load to a 4.5~K-equivalent heat load, we multiply the power by $\sim$3.3 for a rough estimate~\cite{Altarelli:2006zza}, since it is more expensive to remove power at a lower temperature.)
The total power is dominated by the RF pump dissipation in the cavity walls, which is $P  = \omega_0 U_0 / Q_0$, with $Q_0 = G / R_s$ and $U_0 = B_0^2 V/ 2$.
We compute $P \approx 0.4$~kW (at 2~K) and $P \approx 37$~kW (at 4.5~K) in the optimistic and conservative scenarios, respectively.
Combining all of these estimates together, we approximate the cost of the optimistic and conservative scenarios as $\sim$47~MCHF and $\sim$340~MCHF, respectively (see Fig.~\ref{fig:srf_cost}).

We note that part of the refrigeration cost may be saved by reusing existing cryogenic infrastructure.
This cost is around 25\% (30\%) of our estimated total cost under the optimistic (conservative) scenario.
For example, in the optimistic case the $\sim$0.4~kW load at 2~K is a small fraction of a single LCLS-II plant at SLAC, which provides 4~kW at 2~K~\cite{Pflueckhahn:2022zun}.
In the conservative case the $\sim$37~kW 4.5~K-equivalent load is about twice the $\sim$18~kW at 4.5~K of the largest helium plants in operation, those of the LHC and LCLS-II~\cite{Lebrun:2003zm,Fauve:2024lcls}, and thus dedicated facilities would likely be needed.

Note also that~\eqref{eq:cost_SRF} charges the SRF design for its niobium, while the equivalent cost function for lumped-element designs in~\eqref{eq:cost_HB} contains no corresponding term for the superconducting sheath and shield.
This is deliberate and reflects the different roles of the two surfaces.
The SRF cavity is the field source of the heterodyne experiment: it stores $\sim$$0.4$ (7)~MJ of energy in the pump mode in the optimistic (conservative) scenario and is thus the counterpart of the magnet in the lumped-element design.
The SRF walls must sustain peak RF magnetic fields of around 100 (90)~mT and peak RF electric fields of tens of MV/m while dissipating around 0.4 (40)~kW.
These requirements are in practice achieved with bulk high-purity niobium.
The sheath and shield of the lumped-element designs, in contrast, carry only tiny RF currents induced by the signal mode.
We suspect they could  be made of niobium sputtered onto copper, as in, {\it e.g.}, the LEP and LHC cavities~\cite{Benvenuti:1984er,Miyazaki:2018pbz}.
The intrinsic residual resistance of such films has been measured to be $\sim$15~n$\Omega$ at 100~MHz and 4.5~K~\cite{Miyazaki:2018pbz}, which is three times the $\sim$5~n$\Omega$ that the conservative benchmark requires  at a frequency nearly three orders of magnitude lower.
Meeting the resistance requirement thus only requires a modest decrease in resistance with decreasing frequency, while of course the optimistic benchmark is more demanding.

If we price the $\approx$160 (1200)~m$^2$ of sheath and shield of the shielded toroid at the full $c_{\rm Nb}$, this would add $\sim$11 (85)~MCHF to the optimistic (conservative) cost.
In this case the cost-optimized SRF cavities of Sec.~\ref{sec:srf_opt} would cost about the same as the lumped-element designs in the optimistic case and somewhat more in the conservative case.
In reality, however, and for the reasons discussed above, the sheaths and shields of the lumped-element detectors would likely be comparatively cheaper than the SRF niobium surfaces.

\subsubsection{Additional SRF design complications: tuning and vibrations}

We caution that these designs extrapolate SRF technology to volumes far beyond what has been built.
Moreover, scanning $m_a$ over 0.12--1~MHz requires tuning the splitting of the two modes over 880~kHz, a fraction of 1.4\% (4.1\%) of the mode frequency in the optimistic (conservative) case.
One way of achieving this tuning is to change the length of the cavity.
This works because the pump mode is, for small deformations, insensitive to the length of the cavity, while the signal mode frequency is directly proportional to the length.
Achieving the required tunings with a single cavity would be, as far as we are aware, also unprecedented.
First, note that SHADE plans to tune its splitting over $\sim$2~MHz at 1.4~GHz by elastically deforming the cell~\mbox{\cite{Giaccone:2022pke}}.
This corresponds to sub-millimeter deformation on a meter-long cavity (length changes less than $\sim$0.3\% of the total length).
In contrast, in our optimistic (conservative) scenario we need length changes $\sim$1\% ($\sim$4\%), with the total length changing by $\sim$11~cm ($\sim$1~m).
Achieving these deformations with a single cavity is an additional engineering challenge; if multiple cavities are needed, that would rapidly increase the total project cost.

Mechanical vibrations are also a concern for heterodyne axion searches, as they are in the lumped-element scenarios.
The principal challenge is that the pump mode holds a large field and sits only $\sim$MHz away in frequency from the signal mode, which must be unoccupied; this is qualitatively a similar problem to that encountered in the lumped-element case.
Vibrations of the cavity walls at frequency $m_a$ can scatter pump photons into the signal mode, and a tiny fraction of them would already dominate the noise.
Cavities with azimuthal symmetry like SHADE, however, are saved by symmetry for their TM$_{020}$ (pump) and TE$_{011}$ (signal) modes.
Displacements of the wall along its normal couple the two modes in proportion to a surface integral of ${\bm B}_0 \cdot {\bm B}_1$ and ${\bm E}_0 \cdot {\bm E}_1$~\cite{Slater:1946,Berlin:2019ahk,Lasenby:2019prg}, where the subscripts refer to the pump ($0$) and signal ($1$) modes.
In an azimuthally symmetric cavity ${\bm B}_0$ and ${\bm E}_1$ are purely azimuthal, so these dot products vanish at every point on the wall, however the wall moves.\footnote{Note that this requires a perfectly round cavity; see below.}
This is the analog of the statement that ${\bm w} = 0$ for the ideal pickup toroid, so that vibrations of its encasement do not change the flux through the pickup.

There are thus two sets of concerns for SRF cavities with azimuthal symmetry with regard to vibrations: (i) necessary components that break the azimuthal symmetry, such as the readout coupler and other ports; and (ii) static defects in the wall that break azimuthal symmetry on the scale of the wavelength of the wall's bending waves at $m_a$, about a centimeter.
Point (i) may be mitigated, to some extent, by placing the pickup near the node of the pump field on the wall (see Fig.~\ref{fig:srf_design}).
Estimating the thermal motion of the wall as in Sec.~\ref{sec:realistic_noise}, and assuming for definiteness a 4-mm-thick niobium wall, we find that point (ii) requires the out-of-roundness of the cavity at centimeter wavelengths to be at the 10-$\mu$m scale or less over its full 200~m$^2$ in the optimistic scenario, for the thermal motion of the wall to stay a factor of ten below the thermal photon noise in the signal mode.
Both of these constraints impose additional engineering challenges.

\subsubsection{Cost-optimal cavity geometry}
\label{sec:srf_opt}

\begin{figure*}[tp]
\centering
\includegraphics[width=\textwidth]{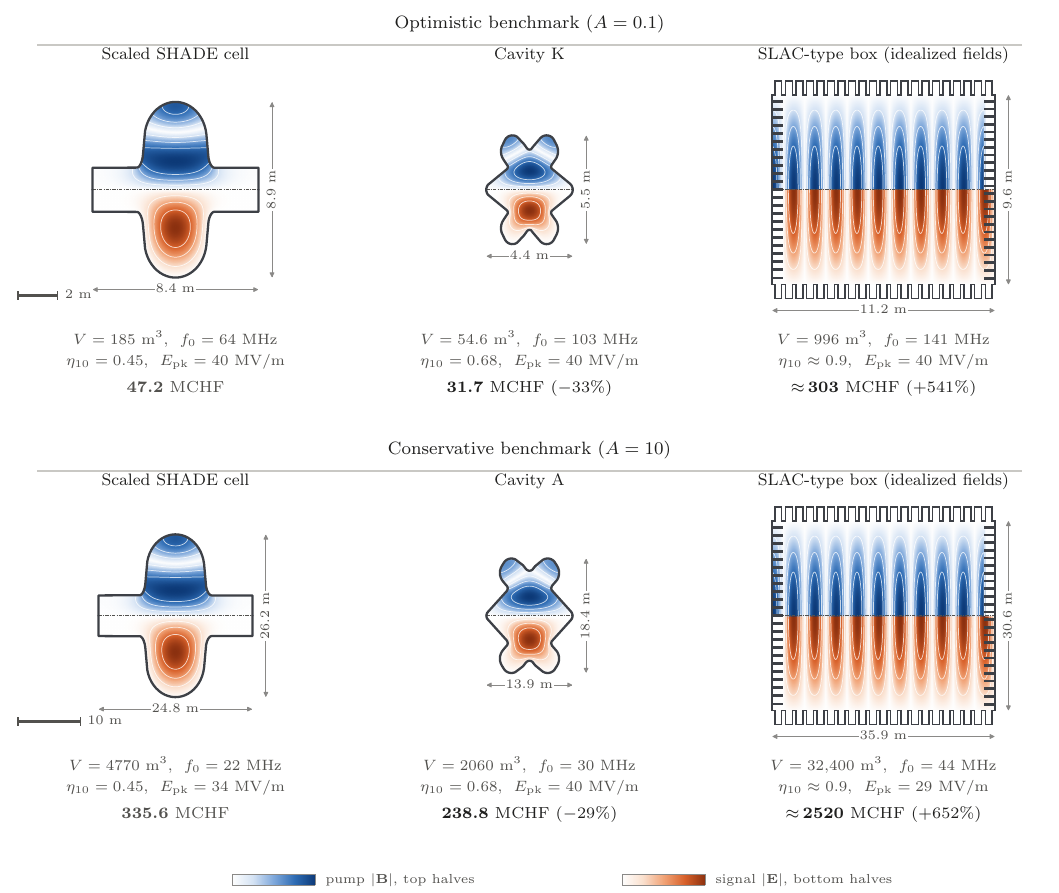}
\caption{
SRF cavities that reach GUT-scale QCD axion sensitivity, as defined through the mass coverage in Tab.~\ref{tab:benchmarks}.
We illustrate the pump fields in the top halves of the cross-sections and the signal mode in the bottom halves.
Note that the scaled SHADE cells and cavities K and A, which we derive in this work, have azimuthal symmetry and are formed by rotation about the symmetry axis given by the dot-dashed lines.
The SLAC-type box, in contrast, is illustrated through its longitudinal cross-section.
The SHADE cell and SLAC-type boxes are scaled in volume with geometric ratios fixed, while cavities K and A are the cost-optimized cavities determined in this work for the optimistic and conservative benchmarks, respectively.
The cost-optimized geometries save around 30\% relative to the scaled SHADE cell.
(See text for details of the optimization procedure.)
}
\label{fig:srf_optimum}
\end{figure*}

The scaled SHADE cavity discussed in the previous subsection did not emerge from an optimization study.
Here, we show that more optimized cavity geometries exist, though we leave a fully exhaustive geometric scan to future work.
First, we note that the SHADE design is not the only one being pursued experimentally at the moment.
At SLAC a corrugated hybrid-mode cavity is being prototyped~\cite{Li:2025pyi}.
That cavity is a rectangular box of 0.474~m $\times$ 0.458~m $\times$ 0.556~m.
Its side walls carry 21 corrugations about a quarter wavelength deep (28 and 34~mm), and its end plates carry quarter-wave fins, oriented orthogonally on the two ends.
The pump and the signal are two orthogonally polarized HE$_{11}$ hybrid modes at 2.86~GHz.
The fins shift the two standing waves by a quarter wavelength, so that the pump magnetic field and the signal electric field line up over most of the box.
The authors estimate $\eta_{10} \approx 0.9$, twice the overlap of the SHADE cell.
We refer to this geometry as the SLAC geometry.
It is illustrated in Fig.~\ref{fig:srf_optimum}.

As for the SHADE geometry, we project sensitivity for the SLAC geometry by scaling it up, keeping the relative dimensions fixed, until the total volume is reached where we hit the required sensitivity to QCD axions.
We find that this geometry is much less optimal than the SHADE geometry, with the optimistic-benchmark cost around 260 MCHF and the conservative benchmark cost over 2,500 MCHF.
One reason for this is that while $\eta_{10}$ is higher for this design than for SHADE, the characteristic frequency $f_0$ is also higher, which hurts the sensitivity and requires a larger total volume $V$.
In addition, the corrugations and fins more than triple the wall area at fixed volume, and at fixed stored energy per unit volume the peak magnetic field on the wall is four times higher than in the SHADE cell.

We attempt to find the optimal cavity designs through the following procedure.
First, let us assume azimuthal symmetry, though below we also comment on non-azimuthally-symmetric designs.
Assume the axis of symmetry of the cavity is the $\bm{\hat z}$ direction, with the unit cell extending from $-L$ to $+L$ in $z$.
For the moment, let us also make the assumption that the cavity is symmetric about $z = 0$, though we also explicitly check designs that break this symmetry.
Under these assumptions, the full geometry is specified by a curve in the $r$-$z$ plane extending from $z = 0$ to $z = L$.
We describe our parametrization of this curve further below.

For a given choice of curve in the $r$-$z$ plane, we use \texttt{axionMultiPhysics} to solve for the modes and compute the quantities needed for the scan rate in~\eqref{eq:SRF_scan}.
We then increase the total volume $V$ until we reach the science goal of each of the two benchmarks.
Given the total volume and the geometry, we may compute the cost according to~\eqref{eq:cost_SRF}.

Our default scan assumes the pump and signal modes are $m=0$ modes, so that they do not depend on the azimuthal coordinate $\phi$. This means that they may be computed efficiently in 2D as opposed to 3D.
Note that the TM$_{mnp}$ and TE$_{mnp}$ modes only exist in a closed cylinder (pillbox).
These modes are not eigenmodes of our cavity, but similar modes still exist, which is why we say, {\it e.g.}, TM$_{020}$-like.
Modes with $m = 0$ split into TE modes, whose electric field is purely azimuthal, and TM modes, whose magnetic field is purely azimuthal.
The integers $n$ and $p$ are the radial and axial mode numbers of the pillbox mode, and we label each mode of our cavities by the pillbox mode with the same pattern of nodes.
Our optimization scan searches for the optimal mode pair in addition to the optimal geometry; our optimized cavities have the TM$_{020}$-like mode as the pump mode and the TE$_{011}$-like mode as the signal mode, as in the SHADE cell.

For a given curve in the $r$-$z$ plane, we stretch or compress the curve in the $z$ direction until the two modes are nearly degenerate; in particular, we require that the pump and signal modes are within one part in $10^4$ of each other.
If this is not possible, we reject the geometry.
We also require that no other mode be within 3.65\% of the pump mode in frequency when the signal and pump are exactly degenerate.
This value is taken from the SHADE geometry and enforces that any new cavity is at least as well isolated as SHADE.
Closer modes may be possible but could be mixed with the pump mode due to imperfections, so we take the SHADE values as a conservative constraint.

We also require that the peak surface electric field of the pump be at most 40~MV/m, to limit field emission.
At high surface electric fields, electrons are emitted from defects on the cavity wall, and this field emission causes both RF loss and noise in the signal mode~\cite{Berlin:2019ahk}.
Modern SRF cavities show little field emission up to peak surface electric fields of $\sim$60~MV/m~\cite{Berlin:2019ahk}, but the cavities considered here have far more wall area than existing cavities, and thus more potential emission sites, so we keep a margin below this value.
For the scaled SHADE cell this requirement lowers the peak wall magnetic field in the optimistic case to about 105~mT.

For geometries that pass the $m = 0$ checks above, we perform 2D solves with \texttt{axionMultiPhysics} for the $m= 0$ degenerate pairs.
We also perform 3D solves for the $m >0$ modes in order to check that there are no $m>0$ modes that are within 3.65\% of the pump mode.
Additionally, we check that the cavity may be tuned over the entire required frequency range.
The splitting between the pump and signal modes is changed by changing the length of the cavity.
As this happens, other modes could approach the pump and signal modes.
We require that over the full tuning range no other mode get within 1\% of the pump or signal mode; the threshold of 1\% is taken from the SHADE design.
Similarly, and motivated by SHADE, we require that on one side of the pump no other mode may lie within 1.2 times the largest splitting of the scan.

We search all pairs of $m = 0$ modes for the pump and signal.
Only a TE and a TM mode with the same $z$-parity have a nonzero overlap $\eta_{10}$, where the $z$-parity is that of the azimuthal field component.
In a cavity with azimuthal symmetry and $z$-symmetry, the $k$-th lowest TM mode of a given parity always lies below the $k$-th lowest TE mode of the same parity, so a TE and a TM mode can only be made degenerate if the TM mode has the higher rank.
We search all such pairs, with either mode as the pump.
The cheapest alternative to the TM$_{020}$-TE$_{011}$ pair, a TM$_{013}$-like pump with a TE$_{012}$-like signal, is more than 20\% more expensive than cavity K (defined below), and at the conservative benchmark we find no alternative pair that satisfies our constraints.
The alternatives need higher frequencies at fixed volume, and their spectra, including the modes with $m > 0$, are denser, which makes it harder to satisfy the mode isolation criterion.

Now that we have described how the cost is computed for a given geometry, we turn to how the geometry itself is parametrized.
We describe the geometry by $2n+1$ nodes, but given the $z$-symmetry it suffices to specify the $n$ nodes with $z > 0$.
(One node is at $z = 0$ and specifies the width in the radial direction at the midplane.)
One of the $n$ nodes lies on the axis, at $r = 0$.
Since the total volume is scaled separately to give the required science reach, and the stretch in the $z$-direction is performed separately to give degenerate pairs, we may fix the height of the axis node and the radius of the midplane node.
The remaining $n - 1$ nodes are free in $r$ and $z$, giving $2(n-1)$ free coordinates.
Beyond the node locations, we describe the geometry by lines connected between nodes with the geometry curved at the nodes themselves.
In particular, each corner is rounded by a circular arc tangent to both of its edges.
The arc begins a distance $t = f\,\ell_{\rm min}$ from the corner, where $\ell_{\rm min}$ is the shorter of the two edges meeting there and $f$ is a separate parameter for each corner.
The radius of each arc is then $\rho = t/\tan(\theta/2)$, with $\theta$ the angle through which the wall turns, so sharp corners receive small arcs and gentle ones large arcs.
The wall may meet the axis at an angle, in which case this corner is rounded by an arc centered on the axis, while the corner at the midplane node is rounded by an arc centered on the midplane, so that the wall stays smooth under reflection.
With $n + 1$ corners, the wall has $3n - 1$ free parameters in total.

Our default scan uses $n = 5$, that is, 11 nodes and 14 free parameters, and we then further refine the optima with $n = 7$ scans starting from the $n=5$ optima; we cannot exclude the possibility that higher-$n$ scans would lead to more substantive improvements.
We perform this scan using the global optimization differential evolution algorithm implemented in \texttt{SciPy}~\cite{Storn:1997,Virtanen:2019joe}.

The optimal geometries found from this procedure are shown in Fig.~\ref{fig:srf_optimum} for the optimistic and conservative geometries; they are around 30\% cheaper than the SHADE-like geometry for the same science reach.
We refer to these geometries as cavities K (optimistic) and A (conservative).
Note that the two geometries differ for the optimistic and conservative benchmarks.
In both cavities the wall meets the axis in a rounded cone and is pinched inward at the midplane, both more pronounced for cavity K.
The costs of these optimal geometries are also given in Fig.~\ref{fig:srf_cost}.
Note that cavity K has a volume of around 55~m$^3$, while cavity A's volume is $\sim$2000~m$^3$.

We additionally check, through more limited searches started from the optima of a simpler 9-node scan, that breaking our symmetry assumptions does not lower the cost.
For example, we also consider breaking $z$-symmetry by letting the upper and lower halves of the wall vary independently.
We additionally allow for $z$-symmetric cavities with non-trivial topology.
Lastly, we perform limited scans of non-axisymmetric cavities, such as nearly circular boxes and cavities with elliptical, fourfold or sixfold deformations, though we note that such cavities would be more prone to vibrational noise.
We find no cavity cheaper than cavities K and A.

\subsection{Comparison to nuclear spin-precession searches}
\label{sec:casper}

All the experiments we have discussed so far use the axion-photon coupling, but it is interesting to compare them to the Cosmic Axion Spin Precession Experiment (CASPEr)~\cite{Graham:2013gfa,Budker:2013hfa,JacksonKimball:2017elr,Aybas:2021nvn}, which uses the axion coupling to gluons directly.
CASPEr makes use of the fact that while the axion sets the static neutron EDM to zero, axion DM gives rise to a residual, oscillating EDM that corresponds to a time-dependent theta angle $\theta(t) = a(t)/f_a$, with amplitude $\theta_0 = \sqrt{2 \rho_{\rm DM}}/(m_a f_a) \approx 3.8 \times 10^{-19}$.
The basic idea behind the experiment is thus the following.
Polarize a sample of nuclear spins along a magnetic field $\bm{B}_0 = B_0 \bm{\hat z}$, so that their Larmor frequency is $\gamma B_0$, and place the sample in an electric field perpendicular to $\bm{\hat z}$.
An EDM then tilts the spins away from $\bm{\hat z}$, but without resonance this tilt time-averages to zero since $\langle \theta(t) \rangle_t = 0$.
On the other hand, if the Larmor frequency equals $m_a$, then the EDM-induced tilt builds through each oscillation cycle instead of canceling, and the sample develops a transverse magnetization that precesses about $\bm{\hat z}$ at the Larmor frequency, where it can be detected with a pickup coil.
This magnetization grows linearly in time until the shorter of the spin and axion coherence times.
In practice the electric field is the large effective field inside a polar crystal, which acts on the nuclear Schiff moment rather than on a bare EDM, as we describe next.

Instead of applying external electric fields, experimental proposals and demonstrations to date use poled lead-based ferroelectric crystals containing $^{207}$Pb, in which the ferroelectric domains are aligned across the sample~\cite{Budker:2013hfa,Aybas:2021nvn}.
This generates an electric field within the sample, though that field is screened at the location of the nucleus~\cite{Schiff:1963zz}.
The nucleus's Schiff moment is not screened, however, and still couples to the electrons around it.
It is conventional to write the resulting energy shifts of the two nuclear spin states along and against the polarization axis as $\pm\Delta\epsilon$, with $\Delta\epsilon = d_n E^*$, as if the $\theta$-induced neutron electric dipole moment, $d_n \approx 2.4 \times 10^{-16}\,\theta\,e\,{\rm cm}$~\cite{Pospelov:1999mv}, sat in an effective field $E^* \approx 0.94$~MV/cm in PbTiO$_3$~\cite{Skripnikov:2016pto,Aybas:2021nvn}.
We emphasize, however, that $E^*$ is not the electric field at the nucleus, but this formula provides a convenient way to compare with EDM searches.

When resonance is achieved, in the frame rotating at the Larmor frequency the magnetization rotates away from $\bm{\hat z}$, about a transverse axis, at the Rabi frequency $\Omega_a = \Delta\epsilon$.
The transverse magnetization therefore grows as $M_\perp(t) = M_0 \Omega_a t$ for small $\Omega_a t$, where $M_0 = n p \gamma/2$ is the prepared longitudinal magnetization of $^{207}$Pb spins with density $n$ and polarization $p$.
This growth continues only while the spins precess in phase with each other and with the driving magnetic field.
First, each nucleus sits in the fluctuating magnetic field of its neighbors, so its precession phase wanders randomly and is lost after the transverse relaxation time $T_2$.
This happens in the same way to every spin, and it gives each spin a Lorentzian response of full width $1/(\pi T_2)$.
Second, spins in different parts of the sample precess at slightly different Larmor frequencies, because $B_0$ is not perfectly uniform and because the surrounding electrons shield $B_0$ by an amount that depends on the orientation of the local crystal axes, the chemical shift anisotropy~\cite{Aybas:2021nvn}.
That is, even if every spin stayed coherent on its own, the spins would eventually fan out in their phases.
For a Lorentzian spread of Larmor frequencies with full width $\Delta\nu_{\rm inh}$, this inhomogeneous broadening makes the net transverse magnetization decay as $e^{-\pi \Delta\nu_{\rm inh} t}$.
The two decay rates add, which defines the effective transverse relaxation time $T_2^*$,
  \es{eq:casper_T2star}{
  {1 \over T_2^*} = {1 \over T_2} + \pi \Delta\nu_{\rm inh} \,,
  }
so that the line of the ensemble has full width $1/(\pi T_2^*)$~\cite{Winter:2026nge}.
As long as the axion field stays coherent over this time, which is the case for all parameters of interest here, dephasing balances the driving force after a time $\sim$$T_2^*$, and the transverse magnetization saturates at $M_1 = M_0 \Omega_a T_2^*$.

Two sources of noise compete with the axion-induced signal.
The first is spin-projection noise: the $N = nV$ spins in a sample of volume $V$ have a random transverse magnetization, since each spin has $\langle S_x^2 \rangle = 1/4$ whatever its polarization~\cite{Aybas:2021cdk,Dror:2022xpi}.
This noise passes through the same spin response as the signal, and it is analogous to thermal noise in, {\it e.g.}, the lumped-element search.
The second noise source is the readout noise of the pickup and SQUID.
At a bias setting with Larmor frequency $\nu_0$, the one-sided power spectral densities of the signal and of the spin-projection noise are
  \es{eq:casper_psd}{
  S_a(\nu) &= {M_1^2 \over 2} \, L(\nu - \nu_0) \, f_a(\nu) \,, \\
  S_{\rm sp}(\nu) &= {\gamma^2 n T_2^* \over 2 V} \, L(\nu - \nu_0) \,,
  }
with $L(x) = [1 + (2\pi x T_2^*)^2]^{-1}$ and $f_a$ the axion line shape, normalized to $\int f_a\, d\nu = 1$~\cite{Dror:2022xpi}.
The readout adds a floor $S_q$, set by the SQUID flux noise referred to the sample through the effective area of the pickup~\cite{Budker:2013hfa} and by the Johnson noise of the pickup.
The readout noise floor thus plays the role of the amplifier noise in the lumped-element searches.
Where spin-projection noise dominates, detuning the bias lowers the signal and the noise together, so a single setting stays sensitive over a band wider than the spin linewidth, $1/(\pi T_2^*)$, by roughly $[S_{\rm sp}(\nu_0)/S_q]^{1/2}$, just as in the lumped-element case.
With $r = S_a/(S_{\rm sp} + S_q)$, the test statistic accumulated over a dwell $t$ at one setting is $\Theta = 2 t \int d\nu \, [\ln(1+r) - r/(1+r)]$, as for the lumped-element designs, which reduces to $t \int d\nu \, r^2$ for $r \ll 1$.
The dwell time per unit frequency needed to reach $\Theta_{95}$ is thus
  \es{eq:casper_scan}{
  {dt \over d\nu_0} = \Theta_{95} \left[ \int \! d\nu_0 \! \int \! d\nu \, 2 \left( \ln(1+r) - {r \over 1+r} \right) \right]^{-1} \,.
  }
This treatment of the signal and the spin-projection noise agrees with Ref.~\cite{Dror:2022xpi}, which found that the signal keeps growing beyond the axion coherence time $\tau_a$ when $\tau_a < T_2$; that regime does not arise here, since $T_2^* < \tau_a$ across our band.
We also emphasize that~\eqref{eq:casper_scan} assumes that each axion frequency is observed for many coherence times.

In the following, let us assume the readout uses a resonant receiver.
This has the advantage that a pickup circuit tuned to $\nu_c \approx \nu_0$ amplifies the signal, the spin-projection noise, and its own Johnson noise by the same gain $G(\nu)$, so the SQUID noise is suppressed by $1/G$ relative to them.
Referred to the sample, the readout floor is then
\es{eq:casper_johnson}{
S_q(\nu) &= {4\, T_c \over \eta\, \omega_0\, Q_c\, V} + {S_{\rm SQ} \over G(\nu)} \,, \\ 
G(\nu) &= {Q_c^2 \over 1 + 4 Q_c^2 (\nu - \nu_c)^2/\nu_c^2} \,,
}
where $\omega_0 = 2\pi\nu_0$, the first term is the Johnson noise, $\eta$ is the filling factor of the pickup, and $S_{\rm SQ}$ is the SQUID noise referred to the sample for an untuned pickup coil.
Near the circuit resonance the Johnson term dominates, and $S_q/S_{\rm sp}(\nu_0) = 8\,T_c/(\eta\,\omega_0\,Q_c\,\gamma^2 n\,T_2^*)$.
Note that raising $Q_c/T_c$ helps only until $S_q \lesssim S_{\rm sp}(\nu_0)$.

We caution that spin-noise-limited readout has not yet been demonstrated for $^{207}$Pb in a ferroelectric or over our frequency range, though it is likely achievable:  Ref.~\cite{Kuenstner:2026fuu} detected the spin-projection noise of $\sim$$5 \times 10^{21}$ $^{19}$F spins in PTFE with a resonant receiver.
Given that, for a fixed quality factor, temperature, and filling factor of the pickup circuit, the ratio of the spin-projection noise to the Johnson noise of the circuit scales as $\omega \gamma^2 n T_2^*$, we compute that for $^{207}$Pb in enriched PbTiO$_3$ this ratio is about 20 times smaller than in Ref.~\cite{Kuenstner:2026fuu} across our mass range, assuming the measured linewidth.
Thus, spin-noise-limited readout could potentially be realized for $^{207}$Pb with a correspondingly higher quality factor or lower temperature.
In the following, however, we adopt the circuit parameters of Ref.~\cite{Kuenstner:2026fuu}, which used a superconducting pickup circuit of quality factor $Q_c = 1.7\times10^4$ and effective temperature $T_c = 670$~mK.

Similarly to the previous sections, we consider an optimistic benchmark, close to projected future performance, and a conservative benchmark, which corresponds to current experimental capabilities.
For the optimistic benchmark we take $p = 1$ and $T_2 = 1$~s from Phase 2 of Ref.~\cite{Budker:2013hfa}.
We also assume PbTiO$_3$ with the lead enriched to 90\% in $^{207}$Pb and $\Delta\nu_{\rm inh} = 10^{-6}\,\nu$.
We use $\Delta\epsilon$ as given above.
In this optimistic case, the resonant readout gives $S_q/S_{\rm sp}(\nu_0) \approx 0.01$--$0.04$.
Our conservative benchmark instead uses the values for polarization and coherence time demonstrated in the laboratory in Ref.~\cite{Aybas:2021nvn}.
We take $p \approx 2 \times 10^{-4}$, the thermal polarization at 4.4~T and 4.2~K of that measurement.
Since our mass range corresponds to bias fields of only 13--110~mT, this assumes that the sample is polarized at high field and that the polarization survives the transfer to the search field.
The measurement used PMN-PT with its natural abundance of $^{207}$Pb, but we keep PbTiO$_3$ enriched to 90\% in $^{207}$Pb, as in the optimistic benchmark.
Ref.~\cite{Aybas:2021nvn} measured $T_2 \approx 16.7$~ms, which we assume, and a linewidth of 78~kHz at 39.7~MHz, or $\Delta\nu_{\rm inh} \approx 2 \times 10^{-3}\,\nu$, which we assume as well.\footnote{Both values were measured in PMN-PT with natural abundance; in enriched PbTiO$_3$ the dipolar couplings are stronger, and $T_2$ is expected to be shorter, $\approx$1.5~ms~\cite{Bouchard:2007tji}.
This would raise the number of required spins $N$ by only $\approx$25\%.}
In this conservative case, the resonant readout gives $S_q/S_{\rm sp}(\nu_0) \approx 20$. We summarize the parameters for each benchmark in Tab.~\ref{tab:casper}.

\begin{table}[!t]
\centering
\caption{Parameters of the optimistic and conservative benchmarks for the spin-precession search, and the resulting number of $^{207}$Pb spins $N$ and sample volume needed to reach the QCD axion over 0.5--4.1~neV. The parameters in the second block are common to both benchmarks.}
\label{tab:casper}
\begin{ruledtabular}
\begin{tabular}{llll}
Quantity & Optimistic & Conservative & Ref. \\
\hline
Polarization $p$ & 1 & $2\times10^{-4}$ & \cite{Budker:2013hfa,Aybas:2021nvn} \\
$T_2$ & 1~s & 16.7~ms & \cite{Budker:2013hfa,Aybas:2021nvn} \\
$\Delta\nu_{\rm inh}/\nu$ & $10^{-6}$ & $2\times10^{-3}$ & \cite{Aybas:2021nvn} \\
$T_2^*$ (at 0.5~neV) & 0.72~s & 1.2~ms & Eq.~\eqref{eq:casper_T2star} \\
\hline
Sample & \multicolumn{2}{l}{PbTiO$_3$, 90\% $^{207}$Pb} & \cite{Mukhamedjanov:2004ki,Budker:2013hfa} \\
$n$ ($^{207}$Pb) & \multicolumn{2}{l}{$1.4\times10^{28}$~m$^{-3}$} & \cite{Nelmes:1985pto} \\
$E^*$ & \multicolumn{2}{l}{0.94~MV/cm} & \cite{Skripnikov:2016pto,Aybas:2021nvn} \\
Bias field $B_0$ & \multicolumn{2}{l}{13--110~mT} & \\
Pickup $Q_c$ & \multicolumn{2}{l}{$1.7\times10^4$} & \cite{Kuenstner:2026fuu} \\
Pickup $T_c$ & \multicolumn{2}{l}{670~mK} & \cite{Kuenstner:2026fuu} \\
Live time & \multicolumn{2}{l}{3~yr} & \\
\hline
Required $N$ & $4\times10^{25}$ & $2\times10^{36}$ & \\
Sample volume & 3~L & $10^8$~m$^3$ & \\
\end{tabular}
\end{ruledtabular}
\end{table}

Under the conservative and optimistic benchmarks we can then calculate the number of spins $N$ that we need to reach the required QCD axion mass sensitivity given in Tab.~\ref{tab:benchmarks} over three years.
We find $N \approx$ $2\times10^{36}$ ($N \approx 4\times10^{25}$) in the conservative (optimistic) scenario, which corresponds to a crystal of volume $10^{8}$~m$^3$ (3 L).
Note that this is a completely infeasible volume in the conservative scenario, so we consider that case unrealistic.
In the optimistic scenario, on the other hand, the entire experiment can be done for a much more modest price than in the cases involving the axion-photon coupling.
The total volume is small, so, {\it e.g.}, cooling costs are negligible compared to those of the axion-photon experiments.
One dominant cost driver is likely the cost of enriching the lead: we estimate this cost at around $2 \times 10^{-2}$~MCHF/kg from the $^{130}$Te enrichment price quoted for ton-scale rare-event experiments~\cite{CUPID:2015tlr}.
Since we need to enrich around 17~kg of lead, the enrichment cost is estimated at around 0.3~MCHF.
Cryogenic costs, using our previously discussed formulae, may add around another 1~MCHF.
This approach thus has the potential to provide a substantially more cost-effective discovery path for the QCD axion, though it rests on assumptions that are more than three orders of magnitude in $p$, and about three in $T_2^*$, away from demonstrated values.

\begin{figure}[!t]
\centering
\includegraphics[width=\columnwidth]{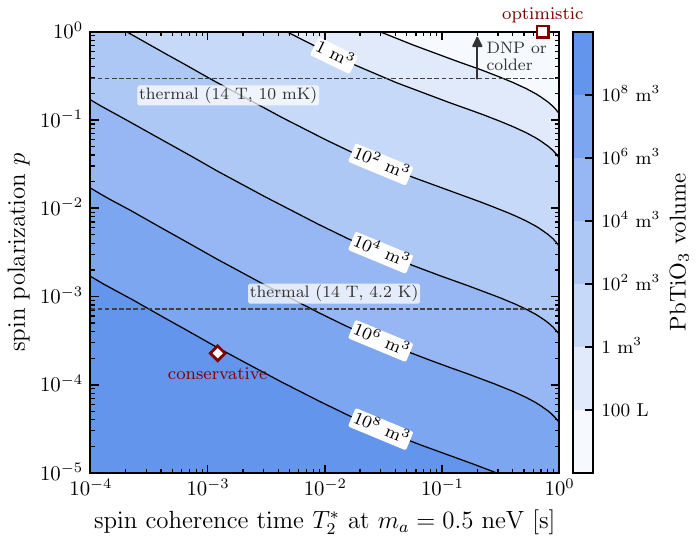}
\caption{Volume of PbTiO$_3$ needed for a spin-precession search to reach the QCD axion over our full mass range in three years, as a function of the spin polarization $p$ and the coherence time $T_2^*$ (at $m_a=0.5$ neV).
Our optimistic and conservative benchmarks are indicated.
We indicate the polarization of the sample achievable by pre-polarizing it thermally in a 14~T field at 10~mK and at 4.2~K (dashed).
Polarizations above the upper line (arrow) require DNP or temperatures below 10~mK.
}
\label{fig:casper_volume}
\end{figure}

The cost also depends strongly on the polarization fraction, as very roughly the required $N$ (and thus the cost) scales as $1/p^2$.
Even $p = 0.1$, which is nearly three orders of magnitude above the polarization demonstrated in this material, would raise the optimistic cost to $\sim$35~MCHF.
Another cost driver is $T_2^*$, as $N \propto  (T_2^*)^{-3/4}$ when spin-projection noise dominates, steepening to $(T_2^*)^{-3/2}$ when the readout noise dominates; these exponents hold for a fixed readout noise, as for the resonant receiver of~\eqref{eq:casper_johnson} and in Fig.~\ref{fig:casper_volume}.
Assuming the fractional inhomogeneous linewidth of $2 \times 10^{-3}$ measured in Ref.~\cite{Aybas:2021nvn} would raise $N$, and thus cost, by a factor of $\sim$2000.

\begin{figure}[!t]
\centering
\includegraphics[width=\columnwidth]{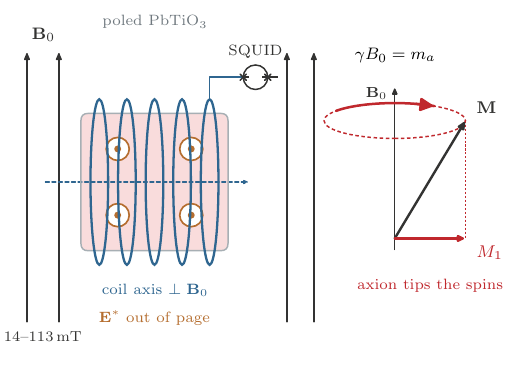}
\caption{A schematic description of the spin-precession search.
The $^{207}$Pb spins in a poled PbTiO$_3$ sample are polarized along the bias field $\bm{B}_0$, and the polarization axis of the ferroelectric, along which $\bm{E}^*$ points, is perpendicular to $\bm{B}_0$ (out of the page).
When $\gamma B_0 = m_a$, the axion tips the magnetization $\bm{M}$ away from $\bm{B}_0$ (the tip angle is exaggerated), and the transverse magnetization $M_1$, which rotates about $\bm{B}_0$ at the Larmor frequency, is read out by a resonant pickup circuit coupled to a dc SQUID.
}
\label{fig:casper_design}
\end{figure}

In Fig.~\ref{fig:casper_volume} we show the dependence of the required sample volume on $p$ and $T_2^*$ for the optimistic and conservative benchmarks.
For reference, polarizing the sample thermally in a 14~T field at 10~mK only gives $p\approx 0.3$.
Achieving a sample of realistic size at the liter scale therefore likely requires both dynamic nuclear polarization (DNP) to $p \sim 1$ and an improvement in $T_2^*$ of more than two orders of magnitude.
We note that no enhanced polarization beyond the thermal value has been demonstrated at present.\footnote{However, a first step towards the former was recently taken in Ref.~\cite{Adam:2026odx}, which showed that illuminating PMN-PT with 405~nm light creates paramagnetic centers that shorten $T_1$ of $^{207}$Pb by a factor of two (such centers are a prerequisite for DNP).}
For $T_2^*$, both the inhomogeneous  and the homogeneous linewidths would need to be significantly improved.
The homogeneous width  is set by dipolar couplings, calculated to be $\approx 230$~Hz in PbTiO$_3$~\cite{Mukhamedjanov:2004ki}, consistent with the measured $T_2 \approx 1.5$~ms in PbTiO$_3$ powder~\cite{Bouchard:2007tji}.
Reaching $T_2 = 1$~s of the optimistic benchmark may require  dynamical decoupling~\cite{Budker:2013hfa}, which has extended nuclear coherence times from milliseconds to tens of seconds in silicon~\cite{Ladd:2005,Dong:2008}.
Such pulse sequences also rescale or average the axion-induced torque, however.
Achieving the optimistic benchmark would also require reducing the inhomogeneous width from the measured value $2\times10^{-3}\,\nu$ down to $10^{-6}\,\nu$.
This calls for a single-domain crystal with sufficiently uniform chemical shifts and a bias field uniform to $10^{-6}$ over a liter-scale sample; at $p\sim1$, the sample's own magnetization, $\mu_0M_0\sim50\,\mu{\rm T}$, must also be made uniform, for example with an ellipsoidal sample.
Finally, at high polarization the back-action of the pickup circuit on the spins (radiation damping) limits the coherence time and must itself be suppressed~\cite{Aybas:2021cdk}.

\section{Discussion}

In this work we ask what is the most cost-effective and robust method to discover a GUT-scale QCD axion, with a mass at or above $m_a \sim 0.5$~neV.
QCD axions with masses at or near 0.5~neV are especially motivated because they are a generic prediction of heterotic string theory~\cite{Svrcek:2006yi,Benabou:2026jtv}, while more broadly extra-dimensional axions that are consistent with grand unification should have masses between $\sim$$10^{-11}$~eV and $10^{-8}$~eV~\cite{Benabou:2025kgx}.

We consider three classes of experiments---lumped-element experiments, heterodyne experiments using SRF cavities, and CASPEr-like experiments using nuclear magnetic resonance---and require that these experiments search an order of magnitude in QCD axion mass at and above $0.5$~neV and reach DFSZ-level sensitivity, as the DFSZ axion-photon coupling is that which arises in theories that undergo grand unification~\cite{Agrawal:2022lsp}.
We conclude that the lumped-element experiments are likely the most cost-effective when restricting to currently available and near-term technology, with the CASPEr approach likely furthest away.
The heterodyne approach, with optimized cavities that we determine in this work, may cost between about the same as and twice as much as the optimized lumped-element detectors, but requires cavities of unprecedented scale and precision, though we see no clear obstruction to constructing such cavities in the future.
We estimate that the magnet and cryogenic systems of lumped-element detectors would cost around 25 (120) MCHF for optimistic (conservative) assumptions for the readout, with the conservative assumptions matching presently available technology.
The full experiment, accounting for all additional expenses such as the rest of the instrumentation and labor, may very roughly cost twice this amount~\cite{Brouwer:2022bwo}.

The magnets needed for the lumped-element detectors are large, with lengths and radii on the order of 10~m in the conservative scenario with field strengths up to 6~T, depending on the topology, but these magnets are far from the biggest magnets built to date, with the ATLAS and CMS magnets, along with fusion reactor magnets, larger.
On the other hand, the heterodyne approach would require an SRF cavity orders of magnitude larger than any existing design and would cost $\sim$30~MCHF ($\sim$240~MCHF) in the optimistic (conservative) scenario for the cost-optimized cavities of Sec.~\ref{sec:srf_opt}, with the cost estimated from the niobium, cryostat, and helium refrigerator only.
In contrast, the CASPEr-like approach, which searches directly for the axion-gluon coupling, is estimated to be infeasible with current technology and would require orders-of-magnitude gains in, {\it e.g.}, the polarization fraction of ferroelectric lead to approach the cost-effectiveness of the lumped-element and heterodyne approaches.

Within the lumped-element category, we consider three detector topologies: the toroid, solenoid, and core.
We show that the toroid and core designs are exact duals to each other and, under ideal conditions, give precisely identical scan rates at fixed cost.
The duality is broken by non-ideal effects; for example, the core design requires a superconducting box that is subject to a large magnetic field.
We require that the field be low enough for the box to be superconducting, which drives the core to larger sizes relative to other designs and requires large iron return yokes to be competitive.
The toroid, on the other hand, is more susceptible to environmental noise and needs to be surrounded by a superconducting shield, which we account for in its cost.
The solenoid geometry has a similar cost to the other designs under ideal conditions but has the disadvantage that the pickup is necessarily in a region of large magnetic field and thus cannot be superconducting without the use of new and expensive materials such as high-$T_c$ superconductors, which we do not consider.
We thus model the solenoid pickup as copper, which lowers its quality factor and requires a lower-temperature stage to not add thermal noise.
In contrast, the pickups of the core and toroid designs can likely be kept at higher temperatures, near 4~K, for example.
With all of these considerations in place, the toroid and core have comparable costs, with the solenoid likely slightly more expensive.

Between the toroid and core designs, the core has the added difficulty of needing large amounts of iron ($\sim$10,000 metric tons in the conservative scenario) and, even with the iron,  keeping the field on its superconducting box below 100~mT makes it 10--20\% more expensive than the toroid with a shield.
Moreover, the iron is modeled as being at room temperature, while the box is, in our ideal design, nearly touching the iron and at 4~K.
Our cost estimate for the core also omits several items that the toroid does not require, at least not to the same extent, such as the additional iron needed once the gradual saturation of real steel is accounted for, the iron support, transportation costs, assembly systems, and the reinforced laboratory space to house such a massive experiment.
A real experiment will likely require more clearance between the 4 K box and the room-temperature iron to accommodate the vacuum vessel, thermal shield, and tuning capacitors, which would also increase the cost, potentially considerably.
Thus, all considered, the toroid appears to be the best-suited design to detect axions at the 0.5-neV mass scale.
The conservative toroid design can be made safe from vibrational noise and electromagnetic interference, given some consideration for how wires and actuators enter the shield and given some level of external shielding for the room containing the experiment.

We release an open-source code package \texttt{axionMultiPhysics}~\href{https://github.com/bsafdi/axionMultiPhysics}{\faGithub} that replicates and improves upon many of the capabilities of COMSOL~\cite{multiphysics1998introduction}.  For the purpose of modeling the electromagnetic properties of axion detectors,
it solves the coupled axion--electromagnetic equations with MQS and full-wave solvers for lumped-element detectors, resonant-cavity haloscopes, and SRF heterodyne cavities.  It also models SQUID readout chains and computes sensitivity reach and cost estimates.
The package models mechanical and thermal responses, and includes a Model Context Protocol server through which AI assistants can build and solve designs from natural-language requests.

The QCD axion with mass near 0.5~neV is one of the most motivated candidates for DM.
We show in this work that this scenario can be discovered or ruled out today for a total cost estimated as roughly 240~MCHF.
This is a small price for understanding the particle nature of DM and gaining insight into the extra-dimensional nature of spacetime.

\section*{Acknowledgments}
We thank Asher Berlin, Karl van Bibber, Giulio Dainelli, Sebastian Ellis, Yoni Kahn, Penelope Quassolo, Nick Rodd, Chiara Salemi, and Kevin Zhou for useful conversations.
We thank Alex Sushkov for guidance on CASPEr benchmarks.
J.B. is supported by the Simons Junior Fellowship.
B.R.S. is supported in part by the DOE award DE-SC0025293.
The research of J.W.F. was supported by the U.S.~Department of Energy under award DE-SC0017647.
This research used the Lawrencium computational cluster resource provided by the IT Division at the Lawrence Berkeley National Laboratory (Supported by the Director, Office of Science, Office of Basic Energy Sciences, of the U.S. Department of Energy under Contract No. DE-AC02-05CH11231).

This work was supported in part by Claude usage credits from Anthropic's AI for Science program.
The authors used the AI models Claude Opus 5.5, Claude Fable 5.1, Claude Fable 5.2, and Claude Sonnet (Anthropic) to help develop the \texttt{axionMultiPhysics}~\href{https://github.com/bsafdi/axionMultiPhysics}{\faGithub} code repository, help run the analyses presented in this work, and proofread human-written text.
The authors reviewed all AI-assisted contributions and take full responsibility for the content of this paper.

\bibliography{refs}

\begin{thebibliography}{150}%
\makeatletter
\providecommand \@ifxundefined [1]{%
 \@ifx{#1\undefined}
}%
\providecommand \@ifnum [1]{%
 \ifnum #1\expandafter \@firstoftwo
 \else \expandafter \@secondoftwo
 \fi
}%
\providecommand \@ifx [1]{%
 \ifx #1\expandafter \@firstoftwo
 \else \expandafter \@secondoftwo
 \fi
}%
\providecommand \natexlab [1]{#1}%
\providecommand \enquote  [1]{``#1''}%
\providecommand \bibnamefont  [1]{#1}%
\providecommand \bibfnamefont [1]{#1}%
\providecommand \citenamefont [1]{#1}%
\providecommand \href@noop [0]{\@secondoftwo}%
\providecommand \href [0]{\begingroup \@sanitize@url \@href}%
\providecommand \@href[1]{\@@startlink{#1}\@@href}%
\providecommand \@@href[1]{\endgroup#1\@@endlink}%
\providecommand \@sanitize@url [0]{\catcode `\\12\catcode `\$12\catcode
  `\&12\catcode `\#12\catcode `\^12\catcode `\_12\catcode `\%12\relax}%
\providecommand \@@startlink[1]{}%
\providecommand \@@endlink[0]{}%
\providecommand \url  [0]{\begingroup\@sanitize@url \@url }%
\providecommand \@url [1]{\endgroup\@href {#1}{\urlprefix }}%
\providecommand \urlprefix  [0]{URL }%
\providecommand \Eprint [0]{\href }%
\providecommand \doibase [0]{http://dx.doi.org/}%
\providecommand \selectlanguage [0]{\@gobble}%
\providecommand \bibinfo  [0]{\@secondoftwo}%
\providecommand \bibfield  [0]{\@secondoftwo}%
\providecommand \translation [1]{[#1]}%
\providecommand \BibitemOpen [0]{}%
\providecommand \bibitemStop [0]{}%
\providecommand \bibitemNoStop [0]{.\EOS\space}%
\providecommand \EOS [0]{\spacefactor3000\relax}%
\providecommand \BibitemShut  [1]{\csname bibitem#1\endcsname}%
\let\auto@bib@innerbib\@empty
\bibitem [{\citenamefont {Peccei}\ and\ \citenamefont
  {Quinn}(1977{\natexlab{a}})}]{Peccei:1977hh}%
  \BibitemOpen
  \bibfield  {author} {\bibinfo {author} {\bibfnamefont {R.~D.}\ \bibnamefont
  {Peccei}}\ and\ \bibinfo {author} {\bibfnamefont {Helen~R.}\ \bibnamefont
  {Quinn}},\ }\bibfield  {title} {\enquote {\bibinfo {title} {{CP Conservation
  in the Presence of Instantons}},}\ }\href {\doibase
  10.1103/PhysRevLett.38.1440} {\bibfield  {journal} {\bibinfo  {journal}
  {Phys. Rev. Lett.}\ }\textbf {\bibinfo {volume} {38}},\ \bibinfo {pages}
  {1440--1443} (\bibinfo {year} {1977}{\natexlab{a}})}\BibitemShut {NoStop}%
\bibitem [{\citenamefont {Peccei}\ and\ \citenamefont
  {Quinn}(1977{\natexlab{b}})}]{Peccei:1977ur}%
  \BibitemOpen
  \bibfield  {author} {\bibinfo {author} {\bibfnamefont {R.~D.}\ \bibnamefont
  {Peccei}}\ and\ \bibinfo {author} {\bibfnamefont {Helen~R.}\ \bibnamefont
  {Quinn}},\ }\bibfield  {title} {\enquote {\bibinfo {title} {{Constraints
  Imposed by CP Conservation in the Presence of Instantons}},}\ }\href
  {\doibase 10.1103/PhysRevD.16.1791} {\bibfield  {journal} {\bibinfo
  {journal} {Phys. Rev.}\ }\textbf {\bibinfo {volume} {D16}},\ \bibinfo {pages}
  {1791--1797} (\bibinfo {year} {1977}{\natexlab{b}})}\BibitemShut {NoStop}%
\bibitem [{\citenamefont {Weinberg}(1978)}]{Weinberg:1977ma}%
  \BibitemOpen
  \bibfield  {author} {\bibinfo {author} {\bibfnamefont {Steven}\ \bibnamefont
  {Weinberg}},\ }\bibfield  {title} {\enquote {\bibinfo {title} {{A New Light
  Boson?}}}\ }\href {\doibase 10.1103/PhysRevLett.40.223} {\bibfield  {journal}
  {\bibinfo  {journal} {Phys. Rev. Lett.}\ }\textbf {\bibinfo {volume} {40}},\
  \bibinfo {pages} {223--226} (\bibinfo {year} {1978})}\BibitemShut {NoStop}%
\bibitem [{\citenamefont {Wilczek}(1978)}]{Wilczek:1977pj}%
  \BibitemOpen
  \bibfield  {author} {\bibinfo {author} {\bibfnamefont {Frank}\ \bibnamefont
  {Wilczek}},\ }\bibfield  {title} {\enquote {\bibinfo {title} {{Problem of
  Strong p and t Invariance in the Presence of Instantons}},}\ }\href {\doibase
  10.1103/PhysRevLett.40.279} {\bibfield  {journal} {\bibinfo  {journal} {Phys.
  Rev. Lett.}\ }\textbf {\bibinfo {volume} {40}},\ \bibinfo {pages} {279--282}
  (\bibinfo {year} {1978})}\BibitemShut {NoStop}%
\bibitem [{\citenamefont {Preskill}\ \emph {et~al.}(1983)\citenamefont
  {Preskill}, \citenamefont {Wise},\ and\ \citenamefont
  {Wilczek}}]{Preskill:1982cy}%
  \BibitemOpen
  \bibfield  {author} {\bibinfo {author} {\bibfnamefont {John}\ \bibnamefont
  {Preskill}}, \bibinfo {author} {\bibfnamefont {Mark~B.}\ \bibnamefont
  {Wise}}, \ and\ \bibinfo {author} {\bibfnamefont {Frank}\ \bibnamefont
  {Wilczek}},\ }\bibfield  {title} {\enquote {\bibinfo {title} {{Cosmology of
  the Invisible Axion}},}\ }\href {\doibase 10.1016/0370-2693(83)90637-8}
  {\bibfield  {journal} {\bibinfo  {journal} {Phys. Lett.}\ }\textbf {\bibinfo
  {volume} {120B}},\ \bibinfo {pages} {127--132} (\bibinfo {year}
  {1983})}\BibitemShut {NoStop}%
\bibitem [{\citenamefont {Abbott}\ and\ \citenamefont
  {Sikivie}(1983)}]{Abbott:1982af}%
  \BibitemOpen
  \bibfield  {author} {\bibinfo {author} {\bibfnamefont {L.~F.}\ \bibnamefont
  {Abbott}}\ and\ \bibinfo {author} {\bibfnamefont {P.}~\bibnamefont
  {Sikivie}},\ }\bibfield  {title} {\enquote {\bibinfo {title} {{A Cosmological
  Bound on the Invisible Axion}},}\ }\href {\doibase
  10.1016/0370-2693(83)90638-X} {\bibfield  {journal} {\bibinfo  {journal}
  {Phys. Lett.}\ }\textbf {\bibinfo {volume} {120B}},\ \bibinfo {pages}
  {133--136} (\bibinfo {year} {1983})}\BibitemShut {NoStop}%
\bibitem [{\citenamefont {Dine}\ and\ \citenamefont
  {Fischler}(1983)}]{Dine:1982ah}%
  \BibitemOpen
  \bibfield  {author} {\bibinfo {author} {\bibfnamefont {Michael}\ \bibnamefont
  {Dine}}\ and\ \bibinfo {author} {\bibfnamefont {Willy}\ \bibnamefont
  {Fischler}},\ }\bibfield  {title} {\enquote {\bibinfo {title} {{The Not So
  Harmless Axion}},}\ }\href {\doibase 10.1016/0370-2693(83)90639-1} {\bibfield
   {journal} {\bibinfo  {journal} {Phys. Lett.}\ }\textbf {\bibinfo {volume}
  {120B}},\ \bibinfo {pages} {137--141} (\bibinfo {year} {1983})}\BibitemShut
  {NoStop}%
\bibitem [{\citenamefont {Witten}(1984)}]{Witten:1984dg}%
  \BibitemOpen
  \bibfield  {author} {\bibinfo {author} {\bibfnamefont {Edward}\ \bibnamefont
  {Witten}},\ }\bibfield  {title} {\enquote {\bibinfo {title} {{Some Properties
  of O(32) Superstrings}},}\ }\href {\doibase 10.1016/0370-2693(84)90422-2}
  {\bibfield  {journal} {\bibinfo  {journal} {Phys. Lett. B}\ }\textbf
  {\bibinfo {volume} {149}},\ \bibinfo {pages} {351--356} (\bibinfo {year}
  {1984})}\BibitemShut {NoStop}%
\bibitem [{\citenamefont {Svrcek}\ and\ \citenamefont
  {Witten}(2006)}]{Svrcek:2006yi}%
  \BibitemOpen
  \bibfield  {author} {\bibinfo {author} {\bibfnamefont {Peter}\ \bibnamefont
  {Svrcek}}\ and\ \bibinfo {author} {\bibfnamefont {Edward}\ \bibnamefont
  {Witten}},\ }\bibfield  {title} {\enquote {\bibinfo {title} {{Axions In
  String Theory}},}\ }\href {\doibase 10.1088/1126-6708/2006/06/051} {\bibfield
   {journal} {\bibinfo  {journal} {JHEP}\ }\textbf {\bibinfo {volume} {06}},\
  \bibinfo {pages} {051} (\bibinfo {year} {2006})},\ \Eprint
  {http://arxiv.org/abs/hep-th/0605206} {arXiv:hep-th/0605206} \BibitemShut
  {NoStop}%
\bibitem [{\citenamefont {Arvanitaki}\ \emph {et~al.}(2010)\citenamefont
  {Arvanitaki}, \citenamefont {Dimopoulos}, \citenamefont {Dubovsky},
  \citenamefont {Kaloper},\ and\ \citenamefont
  {March-Russell}}]{Arvanitaki:2009fg}%
  \BibitemOpen
  \bibfield  {author} {\bibinfo {author} {\bibfnamefont {Asimina}\ \bibnamefont
  {Arvanitaki}}, \bibinfo {author} {\bibfnamefont {Savas}\ \bibnamefont
  {Dimopoulos}}, \bibinfo {author} {\bibfnamefont {Sergei}\ \bibnamefont
  {Dubovsky}}, \bibinfo {author} {\bibfnamefont {Nemanja}\ \bibnamefont
  {Kaloper}}, \ and\ \bibinfo {author} {\bibfnamefont {John}\ \bibnamefont
  {March-Russell}},\ }\bibfield  {title} {\enquote {\bibinfo {title} {{String
  Axiverse}},}\ }\href {\doibase 10.1103/PhysRevD.81.123530} {\bibfield
  {journal} {\bibinfo  {journal} {Phys. Rev. D}\ }\textbf {\bibinfo {volume}
  {81}},\ \bibinfo {pages} {123530} (\bibinfo {year} {2010})},\ \Eprint
  {http://arxiv.org/abs/0905.4720} {arXiv:0905.4720 [hep-th]} \BibitemShut
  {NoStop}%
\bibitem [{\citenamefont {Halverson}\ \emph {et~al.}(2019)\citenamefont
  {Halverson}, \citenamefont {Long}, \citenamefont {Nelson},\ and\
  \citenamefont {Salinas}}]{Halverson:2019cmy}%
  \BibitemOpen
  \bibfield  {author} {\bibinfo {author} {\bibfnamefont {James}\ \bibnamefont
  {Halverson}}, \bibinfo {author} {\bibfnamefont {Cody}\ \bibnamefont {Long}},
  \bibinfo {author} {\bibfnamefont {Brent}\ \bibnamefont {Nelson}}, \ and\
  \bibinfo {author} {\bibfnamefont {Gustavo}\ \bibnamefont {Salinas}},\
  }\bibfield  {title} {\enquote {\bibinfo {title} {{Towards string theory
  expectations for photon couplings to axionlike particles}},}\ }\href
  {\doibase 10.1103/PhysRevD.100.106010} {\bibfield  {journal} {\bibinfo
  {journal} {Phys. Rev. D}\ }\textbf {\bibinfo {volume} {100}},\ \bibinfo
  {pages} {106010} (\bibinfo {year} {2019})},\ \Eprint
  {http://arxiv.org/abs/1909.05257} {arXiv:1909.05257 [hep-th]} \BibitemShut
  {NoStop}%
\bibitem [{\citenamefont {Benabou}\ \emph {et~al.}(2025)\citenamefont
  {Benabou}, \citenamefont {Fraser}, \citenamefont {Reig},\ and\ \citenamefont
  {Safdi}}]{Benabou:2025kgx}%
  \BibitemOpen
  \bibfield  {author} {\bibinfo {author} {\bibfnamefont {Joshua~N.}\
  \bibnamefont {Benabou}}, \bibinfo {author} {\bibfnamefont {Katherine}\
  \bibnamefont {Fraser}}, \bibinfo {author} {\bibfnamefont {Mario}\
  \bibnamefont {Reig}}, \ and\ \bibinfo {author} {\bibfnamefont {Benjamin~R.}\
  \bibnamefont {Safdi}},\ }\bibfield  {title} {\enquote {\bibinfo {title}
  {{String theory and grand unification suggest a submicroelectronvolt QCD
  axion}},}\ }\href {\doibase 10.1103/lthr-97lm} {\bibfield  {journal}
  {\bibinfo  {journal} {Phys. Rev. D}\ }\textbf {\bibinfo {volume} {112}},\
  \bibinfo {pages} {066003} (\bibinfo {year} {2025})},\ \Eprint
  {http://arxiv.org/abs/2505.15884} {arXiv:2505.15884 [hep-ph]} \BibitemShut
  {NoStop}%
\bibitem [{\citenamefont {Choi}\ and\ \citenamefont {Kim}(1985)}]{Choi:1985bz}%
  \BibitemOpen
  \bibfield  {author} {\bibinfo {author} {\bibfnamefont {Kiwoon}\ \bibnamefont
  {Choi}}\ and\ \bibinfo {author} {\bibfnamefont {Jihn~E.}\ \bibnamefont
  {Kim}},\ }\bibfield  {title} {\enquote {\bibinfo {title} {{Compactification
  and Axions in E(8) x E(8)-prime Superstring Models}},}\ }\href {\doibase
  10.1016/0370-2693(85)90693-8} {\bibfield  {journal} {\bibinfo  {journal}
  {Phys. Lett. B}\ }\textbf {\bibinfo {volume} {165}},\ \bibinfo {pages}
  {71--75} (\bibinfo {year} {1985})}\BibitemShut {NoStop}%
\bibitem [{\citenamefont {Benabou}\ \emph
  {et~al.}(2026{\natexlab{a}})\citenamefont {Benabou}, \citenamefont
  {Dainelli}, \citenamefont {Reig},\ and\ \citenamefont
  {Safdi}}]{Benabou:2026jtv}%
  \BibitemOpen
  \bibfield  {author} {\bibinfo {author} {\bibfnamefont {Joshua~N.}\
  \bibnamefont {Benabou}}, \bibinfo {author} {\bibfnamefont {Giulio~Alvise}\
  \bibnamefont {Dainelli}}, \bibinfo {author} {\bibfnamefont {Mario}\
  \bibnamefont {Reig}}, \ and\ \bibinfo {author} {\bibfnamefont {Benjamin~R.}\
  \bibnamefont {Safdi}},\ }\bibfield  {title} {\enquote {\bibinfo {title}
  {{Heterotic String Theory Suggests a QCD Axion Near 0.5 neV}},}\ }\href@noop
  {} {\  (\bibinfo {year} {2026}{\natexlab{a}})},\ \bibinfo {note}
  {submitted},\ \Eprint {http://arxiv.org/abs/2605.04142} {arXiv:2605.04142
  [hep-ph]} \BibitemShut {NoStop}%
\bibitem [{\citenamefont {Benabou}\ \emph
  {et~al.}(2026{\natexlab{b}})\citenamefont {Benabou}, \citenamefont
  {Dainelli}, \citenamefont {Harvey}, \citenamefont {Reig},\ and\ \citenamefont
  {Safdi}}]{Benabou:2026inprep}%
  \BibitemOpen
  \bibfield  {author} {\bibinfo {author} {\bibfnamefont {Joshua~N.}\
  \bibnamefont {Benabou}}, \bibinfo {author} {\bibfnamefont {Giulio~Alvise}\
  \bibnamefont {Dainelli}}, \bibinfo {author} {\bibfnamefont {Thomas~R.}\
  \bibnamefont {Harvey}}, \bibinfo {author} {\bibfnamefont {Mario}\
  \bibnamefont {Reig}}, \ and\ \bibinfo {author} {\bibfnamefont {Benjamin~R.}\
  \bibnamefont {Safdi}},\ }\href@noop {} {\enquote {\bibinfo {title} {{in
  preparation}},}\ } (\bibinfo {year} {2026}{\natexlab{b}})\BibitemShut
  {NoStop}%
\bibitem [{\citenamefont {Sikivie}\ \emph {et~al.}(2014)\citenamefont
  {Sikivie}, \citenamefont {Sullivan},\ and\ \citenamefont
  {Tanner}}]{Sikivie:2013laa}%
  \BibitemOpen
  \bibfield  {author} {\bibinfo {author} {\bibfnamefont {P.}~\bibnamefont
  {Sikivie}}, \bibinfo {author} {\bibfnamefont {N.}~\bibnamefont {Sullivan}}, \
  and\ \bibinfo {author} {\bibfnamefont {D.~B.}\ \bibnamefont {Tanner}},\
  }\bibfield  {title} {\enquote {\bibinfo {title} {{Proposal for Axion Dark
  Matter Detection Using an LC Circuit}},}\ }\href {\doibase
  10.1103/PhysRevLett.112.131301} {\bibfield  {journal} {\bibinfo  {journal}
  {Phys. Rev. Lett.}\ }\textbf {\bibinfo {volume} {112}},\ \bibinfo {pages}
  {131301} (\bibinfo {year} {2014})},\ \Eprint {http://arxiv.org/abs/1310.8545}
  {arXiv:1310.8545 [hep-ph]} \BibitemShut {NoStop}%
\bibitem [{\citenamefont {Kahn}\ \emph {et~al.}(2016)\citenamefont {Kahn},
  \citenamefont {Safdi},\ and\ \citenamefont {Thaler}}]{Kahn:2016aff}%
  \BibitemOpen
  \bibfield  {author} {\bibinfo {author} {\bibfnamefont {Yonatan}\ \bibnamefont
  {Kahn}}, \bibinfo {author} {\bibfnamefont {Benjamin~R.}\ \bibnamefont
  {Safdi}}, \ and\ \bibinfo {author} {\bibfnamefont {Jesse}\ \bibnamefont
  {Thaler}},\ }\bibfield  {title} {\enquote {\bibinfo {title} {{Broadband and
  Resonant Approaches to Axion Dark Matter Detection}},}\ }\href {\doibase
  10.1103/PhysRevLett.117.141801} {\bibfield  {journal} {\bibinfo  {journal}
  {Phys. Rev. Lett.}\ }\textbf {\bibinfo {volume} {117}},\ \bibinfo {pages}
  {141801} (\bibinfo {year} {2016})},\ \Eprint
  {http://arxiv.org/abs/1602.01086} {arXiv:1602.01086 [hep-ph]} \BibitemShut
  {NoStop}%
\bibitem [{\citenamefont {Berlin}\ \emph {et~al.}(2020)\citenamefont {Berlin},
  \citenamefont {D'Agnolo}, \citenamefont {Ellis}, \citenamefont {Nantista},
  \citenamefont {Neilson}, \citenamefont {Schuster}, \citenamefont {Tantawi},
  \citenamefont {Toro},\ and\ \citenamefont {Zhou}}]{Berlin:2019ahk}%
  \BibitemOpen
  \bibfield  {author} {\bibinfo {author} {\bibfnamefont {Asher}\ \bibnamefont
  {Berlin}}, \bibinfo {author} {\bibfnamefont {Raffaele~Tito}\ \bibnamefont
  {D'Agnolo}}, \bibinfo {author} {\bibfnamefont {Sebastian A.~R.}\ \bibnamefont
  {Ellis}}, \bibinfo {author} {\bibfnamefont {Christopher}\ \bibnamefont
  {Nantista}}, \bibinfo {author} {\bibfnamefont {Jeffrey}\ \bibnamefont
  {Neilson}}, \bibinfo {author} {\bibfnamefont {Philip}\ \bibnamefont
  {Schuster}}, \bibinfo {author} {\bibfnamefont {Sami}\ \bibnamefont
  {Tantawi}}, \bibinfo {author} {\bibfnamefont {Natalia}\ \bibnamefont {Toro}},
  \ and\ \bibinfo {author} {\bibfnamefont {Kevin}\ \bibnamefont {Zhou}},\
  }\bibfield  {title} {\enquote {\bibinfo {title} {{Axion Dark Matter Detection
  by Superconducting Resonant Frequency Conversion}},}\ }\href {\doibase
  10.1007/JHEP07(2020)088} {\bibfield  {journal} {\bibinfo  {journal} {JHEP}\
  }\textbf {\bibinfo {volume} {07}},\ \bibinfo {pages} {088} (\bibinfo {year}
  {2020})},\ \Eprint {http://arxiv.org/abs/1912.11048} {arXiv:1912.11048
  [hep-ph]} \BibitemShut {NoStop}%
\bibitem [{\citenamefont {Berlin}\ \emph {et~al.}(2021)\citenamefont {Berlin},
  \citenamefont {D'Agnolo}, \citenamefont {Ellis},\ and\ \citenamefont
  {Zhou}}]{Berlin:2020vrk}%
  \BibitemOpen
  \bibfield  {author} {\bibinfo {author} {\bibfnamefont {Asher}\ \bibnamefont
  {Berlin}}, \bibinfo {author} {\bibfnamefont {Raffaele~Tito}\ \bibnamefont
  {D'Agnolo}}, \bibinfo {author} {\bibfnamefont {Sebastian A.~R.}\ \bibnamefont
  {Ellis}}, \ and\ \bibinfo {author} {\bibfnamefont {Kevin}\ \bibnamefont
  {Zhou}},\ }\bibfield  {title} {\enquote {\bibinfo {title} {{Heterodyne
  broadband detection of axion dark matter}},}\ }\href {\doibase
  10.1103/PhysRevD.104.L111701} {\bibfield  {journal} {\bibinfo  {journal}
  {Phys. Rev. D}\ }\textbf {\bibinfo {volume} {104}},\ \bibinfo {pages}
  {L111701} (\bibinfo {year} {2021})},\ \Eprint
  {http://arxiv.org/abs/2007.15656} {arXiv:2007.15656 [hep-ph]} \BibitemShut
  {NoStop}%
\bibitem [{\citenamefont {Lasenby}(2020)}]{Lasenby:2019prg}%
  \BibitemOpen
  \bibfield  {author} {\bibinfo {author} {\bibfnamefont {Robert}\ \bibnamefont
  {Lasenby}},\ }\bibfield  {title} {\enquote {\bibinfo {title} {{Microwave
  cavity searches for low-frequency axion dark matter}},}\ }\href {\doibase
  10.1103/PhysRevD.102.015008} {\bibfield  {journal} {\bibinfo  {journal}
  {Phys. Rev. D}\ }\textbf {\bibinfo {volume} {102}},\ \bibinfo {pages}
  {015008} (\bibinfo {year} {2020})},\ \Eprint
  {http://arxiv.org/abs/1912.11056} {arXiv:1912.11056 [hep-ph]} \BibitemShut
  {NoStop}%
\bibitem [{\citenamefont {Graham}\ and\ \citenamefont
  {Rajendran}(2013)}]{Graham:2013gfa}%
  \BibitemOpen
  \bibfield  {author} {\bibinfo {author} {\bibfnamefont {Peter~W.}\
  \bibnamefont {Graham}}\ and\ \bibinfo {author} {\bibfnamefont {Surjeet}\
  \bibnamefont {Rajendran}},\ }\bibfield  {title} {\enquote {\bibinfo {title}
  {{New Observables for Direct Detection of Axion Dark Matter}},}\ }\href
  {\doibase 10.1103/PhysRevD.88.035023} {\bibfield  {journal} {\bibinfo
  {journal} {Phys. Rev. D}\ }\textbf {\bibinfo {volume} {88}},\ \bibinfo
  {pages} {035023} (\bibinfo {year} {2013})},\ \Eprint
  {http://arxiv.org/abs/1306.6088} {arXiv:1306.6088 [hep-ph]} \BibitemShut
  {NoStop}%
\bibitem [{\citenamefont {Budker}\ \emph {et~al.}(2014)\citenamefont {Budker},
  \citenamefont {Graham}, \citenamefont {Ledbetter}, \citenamefont
  {Rajendran},\ and\ \citenamefont {Sushkov}}]{Budker:2013hfa}%
  \BibitemOpen
  \bibfield  {author} {\bibinfo {author} {\bibfnamefont {Dmitry}\ \bibnamefont
  {Budker}}, \bibinfo {author} {\bibfnamefont {Peter~W.}\ \bibnamefont
  {Graham}}, \bibinfo {author} {\bibfnamefont {Micah}\ \bibnamefont
  {Ledbetter}}, \bibinfo {author} {\bibfnamefont {Surjeet}\ \bibnamefont
  {Rajendran}}, \ and\ \bibinfo {author} {\bibfnamefont {Alex}\ \bibnamefont
  {Sushkov}},\ }\bibfield  {title} {\enquote {\bibinfo {title} {{Proposal for a
  Cosmic Axion Spin Precession Experiment (CASPEr)}},}\ }\href {\doibase
  10.1103/PhysRevX.4.021030} {\bibfield  {journal} {\bibinfo  {journal} {Phys.
  Rev. X}\ }\textbf {\bibinfo {volume} {4}},\ \bibinfo {pages} {021030}
  (\bibinfo {year} {2014})},\ \Eprint {http://arxiv.org/abs/1306.6089}
  {arXiv:1306.6089 [hep-ph]} \BibitemShut {NoStop}%
\bibitem [{\citenamefont {Jackson~Kimball}\ \emph {et~al.}(2020)\citenamefont
  {Jackson~Kimball} \emph {et~al.}}]{JacksonKimball:2017elr}%
  \BibitemOpen
  \bibfield  {author} {\bibinfo {author} {\bibfnamefont {Derek~F.}\
  \bibnamefont {Jackson~Kimball}} \emph {et~al.},\ }\bibfield  {title}
  {\enquote {\bibinfo {title} {{Overview of the Cosmic Axion Spin Precession
  Experiment (CASPEr)}},}\ }\href {\doibase 10.1007/978-3-030-43761-9_13}
  {\bibfield  {journal} {\bibinfo  {journal} {Springer Proc. Phys.}\ }\textbf
  {\bibinfo {volume} {245}},\ \bibinfo {pages} {105--121} (\bibinfo {year}
  {2020})},\ \Eprint {http://arxiv.org/abs/1711.08999} {arXiv:1711.08999
  [physics.ins-det]} \BibitemShut {NoStop}%
\bibitem [{\citenamefont {Aybas}\ \emph
  {et~al.}(2021{\natexlab{a}})\citenamefont {Aybas} \emph
  {et~al.}}]{Aybas:2021nvn}%
  \BibitemOpen
  \bibfield  {author} {\bibinfo {author} {\bibfnamefont {Deniz}\ \bibnamefont
  {Aybas}} \emph {et~al.},\ }\bibfield  {title} {\enquote {\bibinfo {title}
  {{Search for Axionlike Dark Matter Using Solid-State Nuclear Magnetic
  Resonance}},}\ }\href {\doibase 10.1103/PhysRevLett.126.141802} {\bibfield
  {journal} {\bibinfo  {journal} {Phys. Rev. Lett.}\ }\textbf {\bibinfo
  {volume} {126}},\ \bibinfo {pages} {141802} (\bibinfo {year}
  {2021}{\natexlab{a}})},\ \Eprint {http://arxiv.org/abs/2101.01241}
  {arXiv:2101.01241 [hep-ex]} \BibitemShut {NoStop}%
\bibitem [{\citenamefont {Benabou}\ \emph {et~al.}(2023)\citenamefont
  {Benabou}, \citenamefont {Foster}, \citenamefont {Kahn}, \citenamefont
  {Safdi},\ and\ \citenamefont {Salemi}}]{Benabou:2022qpv}%
  \BibitemOpen
  \bibfield  {author} {\bibinfo {author} {\bibfnamefont {Joshua~N.}\
  \bibnamefont {Benabou}}, \bibinfo {author} {\bibfnamefont {Joshua~W.}\
  \bibnamefont {Foster}}, \bibinfo {author} {\bibfnamefont {Yonatan}\
  \bibnamefont {Kahn}}, \bibinfo {author} {\bibfnamefont {Benjamin~R.}\
  \bibnamefont {Safdi}}, \ and\ \bibinfo {author} {\bibfnamefont {Chiara~P.}\
  \bibnamefont {Salemi}},\ }\bibfield  {title} {\enquote {\bibinfo {title}
  {{Lumped-element axion dark matter detection beyond the magnetoquasistatic
  limit}},}\ }\href {\doibase 10.1103/PhysRevD.108.035009} {\bibfield
  {journal} {\bibinfo  {journal} {Phys. Rev. D}\ }\textbf {\bibinfo {volume}
  {108}},\ \bibinfo {pages} {035009} (\bibinfo {year} {2023})},\ \Eprint
  {http://arxiv.org/abs/2211.00008} {arXiv:2211.00008 [hep-ph]} \BibitemShut
  {NoStop}%
\bibitem [{\citenamefont {AlShirawi}\ \emph {et~al.}(2023)\citenamefont
  {AlShirawi} \emph {et~al.}}]{DMRadio:2023igr}%
  \BibitemOpen
  \bibfield  {author} {\bibinfo {author} {\bibfnamefont {A.}~\bibnamefont
  {AlShirawi}} \emph {et~al.} (\bibinfo {collaboration} {DMRadio}),\ }\bibfield
   {title} {\enquote {\bibinfo {title} {{Electromagnetic modeling and science
  reach of DMRadio-m$^3$}},}\ }\href@noop {} {\  (\bibinfo {year} {2023})},\
  \Eprint {http://arxiv.org/abs/2302.14084} {arXiv:2302.14084 [hep-ex]}
  \BibitemShut {NoStop}%
\bibitem [{\citenamefont {Ouellet}\ \emph
  {et~al.}(2019{\natexlab{a}})\citenamefont {Ouellet} \emph
  {et~al.}}]{Ouellet:2019tlz}%
  \BibitemOpen
  \bibfield  {author} {\bibinfo {author} {\bibfnamefont {Jonathan~L.}\
  \bibnamefont {Ouellet}} \emph {et~al.},\ }\bibfield  {title} {\enquote
  {\bibinfo {title} {{Design and implementation of the ABRACADABRA-10 cm axion
  dark matter search}},}\ }\href {\doibase 10.1103/PhysRevD.99.052012}
  {\bibfield  {journal} {\bibinfo  {journal} {Phys. Rev. D}\ }\textbf {\bibinfo
  {volume} {99}},\ \bibinfo {pages} {052012} (\bibinfo {year}
  {2019}{\natexlab{a}})},\ \Eprint {http://arxiv.org/abs/1901.10652}
  {arXiv:1901.10652 [physics.ins-det]} \BibitemShut {NoStop}%
\bibitem [{\citenamefont {Ouellet}\ \emph
  {et~al.}(2019{\natexlab{b}})\citenamefont {Ouellet} \emph
  {et~al.}}]{Ouellet:2018beu}%
  \BibitemOpen
  \bibfield  {author} {\bibinfo {author} {\bibfnamefont {Jonathan~L.}\
  \bibnamefont {Ouellet}} \emph {et~al.},\ }\bibfield  {title} {\enquote
  {\bibinfo {title} {{First Results from ABRACADABRA-10 cm: A Search for
  Sub-$\mu$eV Axion Dark Matter}},}\ }\href {\doibase
  10.1103/PhysRevLett.122.121802} {\bibfield  {journal} {\bibinfo  {journal}
  {Phys. Rev. Lett.}\ }\textbf {\bibinfo {volume} {122}},\ \bibinfo {pages}
  {121802} (\bibinfo {year} {2019}{\natexlab{b}})},\ \Eprint
  {http://arxiv.org/abs/1810.12257} {arXiv:1810.12257 [hep-ex]} \BibitemShut
  {NoStop}%
\bibitem [{\citenamefont {Salemi}\ \emph {et~al.}(2021)\citenamefont {Salemi}
  \emph {et~al.}}]{Salemi:2021gck}%
  \BibitemOpen
  \bibfield  {author} {\bibinfo {author} {\bibfnamefont {Chiara~P.}\
  \bibnamefont {Salemi}} \emph {et~al.},\ }\bibfield  {title} {\enquote
  {\bibinfo {title} {{Search for Low-Mass Axion Dark Matter with
  ABRACADABRA-10~cm}},}\ }\href {\doibase 10.1103/PhysRevLett.127.081801}
  {\bibfield  {journal} {\bibinfo  {journal} {Phys. Rev. Lett.}\ }\textbf
  {\bibinfo {volume} {127}},\ \bibinfo {pages} {081801} (\bibinfo {year}
  {2021})},\ \Eprint {http://arxiv.org/abs/2102.06722} {arXiv:2102.06722
  [hep-ex]} \BibitemShut {NoStop}%
\bibitem [{\citenamefont {Rapidis}(2023)}]{Rapidis:2022gti}%
  \BibitemOpen
  \bibfield  {author} {\bibinfo {author} {\bibfnamefont {Nicholas~Marcos}\
  \bibnamefont {Rapidis}} (\bibinfo {collaboration} {DMRadio}),\ }\bibfield
  {title} {\enquote {\bibinfo {title} {{Status of DMRadio-50L and
  DMRadio-m$^3$}},}\ }\href {\doibase 10.21468/SciPostPhysProc.12.036}
  {\bibfield  {journal} {\bibinfo  {journal} {SciPost Phys. Proc.}\ }\textbf
  {\bibinfo {volume} {12}},\ \bibinfo {pages} {036} (\bibinfo {year} {2023})},\
  \Eprint {http://arxiv.org/abs/2210.07215} {arXiv:2210.07215 [hep-ex]}
  \BibitemShut {NoStop}%
\bibitem [{\citenamefont {Ankel}\ \emph
  {et~al.}(2026{\natexlab{a}})\citenamefont {Ankel} \emph
  {et~al.}}]{Ankel:2026cef}%
  \BibitemOpen
  \bibfield  {author} {\bibinfo {author} {\bibfnamefont {V.}~\bibnamefont
  {Ankel}} \emph {et~al.},\ }\bibfield  {title} {\enquote {\bibinfo {title}
  {{The Cryogenic System of DMRadio-50L}},}\ }\href@noop {} {\  (\bibinfo
  {year} {2026}{\natexlab{a}})},\ \Eprint {http://arxiv.org/abs/2607.05771}
  {arXiv:2607.05771 [physics.ins-det]} \BibitemShut {NoStop}%
\bibitem [{\citenamefont {Brouwer}\ \emph
  {et~al.}(2022{\natexlab{a}})\citenamefont {Brouwer} \emph
  {et~al.}}]{DMRadio:2022pkf}%
  \BibitemOpen
  \bibfield  {author} {\bibinfo {author} {\bibfnamefont {L.}~\bibnamefont
  {Brouwer}} \emph {et~al.} (\bibinfo {collaboration} {DMRadio}),\ }\bibfield
  {title} {\enquote {\bibinfo {title} {{DMRadio-m$^3$: A Search for the QCD
  Axion Below $1\,\mu$eV}},}\ }\href@noop {} {\  (\bibinfo {year}
  {2022}{\natexlab{a}})},\ \Eprint {http://arxiv.org/abs/2204.13781}
  {arXiv:2204.13781 [hep-ex]} \BibitemShut {NoStop}%
\bibitem [{\citenamefont {Puranam}(2026)}]{Puranam:2026cpad}%
  \BibitemOpen
  \bibfield  {author} {\bibinfo {author} {\bibfnamefont {Shreya}\ \bibnamefont
  {Puranam}},\ }\href@noop {} {\enquote {\bibinfo {title} {{The CAL-Pathfinder
  Axion Detector}},}\ }\bibinfo {howpublished}
  {\url{https://indico.global/event/17230/contributions/170455/}} (\bibinfo
  {year} {2026}),\ \bibinfo {note} {talk at the CPAD 2026 Workshop (APS DPF
  Coordinating Panel on Advanced Detectors), University of Washington, Seattle,
  21 October 2026}\BibitemShut {NoStop}%
\bibitem [{\citenamefont {Brouwer}\ \emph
  {et~al.}(2022{\natexlab{b}})\citenamefont {Brouwer} \emph
  {et~al.}}]{Brouwer:2022bwo}%
  \BibitemOpen
  \bibfield  {author} {\bibinfo {author} {\bibfnamefont {L.}~\bibnamefont
  {Brouwer}} \emph {et~al.},\ }\bibfield  {title} {\enquote {\bibinfo {title}
  {{Introducing DMRadio-GUT, a search for GUT-scale QCD axions}},}\ }\href@noop
  {} {\  (\bibinfo {year} {2022}{\natexlab{b}})},\ \Eprint
  {http://arxiv.org/abs/2203.11246} {arXiv:2203.11246 [hep-ex]} \BibitemShut
  {NoStop}%
\bibitem [{\citenamefont {Ankel}\ \emph
  {et~al.}(2026{\natexlab{b}})\citenamefont {Ankel} \emph
  {et~al.}}]{Ankel:2026zrv}%
  \BibitemOpen
  \bibfield  {author} {\bibinfo {author} {\bibfnamefont {V.}~\bibnamefont
  {Ankel}} \emph {et~al.},\ }\bibfield  {title} {\enquote {\bibinfo {title}
  {{DMRadio-Core: A new approach for GUT-scale axion searches}},}\ }\href@noop
  {} {\  (\bibinfo {year} {2026}{\natexlab{b}})},\ \Eprint
  {http://arxiv.org/abs/2604.16602} {arXiv:2604.16602 [hep-ex]} \BibitemShut
  {NoStop}%
\bibitem [{axi(2026)}]{axionMultiPhysics}%
  \BibitemOpen
  \href {https://github.com/bsafdi/axionMultiPhysics} {\enquote {\bibinfo
  {title} {{axionMultiPhysics: finite-element workbench for lumped-element
  axion dark-matter detectors}},}\ } (\bibinfo {year} {2026})\BibitemShut
  {NoStop}%
\bibitem [{mul()}]{multiphysics1998introduction}%
  \BibitemOpen
  \href@noop {} {\enquote {\bibinfo {title} {{COMSOL
  Multiphysics\textregistered{} v.~6.4}},}\ }\bibinfo {howpublished}
  {\url{www.comsol.com}},\ \bibinfo {note} {{COMSOL AB}, Stockholm,
  Sweden}\BibitemShut {NoStop}%
\bibitem [{\citenamefont {Anderson}\ \emph {et~al.}(2021)\citenamefont
  {Anderson}, \citenamefont {Andrej}, \citenamefont {Barker}, \citenamefont
  {Bramwell}, \citenamefont {Camier}, \citenamefont {Cerveny}, \citenamefont
  {Dobrev}, \citenamefont {Dudouit}, \citenamefont {Fisher}, \citenamefont
  {Kolev}, \citenamefont {Pazner}, \citenamefont {Stowell}, \citenamefont
  {Tomov}, \citenamefont {Akkerman}, \citenamefont {Dahm}, \citenamefont
  {Medina},\ and\ \citenamefont {Zampini}}]{mfem}%
  \BibitemOpen
  \bibfield  {author} {\bibinfo {author} {\bibfnamefont {Robert}\ \bibnamefont
  {Anderson}}, \bibinfo {author} {\bibfnamefont {Julian}\ \bibnamefont
  {Andrej}}, \bibinfo {author} {\bibfnamefont {Andrew}\ \bibnamefont {Barker}},
  \bibinfo {author} {\bibfnamefont {Jamie}\ \bibnamefont {Bramwell}}, \bibinfo
  {author} {\bibfnamefont {Jean-Sylvain}\ \bibnamefont {Camier}}, \bibinfo
  {author} {\bibfnamefont {Jakub}\ \bibnamefont {Cerveny}}, \bibinfo {author}
  {\bibfnamefont {Veselin}\ \bibnamefont {Dobrev}}, \bibinfo {author}
  {\bibfnamefont {Yohann}\ \bibnamefont {Dudouit}}, \bibinfo {author}
  {\bibfnamefont {Aaron}\ \bibnamefont {Fisher}}, \bibinfo {author}
  {\bibfnamefont {Tzanio}\ \bibnamefont {Kolev}}, \bibinfo {author}
  {\bibfnamefont {Will}\ \bibnamefont {Pazner}}, \bibinfo {author}
  {\bibfnamefont {Mark}\ \bibnamefont {Stowell}}, \bibinfo {author}
  {\bibfnamefont {Vladimir}\ \bibnamefont {Tomov}}, \bibinfo {author}
  {\bibfnamefont {Ido}\ \bibnamefont {Akkerman}}, \bibinfo {author}
  {\bibfnamefont {Johann}\ \bibnamefont {Dahm}}, \bibinfo {author}
  {\bibfnamefont {David}\ \bibnamefont {Medina}}, \ and\ \bibinfo {author}
  {\bibfnamefont {Stefano}\ \bibnamefont {Zampini}},\ }\bibfield  {title}
  {\enquote {\bibinfo {title} {{MFEM: A modular finite element methods
  library}},}\ }\href {\doibase 10.1016/j.camwa.2020.06.009} {\bibfield
  {journal} {\bibinfo  {journal} {Comput. Math. Appl.}\ }\textbf {\bibinfo
  {volume} {81}},\ \bibinfo {pages} {42--74} (\bibinfo {year} {2021})},\
  \Eprint {http://arxiv.org/abs/1911.09220} {arXiv:1911.09220} \BibitemShut
  {NoStop}%
\bibitem [{\citenamefont {Foster}\ \emph {et~al.}(2018)\citenamefont {Foster},
  \citenamefont {Rodd},\ and\ \citenamefont {Safdi}}]{Foster:2017hbq}%
  \BibitemOpen
  \bibfield  {author} {\bibinfo {author} {\bibfnamefont {Joshua~W.}\
  \bibnamefont {Foster}}, \bibinfo {author} {\bibfnamefont {Nicholas~L.}\
  \bibnamefont {Rodd}}, \ and\ \bibinfo {author} {\bibfnamefont {Benjamin~R.}\
  \bibnamefont {Safdi}},\ }\bibfield  {title} {\enquote {\bibinfo {title}
  {{Revealing the Dark Matter Halo with Axion Direct Detection}},}\ }\href
  {\doibase 10.1103/PhysRevD.97.123006} {\bibfield  {journal} {\bibinfo
  {journal} {Phys. Rev. D}\ }\textbf {\bibinfo {volume} {97}},\ \bibinfo
  {pages} {123006} (\bibinfo {year} {2018})},\ \Eprint
  {http://arxiv.org/abs/1711.10489} {arXiv:1711.10489 [astro-ph.CO]}
  \BibitemShut {NoStop}%
\bibitem [{\citenamefont {Chaudhuri}\ \emph {et~al.}(2018)\citenamefont
  {Chaudhuri}, \citenamefont {Irwin}, \citenamefont {Graham},\ and\
  \citenamefont {Mardon}}]{Chaudhuri:2018rqn}%
  \BibitemOpen
  \bibfield  {author} {\bibinfo {author} {\bibfnamefont {Saptarshi}\
  \bibnamefont {Chaudhuri}}, \bibinfo {author} {\bibfnamefont {Kent}\
  \bibnamefont {Irwin}}, \bibinfo {author} {\bibfnamefont {Peter~W.}\
  \bibnamefont {Graham}}, \ and\ \bibinfo {author} {\bibfnamefont {Jeremy}\
  \bibnamefont {Mardon}},\ }\bibfield  {title} {\enquote {\bibinfo {title}
  {{Optimal Impedance Matching and Quantum Limits of Electromagnetic Axion and
  Hidden-Photon Dark Matter Searches}},}\ }\href@noop {} {\  (\bibinfo {year}
  {2018})},\ \Eprint {http://arxiv.org/abs/1803.01627} {arXiv:1803.01627
  [hep-ph]} \BibitemShut {NoStop}%
\bibitem [{\citenamefont {Cowan}\ \emph {et~al.}(2011)\citenamefont {Cowan},
  \citenamefont {Cranmer}, \citenamefont {Gross},\ and\ \citenamefont
  {Vitells}}]{Cowan:2010js}%
  \BibitemOpen
  \bibfield  {author} {\bibinfo {author} {\bibfnamefont {Glen}\ \bibnamefont
  {Cowan}}, \bibinfo {author} {\bibfnamefont {Kyle}\ \bibnamefont {Cranmer}},
  \bibinfo {author} {\bibfnamefont {Eilam}\ \bibnamefont {Gross}}, \ and\
  \bibinfo {author} {\bibfnamefont {Ofer}\ \bibnamefont {Vitells}},\ }\bibfield
   {title} {\enquote {\bibinfo {title} {{Asymptotic formulae for
  likelihood-based tests of new physics}},}\ }\href {\doibase
  10.1140/epjc/s10052-011-1554-0} {\bibfield  {journal} {\bibinfo  {journal}
  {Eur. Phys. J. C}\ }\textbf {\bibinfo {volume} {71}},\ \bibinfo {pages}
  {1554} (\bibinfo {year} {2011})},\ \bibinfo {note} {[Erratum: Eur.Phys.J.C
  73, 2501 (2013)]},\ \Eprint {http://arxiv.org/abs/1007.1727} {arXiv:1007.1727
  [physics.data-an]} \BibitemShut {NoStop}%
\bibitem [{\citenamefont {Chaudhuri}\ \emph {et~al.}(2019)\citenamefont
  {Chaudhuri}, \citenamefont {Irwin}, \citenamefont {Graham},\ and\
  \citenamefont {Mardon}}]{Chaudhuri:2019ntz}%
  \BibitemOpen
  \bibfield  {author} {\bibinfo {author} {\bibfnamefont {Saptarshi}\
  \bibnamefont {Chaudhuri}}, \bibinfo {author} {\bibfnamefont {Kent~D.}\
  \bibnamefont {Irwin}}, \bibinfo {author} {\bibfnamefont {Peter~W.}\
  \bibnamefont {Graham}}, \ and\ \bibinfo {author} {\bibfnamefont {Jeremy}\
  \bibnamefont {Mardon}},\ }\bibfield  {title} {\enquote {\bibinfo {title}
  {{Optimal Electromagnetic Searches for Axion and Hidden-Photon Dark
  Matter}},}\ }\href@noop {} {\  (\bibinfo {year} {2019})},\ \Eprint
  {http://arxiv.org/abs/1904.05806} {arXiv:1904.05806 [hep-ex]} \BibitemShut
  {NoStop}%
\bibitem [{\citenamefont {Abe}\ \emph {et~al.}(2009)\citenamefont {Abe} \emph
  {et~al.}}]{ISSDetectorWorkingGroup:2007uvu}%
  \BibitemOpen
  \bibfield  {author} {\bibinfo {author} {\bibfnamefont {T.}~\bibnamefont
  {Abe}} \emph {et~al.} (\bibinfo {collaboration} {ISS Detector Working
  Group}),\ }\bibfield  {title} {\enquote {\bibinfo {title} {{Detectors and
  flux instrumentation for future neutrino facilities}},}\ }\href {\doibase
  10.1088/1748-0221/4/05/T05001} {\bibfield  {journal} {\bibinfo  {journal}
  {JINST}\ }\textbf {\bibinfo {volume} {4}},\ \bibinfo {pages} {T05001}
  (\bibinfo {year} {2009})},\ \Eprint {http://arxiv.org/abs/0712.4129}
  {arXiv:0712.4129 [physics.ins-det]} \BibitemShut {NoStop}%
\bibitem [{\citenamefont {Adey}\ \emph {et~al.}(2013)\citenamefont {Adey} \emph
  {et~al.}}]{nuSTORM:2013cqr}%
  \BibitemOpen
  \bibfield  {author} {\bibinfo {author} {\bibfnamefont {D.}~\bibnamefont
  {Adey}} \emph {et~al.} (\bibinfo {collaboration} {nuSTORM}),\ }\bibfield
  {title} {\enquote {\bibinfo {title} {{nuSTORM - Neutrinos from STORed Muons:
  Proposal to the Fermilab PAC}},}\ }\href@noop {} {\  (\bibinfo {year}
  {2013})},\ \Eprint {http://arxiv.org/abs/1308.6822} {arXiv:1308.6822
  [physics.acc-ph]} \BibitemShut {NoStop}%
\bibitem [{\citenamefont {Green}\ and\ \citenamefont
  {Strauss}(2008)}]{Green:2008cost}%
  \BibitemOpen
  \bibfield  {author} {\bibinfo {author} {\bibfnamefont {M.~A.}\ \bibnamefont
  {Green}}\ and\ \bibinfo {author} {\bibfnamefont {B.~P.}\ \bibnamefont
  {Strauss}},\ }\bibfield  {title} {\enquote {\bibinfo {title} {{The Cost of
  Superconducting Magnets as a Function of Stored Energy and Design Magnetic
  Induction Times the Field Volume}},}\ }\href {\doibase
  10.1109/TASC.2008.921279} {\bibfield  {journal} {\bibinfo  {journal} {IEEE
  Trans. Appl. Supercond.}\ }\textbf {\bibinfo {volume} {18}},\ \bibinfo
  {pages} {248--251} (\bibinfo {year} {2008})}\BibitemShut {NoStop}%
\bibitem [{\citenamefont {Herv{\'e}}\ \emph {et~al.}(2004)\citenamefont
  {Herv{\'e}} \emph {et~al.}}]{Herve:2004cms}%
  \BibitemOpen
  \bibfield  {author} {\bibinfo {author} {\bibfnamefont {A.}~\bibnamefont
  {Herv{\'e}}} \emph {et~al.},\ }\bibfield  {title} {\enquote {\bibinfo {title}
  {{Status of the Construction of the CMS Magnet}},}\ }\href {\doibase
  10.1109/TASC.2004.829715} {\bibfield  {journal} {\bibinfo  {journal} {IEEE
  Trans. Appl. Supercond.}\ }\textbf {\bibinfo {volume} {14}},\ \bibinfo
  {pages} {542--547} (\bibinfo {year} {2004})}\BibitemShut {NoStop}%
\bibitem [{\citenamefont {ten Kate}(2008)}]{tenKate:2008lhec}%
  \BibitemOpen
  \bibfield  {author} {\bibinfo {author} {\bibfnamefont {H.~H.~J.}\
  \bibnamefont {ten Kate}},\ }\href
  {https://indico.cern.ch/event/31463/contributions/726195/} {\enquote
  {\bibinfo {title} {{Design and realization of the ATLAS Magnets -- options
  for LHeC}},}\ }\bibinfo {howpublished} {Talk at the LHeC Workshop,
  Divonne-les-Bains, 2 September 2008} (\bibinfo {year} {2008})\BibitemShut
  {NoStop}%
\bibitem [{\citenamefont {De~Palma}\ \emph {et~al.}(1983)\citenamefont
  {De~Palma} \emph {et~al.}}]{ALEPH:1983bkx}%
  \BibitemOpen
  \bibfield  {author} {\bibinfo {author} {\bibfnamefont {M.}~\bibnamefont
  {De~Palma}} \emph {et~al.} (\bibinfo {collaboration} {ALEPH}),\ }\href@noop
  {} {\emph {\bibinfo {title} {{ALEPH: Technical Report 1983}}}},\ \bibinfo
  {type} {Tech. Rep.}\ \bibinfo {number} {CERN-LEPC-83-2, LEPC-P1}\ (\bibinfo
  {institution} {CERN},\ \bibinfo {year} {1983})\ \bibinfo {note} {table
  2.3}\BibitemShut {NoStop}%
\bibitem [{\citenamefont {Bartl}\ \emph {et~al.}(1983)\citenamefont {Bartl}
  \emph {et~al.}}]{Bartl:1983zq}%
  \BibitemOpen
  \bibfield  {author} {\bibinfo {author} {\bibfnamefont {W.}~\bibnamefont
  {Bartl}} \emph {et~al.} (\bibinfo {collaboration} {DELPHI}),\ }\href@noop {}
  {\emph {\bibinfo {title} {{DELPHI: Technical Proposal}}}},\ \bibinfo {type}
  {Tech. Rep.}\ \bibinfo {number} {DELPHI-83-66, CERN-LEPC-83-3, LEPC-P-2}\
  (\bibinfo  {institution} {CERN},\ \bibinfo {year} {1983})\ \bibinfo {note}
  {sec.~1.5.6}\BibitemShut {NoStop}%
\bibitem [{\citenamefont {{H1 Collaboration}}(1986)}]{H1:1986ab}%
  \BibitemOpen
  \bibfield  {author} {\bibinfo {author} {\bibnamefont {{H1 Collaboration}}},\
  }\href@noop {} {\emph {\bibinfo {title} {{Technical proposal for the H1
  detector}}}},\ \bibinfo {type} {Tech. Rep.}\ (\bibinfo  {institution}
  {DESY},\ \bibinfo {address} {Hamburg},\ \bibinfo {year} {1986})\ \bibinfo
  {note} {table 15.2b}\BibitemShut {NoStop}%
\bibitem [{\citenamefont {Armengaud}\ \emph {et~al.}(2014)\citenamefont
  {Armengaud} \emph {et~al.}}]{Armengaud:2014gea}%
  \BibitemOpen
  \bibfield  {author} {\bibinfo {author} {\bibfnamefont {E.}~\bibnamefont
  {Armengaud}} \emph {et~al.},\ }\bibfield  {title} {\enquote {\bibinfo {title}
  {{Conceptual Design of the International Axion Observatory (IAXO)}},}\ }\href
  {\doibase 10.1088/1748-0221/9/05/T05002} {\bibfield  {journal} {\bibinfo
  {journal} {JINST}\ }\textbf {\bibinfo {volume} {9}},\ \bibinfo {pages}
  {T05002} (\bibinfo {year} {2014})},\ \Eprint {http://arxiv.org/abs/1401.3233}
  {arXiv:1401.3233 [physics.ins-det]} \BibitemShut {NoStop}%
\bibitem [{\citenamefont {{ITER Joint Central Team and ITER Home
  Teams}}(2001)}]{ITER:FDR2001}%
  \BibitemOpen
  \bibfield  {author} {\bibinfo {author} {\bibnamefont {{ITER Joint Central
  Team and ITER Home Teams}}},\ }\href@noop {} {\emph {\bibinfo {title}
  {{Summary of the ITER Final Design Report}}}},\ \bibinfo {type} {ITER EDA
  Documentation Series}\ \bibinfo {number} {22}\ (\bibinfo  {institution}
  {International Atomic Energy Agency},\ \bibinfo {address} {Vienna},\ \bibinfo
  {year} {2001})\ \bibinfo {note} {table 9.1-2,
  \url{https://www-pub.iaea.org/MTCD/Publications/PDF/ITER-EDA-DS-22.pdf}}\BibitemShut
  {NoStop}%
\bibitem [{\citenamefont {Sauthoff}(2003)}]{Sauthoff:2003fesac}%
  \BibitemOpen
  \bibfield  {author} {\bibinfo {author} {\bibfnamefont {N.}~\bibnamefont
  {Sauthoff}},\ }\href@noop {} {\enquote {\bibinfo {title} {{Report on
  Developing Industrial Cost Estimates for ITER Systems of Possible Interest to
  the US}},}\ }\bibinfo {howpublished} {Presentation to the Fusion Energy
  Sciences Advisory Committee,
  \url{https://science.osti.gov/-/media/fes/fesac/pdf/2003/Fesac_sauthoff_march_2003.pdf}}
  (\bibinfo {year} {2003})\BibitemShut {NoStop}%
\bibitem [{\citenamefont {Klinger}(2011)}]{Klinger:W7Xiter}%
  \BibitemOpen
  \bibfield  {author} {\bibinfo {author} {\bibfnamefont {T.}~\bibnamefont
  {Klinger}},\ }\href@noop {} {\enquote {\bibinfo {title} {{Stellarators
  difficult to build? The construction of Wendelstein 7-X}},}\ }\bibinfo
  {howpublished} {ITER Seminar, Cadarache, 14 April 2011,
  \url{https://www.iter.org/sites/default/files/media/680/iter_w7x.pdf}}
  (\bibinfo {year} {2011})\BibitemShut {NoStop}%
\bibitem [{ATL(1997)}]{ATLAS:1997magnetTDR}%
  \BibitemOpen
  \href@noop {} {\emph {\bibinfo {title} {{ATLAS Magnet System: Technical
  Design Report, Volume 1}}}},\ \bibinfo {type} {Tech. Rep.}\ \bibinfo {number}
  {CERN-LHCC-97-018}\ (\bibinfo  {institution} {CERN},\ \bibinfo {year}
  {1997})\ \bibinfo {note} {table 4-2}\BibitemShut {NoStop}%
\bibitem [{\citenamefont {Jenni}(2002)}]{ATLAS:RRB2002062}%
  \BibitemOpen
  \bibfield  {author} {\bibinfo {author} {\bibfnamefont {P.}~\bibnamefont
  {Jenni}} (\bibinfo {collaboration} {ATLAS}),\ }\href@noop {} {\emph {\bibinfo
  {title} {{Completion and Pre-Operation Costs for the Initial Staged ATLAS
  Detector}}}},\ \bibinfo {type} {Tech. Rep.}\ \bibinfo {number}
  {CERN-RRB-2002-062}\ (\bibinfo  {institution} {CERN},\ \bibinfo {year}
  {2002})\BibitemShut {NoStop}%
\bibitem [{ATL(2006)}]{ATLAS:RRB2006069}%
  \BibitemOpen
  \href@noop {} {\emph {\bibinfo {title} {{Updated ATLAS Progress Report for
  the October 2006 RRB}}}},\ \bibinfo {type} {Tech. Rep.}\ \bibinfo {number}
  {CERN-RRB-2006-069}\ (\bibinfo  {institution} {CERN},\ \bibinfo {year}
  {2006})\BibitemShut {NoStop}%
\bibitem [{\citenamefont {Dhakal}\ \emph {et~al.}(2012)\citenamefont {Dhakal},
  \citenamefont {Ciovati}, \citenamefont {Kneisel},\ and\ \citenamefont
  {Myneni}}]{Dhakal:2012iq}%
  \BibitemOpen
  \bibfield  {author} {\bibinfo {author} {\bibfnamefont {Pashupati}\
  \bibnamefont {Dhakal}}, \bibinfo {author} {\bibfnamefont {Gianluigi}\
  \bibnamefont {Ciovati}}, \bibinfo {author} {\bibfnamefont {Peter}\
  \bibnamefont {Kneisel}}, \ and\ \bibinfo {author} {\bibfnamefont
  {Ganapati~Rao}\ \bibnamefont {Myneni}},\ }\bibfield  {title} {\enquote
  {\bibinfo {title} {{Superconducting DC and RF Properties of Ingot
  Niobium}},}\ }\href@noop {} {\  (\bibinfo {year} {2012})},\ \Eprint
  {http://arxiv.org/abs/1202.0811} {arXiv:1202.0811 [physics.acc-ph]}
  \BibitemShut {NoStop}%
\bibitem [{\citenamefont {Brown}\ \emph {et~al.}(2026)\citenamefont {Brown},
  \citenamefont {Kim}, \citenamefont {Hartung},\ and\ \citenamefont
  {Xu}}]{Brown:2026cqz}%
  \BibitemOpen
  \bibfield  {author} {\bibinfo {author} {\bibfnamefont {Jacob}\ \bibnamefont
  {Brown}}, \bibinfo {author} {\bibfnamefont {Sang-hoon}\ \bibnamefont {Kim}},
  \bibinfo {author} {\bibfnamefont {Walter}\ \bibnamefont {Hartung}}, \ and\
  \bibinfo {author} {\bibfnamefont {Ting}\ \bibnamefont {Xu}},\ }\bibfield
  {title} {\enquote {\bibinfo {title} {{Impact of heat treatments on the
  performance of low-frequency superconducting quarter-wave resonators at 4.3
  K}},}\ }\href@noop {} {\  (\bibinfo {year} {2026})},\ \Eprint
  {http://arxiv.org/abs/2603.00388} {arXiv:2603.00388 [physics.acc-ph]}
  \BibitemShut {NoStop}%
\bibitem [{\citenamefont {Cooley}\ \emph {et~al.}(2005)\citenamefont {Cooley},
  \citenamefont {Ghosh},\ and\ \citenamefont {Scanlan}}]{Cooley:2005ize}%
  \BibitemOpen
  \bibfield  {author} {\bibinfo {author} {\bibfnamefont {L.~D.}\ \bibnamefont
  {Cooley}}, \bibinfo {author} {\bibfnamefont {A.~K.}\ \bibnamefont {Ghosh}}, \
  and\ \bibinfo {author} {\bibfnamefont {R.~M.}\ \bibnamefont {Scanlan}},\
  }\bibfield  {title} {\enquote {\bibinfo {title} {{Costs of high-field
  superconducting strands for particle accelerator magnets}},}\ }\href
  {\doibase 10.1088/0953-2048/18/4/r01} {\bibfield  {journal} {\bibinfo
  {journal} {Supercond. Sci. Technol.}\ }\textbf {\bibinfo {volume} {18}},\
  \bibinfo {pages} {R51--R65} (\bibinfo {year} {2005})}\BibitemShut {NoStop}%
\bibitem [{\citenamefont {Abramowicz}\ \emph {et~al.}(2020)\citenamefont
  {Abramowicz} \emph {et~al.}}]{ILDConceptGroup:2020sfq}%
  \BibitemOpen
  \bibfield  {author} {\bibinfo {author} {\bibfnamefont {Halina}\ \bibnamefont
  {Abramowicz}} \emph {et~al.} (\bibinfo {collaboration} {ILD Concept Group}),\
  }\bibfield  {title} {\enquote {\bibinfo {title} {{International Large
  Detector: Interim Design Report}},}\ }\href@noop {} {\  (\bibinfo {year}
  {2020})},\ \Eprint {http://arxiv.org/abs/2003.01116} {arXiv:2003.01116
  [physics.ins-det]} \BibitemShut {NoStop}%
\bibitem [{\citenamefont {Ankel}\ \emph {et~al.}(2025)\citenamefont {Ankel}
  \emph {et~al.}}]{Ankel:2025nld}%
  \BibitemOpen
  \bibfield  {author} {\bibinfo {author} {\bibfnamefont {V.}~\bibnamefont
  {Ankel}} \emph {et~al.},\ }\bibfield  {title} {\enquote {\bibinfo {title}
  {{Noise limits for dc SQUID readout of high-Q resonators below
  300{\,}MHz}},}\ }\href {\doibase 10.1063/5.0280831} {\bibfield  {journal}
  {\bibinfo  {journal} {J. Appl. Phys.}\ }\textbf {\bibinfo {volume} {138}},\
  \bibinfo {pages} {094505} (\bibinfo {year} {2025})},\ \Eprint
  {http://arxiv.org/abs/2504.20398} {arXiv:2504.20398 [quant-ph]} \BibitemShut
  {NoStop}%
\bibitem [{\citenamefont {Baxter}\ \emph {et~al.}(2021)\citenamefont {Baxter}
  \emph {et~al.}}]{Baxter:2021pqo}%
  \BibitemOpen
  \bibfield  {author} {\bibinfo {author} {\bibfnamefont {D.}~\bibnamefont
  {Baxter}} \emph {et~al.},\ }\bibfield  {title} {\enquote {\bibinfo {title}
  {{Recommended conventions for reporting results from direct dark matter
  searches}},}\ }\href {\doibase 10.1140/epjc/s10052-021-09655-y} {\bibfield
  {journal} {\bibinfo  {journal} {Eur. Phys. J. C}\ }\textbf {\bibinfo {volume}
  {81}},\ \bibinfo {pages} {907} (\bibinfo {year} {2021})},\ \Eprint
  {http://arxiv.org/abs/2105.00599} {arXiv:2105.00599 [hep-ex]} \BibitemShut
  {NoStop}%
\bibitem [{\citenamefont {Brandt}\ \emph {et~al.}(2026)\citenamefont {Brandt},
  \citenamefont {Endr{\H{o}}di}, \citenamefont {Hern{\'a}ndez~Hern{\'a}ndez},
  \citenamefont {Mark{\'o}},\ and\ \citenamefont {Pannullo}}]{Brandt:2026wir}%
  \BibitemOpen
  \bibfield  {author} {\bibinfo {author} {\bibfnamefont {Bastian~B.}\
  \bibnamefont {Brandt}}, \bibinfo {author} {\bibfnamefont {Gergely}\
  \bibnamefont {Endr{\H{o}}di}}, \bibinfo {author} {\bibfnamefont
  {Jos{\'e}~Javier}\ \bibnamefont {Hern{\'a}ndez~Hern{\'a}ndez}}, \bibinfo
  {author} {\bibfnamefont {Gergely}\ \bibnamefont {Mark{\'o}}}, \ and\ \bibinfo
  {author} {\bibfnamefont {Laurin}\ \bibnamefont {Pannullo}},\ }\bibfield
  {title} {\enquote {\bibinfo {title} {{The axion-photon coupling from lattice
  Quantum Chromodynamics}},}\ }\href@noop {} {\  (\bibinfo {year} {2026})},\
  \Eprint {http://arxiv.org/abs/2603.29153} {arXiv:2603.29153 [hep-lat]}
  \BibitemShut {NoStop}%
\bibitem [{\citenamefont {Grilli~di Cortona}\ \emph {et~al.}(2016)\citenamefont
  {Grilli~di Cortona}, \citenamefont {Hardy}, \citenamefont {Pardo~Vega},\ and\
  \citenamefont {Villadoro}}]{diCortona:2015ldu}%
  \BibitemOpen
  \bibfield  {author} {\bibinfo {author} {\bibfnamefont {Giovanni}\
  \bibnamefont {Grilli~di Cortona}}, \bibinfo {author} {\bibfnamefont {Edward}\
  \bibnamefont {Hardy}}, \bibinfo {author} {\bibfnamefont {Javier}\
  \bibnamefont {Pardo~Vega}}, \ and\ \bibinfo {author} {\bibfnamefont
  {Giovanni}\ \bibnamefont {Villadoro}},\ }\bibfield  {title} {\enquote
  {\bibinfo {title} {{The QCD axion, precisely}},}\ }\href {\doibase
  10.1007/JHEP01(2016)034} {\bibfield  {journal} {\bibinfo  {journal} {JHEP}\
  }\textbf {\bibinfo {volume} {01}},\ \bibinfo {pages} {034} (\bibinfo {year}
  {2016})},\ \Eprint {http://arxiv.org/abs/1511.02867} {arXiv:1511.02867
  [hep-ph]} \BibitemShut {NoStop}%
\bibitem [{\citenamefont {Kolevatov}\ \emph {et~al.}(2026)\citenamefont
  {Kolevatov}, \citenamefont {Chaudhuri},\ and\ \citenamefont
  {Page}}]{Kolevatov:2025lzx}%
  \BibitemOpen
  \bibfield  {author} {\bibinfo {author} {\bibfnamefont {Roman}\ \bibnamefont
  {Kolevatov}}, \bibinfo {author} {\bibfnamefont {Saptarshi}\ \bibnamefont
  {Chaudhuri}}, \ and\ \bibinfo {author} {\bibfnamefont {Lyman}\ \bibnamefont
  {Page}},\ }\bibfield  {title} {\enquote {\bibinfo {title} {{High-Q
  superconducting lumped-element resonators for low-mass axion searches}},}\
  }\href {\doibase 10.1063/5.0311286} {\bibfield  {journal} {\bibinfo
  {journal} {Rev. Sci. Instrum.}\ }\textbf {\bibinfo {volume} {97}},\ \bibinfo
  {pages} {024503} (\bibinfo {year} {2026})},\ \Eprint
  {http://arxiv.org/abs/2511.03639} {arXiv:2511.03639 [physics.ins-det]}
  \BibitemShut {NoStop}%
\bibitem [{\citenamefont {Kuenstner}\ \emph {et~al.}(2025)\citenamefont
  {Kuenstner} \emph {et~al.}}]{Kuenstner:2022gyc}%
  \BibitemOpen
  \bibfield  {author} {\bibinfo {author} {\bibfnamefont {Stephen~E.}\
  \bibnamefont {Kuenstner}} \emph {et~al.},\ }\bibfield  {title} {\enquote
  {\bibinfo {title} {{Quantum metrology of low-frequency electromagnetic modes
  with frequency upconverters}},}\ }\href {\doibase
  10.1103/PhysRevResearch.7.013281} {\bibfield  {journal} {\bibinfo  {journal}
  {Phys. Rev. Res.}\ }\textbf {\bibinfo {volume} {7}},\ \bibinfo {pages}
  {013281} (\bibinfo {year} {2025})},\ \Eprint
  {http://arxiv.org/abs/2210.05576} {arXiv:2210.05576 [quant-ph]} \BibitemShut
  {NoStop}%
\bibitem [{\citenamefont {de~Salas}\ and\ \citenamefont
  {Widmark}(2021)}]{deSalas:2020hbh}%
  \BibitemOpen
  \bibfield  {author} {\bibinfo {author} {\bibfnamefont {Pablo~F.}\
  \bibnamefont {de~Salas}}\ and\ \bibinfo {author} {\bibfnamefont {Axel}\
  \bibnamefont {Widmark}},\ }\bibfield  {title} {\enquote {\bibinfo {title}
  {{Dark matter local density determination: recent observations and future
  prospects}},}\ }\href {\doibase 10.1088/1361-6633/ac24e7} {\bibfield
  {journal} {\bibinfo  {journal} {Rept. Prog. Phys.}\ }\textbf {\bibinfo
  {volume} {84}},\ \bibinfo {pages} {104901} (\bibinfo {year} {2021})},\
  \Eprint {http://arxiv.org/abs/2012.11477} {arXiv:2012.11477 [astro-ph.GA]}
  \BibitemShut {NoStop}%
\bibitem [{\citenamefont {Lin}\ \emph {et~al.}(2026)\citenamefont {Lin},
  \citenamefont {Cheng}, \citenamefont {Feng}, \citenamefont {Jin},
  \citenamefont {Liu},\ and\ \citenamefont {Luo}}]{Lin:2026hbw}%
  \BibitemOpen
  \bibfield  {author} {\bibinfo {author} {\bibfnamefont {Tian}\ \bibnamefont
  {Lin}}, \bibinfo {author} {\bibfnamefont {Hao-Yang}\ \bibnamefont {Cheng}},
  \bibinfo {author} {\bibfnamefont {Xu}~\bibnamefont {Feng}}, \bibinfo {author}
  {\bibfnamefont {Lu-Chang}\ \bibnamefont {Jin}}, \bibinfo {author}
  {\bibfnamefont {Chuan}\ \bibnamefont {Liu}}, \ and\ \bibinfo {author}
  {\bibfnamefont {Qi-Yuan}\ \bibnamefont {Luo}},\ }\bibfield  {title} {\enquote
  {\bibinfo {title} {{First-Principles Determination of the QCD Contribution to
  the Axion-Photon Coupling Using Domain-Wall Fermions}},}\ }\href@noop {} {\
  (\bibinfo {year} {2026})},\ \Eprint {http://arxiv.org/abs/2609.22698}
  {arXiv:2609.22698 [hep-lat]} \BibitemShut {NoStop}%
\bibitem [{\citenamefont {Agrawal}\ \emph {et~al.}(2022)\citenamefont
  {Agrawal}, \citenamefont {Nee},\ and\ \citenamefont
  {Reig}}]{Agrawal:2022lsp}%
  \BibitemOpen
  \bibfield  {author} {\bibinfo {author} {\bibfnamefont {Prateek}\ \bibnamefont
  {Agrawal}}, \bibinfo {author} {\bibfnamefont {Michael}\ \bibnamefont {Nee}},
  \ and\ \bibinfo {author} {\bibfnamefont {Mario}\ \bibnamefont {Reig}},\
  }\bibfield  {title} {\enquote {\bibinfo {title} {{Axion couplings in grand
  unified theories}},}\ }\href {\doibase 10.1007/JHEP10(2022)141} {\bibfield
  {journal} {\bibinfo  {journal} {JHEP}\ }\textbf {\bibinfo {volume} {10}},\
  \bibinfo {pages} {141} (\bibinfo {year} {2022})},\ \Eprint
  {http://arxiv.org/abs/2206.07053} {arXiv:2206.07053 [hep-ph]} \BibitemShut
  {NoStop}%
\bibitem [{\citenamefont {Storn}\ and\ \citenamefont
  {Price}(1997)}]{Storn:1997}%
  \BibitemOpen
  \bibfield  {author} {\bibinfo {author} {\bibfnamefont {Rainer}\ \bibnamefont
  {Storn}}\ and\ \bibinfo {author} {\bibfnamefont {Kenneth}\ \bibnamefont
  {Price}},\ }\bibfield  {title} {\enquote {\bibinfo {title} {{Differential
  Evolution -- A Simple and Efficient Heuristic for Global Optimization over
  Continuous Spaces}},}\ }\href {\doibase 10.1023/A:1008202821328} {\bibfield
  {journal} {\bibinfo  {journal} {J. Global Optim.}\ }\textbf {\bibinfo
  {volume} {11}},\ \bibinfo {pages} {341--359} (\bibinfo {year}
  {1997})}\BibitemShut {NoStop}%
\bibitem [{\citenamefont {Virtanen}\ \emph {et~al.}(2020)\citenamefont
  {Virtanen} \emph {et~al.}}]{Virtanen:2019joe}%
  \BibitemOpen
  \bibfield  {author} {\bibinfo {author} {\bibfnamefont {Pauli}\ \bibnamefont
  {Virtanen}} \emph {et~al.},\ }\bibfield  {title} {\enquote {\bibinfo {title}
  {{SciPy 1.0--Fundamental Algorithms for Scientific Computing in Python}},}\
  }\href {\doibase 10.1038/s41592-019-0686-2} {\bibfield  {journal} {\bibinfo
  {journal} {Nature Meth.}\ }\textbf {\bibinfo {volume} {17}},\ \bibinfo
  {pages} {261--272} (\bibinfo {year} {2020})},\ \Eprint
  {http://arxiv.org/abs/1907.10121} {arXiv:1907.10121 [cs.MS]} \BibitemShut
  {NoStop}%
\bibitem [{\citenamefont {Gramolin}\ \emph {et~al.}(2021)\citenamefont
  {Gramolin}, \citenamefont {Aybas}, \citenamefont {Johnson}, \citenamefont
  {Adam},\ and\ \citenamefont {Sushkov}}]{Gramolin:2020ict}%
  \BibitemOpen
  \bibfield  {author} {\bibinfo {author} {\bibfnamefont {Alexander~V.}\
  \bibnamefont {Gramolin}}, \bibinfo {author} {\bibfnamefont {Deniz}\
  \bibnamefont {Aybas}}, \bibinfo {author} {\bibfnamefont {Dorian}\
  \bibnamefont {Johnson}}, \bibinfo {author} {\bibfnamefont {Janos}\
  \bibnamefont {Adam}}, \ and\ \bibinfo {author} {\bibfnamefont {Alexander~O.}\
  \bibnamefont {Sushkov}},\ }\bibfield  {title} {\enquote {\bibinfo {title}
  {{Search for axion-like dark matter with ferromagnets}},}\ }\href {\doibase
  10.1038/s41567-020-1006-6} {\bibfield  {journal} {\bibinfo  {journal} {Nature
  Phys.}\ }\textbf {\bibinfo {volume} {17}},\ \bibinfo {pages} {79--84}
  (\bibinfo {year} {2021})},\ \Eprint {http://arxiv.org/abs/2003.03348}
  {arXiv:2003.03348 [hep-ex]} \BibitemShut {NoStop}%
\bibitem [{\citenamefont {Alduino}\ \emph {et~al.}(2019)\citenamefont {Alduino}
  \emph {et~al.}}]{Alduino:2019xia}%
  \BibitemOpen
  \bibfield  {author} {\bibinfo {author} {\bibfnamefont {C.}~\bibnamefont
  {Alduino}} \emph {et~al.},\ }\bibfield  {title} {\enquote {\bibinfo {title}
  {{The CUORE cryostat: An infrastructure for rare event searches at
  millikelvin temperatures}},}\ }\href {\doibase
  10.1016/j.cryogenics.2019.06.011} {\bibfield  {journal} {\bibinfo  {journal}
  {Cryogenics}\ }\textbf {\bibinfo {volume} {102}},\ \bibinfo {pages} {9--21}
  (\bibinfo {year} {2019})},\ \Eprint {http://arxiv.org/abs/1904.05745}
  {arXiv:1904.05745 [physics.ins-det]} \BibitemShut {NoStop}%
\bibitem [{\citenamefont {Droster}(2025)}]{Droster:2025utl}%
  \BibitemOpen
  \bibfield  {author} {\bibinfo {author} {\bibfnamefont {Alex}\ \bibnamefont
  {Droster}},\ }\emph {\bibinfo {title} {{Searching for Sub-{\textmu}eV Axions
  With DMRadio}}},\ \href@noop {} {Ph.D. thesis},\ \bibinfo  {school} {UC,
  Berkeley} (\bibinfo {year} {2025})\BibitemShut {NoStop}%
\bibitem [{\citenamefont {Di~Ubaldo}\ \emph {et~al.}(2026)\citenamefont
  {Di~Ubaldo}, \citenamefont {Iliesiu}, \citenamefont {Lin},\ and\
  \citenamefont {Yan}}]{DiUbaldo:2026rly}%
  \BibitemOpen
  \bibfield  {author} {\bibinfo {author} {\bibfnamefont {Gabriele}\
  \bibnamefont {Di~Ubaldo}}, \bibinfo {author} {\bibfnamefont {Luca~V.}\
  \bibnamefont {Iliesiu}}, \bibinfo {author} {\bibfnamefont {Henry~W.}\
  \bibnamefont {Lin}}, \ and\ \bibinfo {author} {\bibfnamefont {Cynthia}\
  \bibnamefont {Yan}},\ }\bibfield  {title} {\enquote {\bibinfo {title}
  {{Positivity of the gravitational path integral implies the axionic weak
  gravity conjecture}},}\ }\href@noop {} {\  (\bibinfo {year} {2026})},\
  \Eprint {http://arxiv.org/abs/2605.05305} {arXiv:2605.05305 [hep-th]}
  \BibitemShut {NoStop}%
\bibitem [{\citenamefont {Arvanitaki}\ and\ \citenamefont
  {Dubovsky}(2011)}]{Arvanitaki:2010sy}%
  \BibitemOpen
  \bibfield  {author} {\bibinfo {author} {\bibfnamefont {Asimina}\ \bibnamefont
  {Arvanitaki}}\ and\ \bibinfo {author} {\bibfnamefont {Sergei}\ \bibnamefont
  {Dubovsky}},\ }\bibfield  {title} {\enquote {\bibinfo {title} {{Exploring the
  String Axiverse with Precision Black Hole Physics}},}\ }\href {\doibase
  10.1103/PhysRevD.83.044026} {\bibfield  {journal} {\bibinfo  {journal} {Phys.
  Rev. D}\ }\textbf {\bibinfo {volume} {83}},\ \bibinfo {pages} {044026}
  (\bibinfo {year} {2011})},\ \Eprint {http://arxiv.org/abs/1004.3558}
  {arXiv:1004.3558 [hep-th]} \BibitemShut {NoStop}%
\bibitem [{\citenamefont {Arvanitaki}\ \emph {et~al.}(2015)\citenamefont
  {Arvanitaki}, \citenamefont {Baryakhtar},\ and\ \citenamefont
  {Huang}}]{Arvanitaki:2014wva}%
  \BibitemOpen
  \bibfield  {author} {\bibinfo {author} {\bibfnamefont {Asimina}\ \bibnamefont
  {Arvanitaki}}, \bibinfo {author} {\bibfnamefont {Masha}\ \bibnamefont
  {Baryakhtar}}, \ and\ \bibinfo {author} {\bibfnamefont {Xinlu}\ \bibnamefont
  {Huang}},\ }\bibfield  {title} {\enquote {\bibinfo {title} {{Discovering the
  QCD Axion with Black Holes and Gravitational Waves}},}\ }\href {\doibase
  10.1103/PhysRevD.91.084011} {\bibfield  {journal} {\bibinfo  {journal} {Phys.
  Rev. D}\ }\textbf {\bibinfo {volume} {91}},\ \bibinfo {pages} {084011}
  (\bibinfo {year} {2015})},\ \Eprint {http://arxiv.org/abs/1411.2263}
  {arXiv:1411.2263 [hep-ph]} \BibitemShut {NoStop}%
\bibitem [{\citenamefont {Baryakhtar}\ \emph {et~al.}(2021)\citenamefont
  {Baryakhtar}, \citenamefont {Galanis}, \citenamefont {Lasenby},\ and\
  \citenamefont {Simon}}]{Baryakhtar:2020gao}%
  \BibitemOpen
  \bibfield  {author} {\bibinfo {author} {\bibfnamefont {Masha}\ \bibnamefont
  {Baryakhtar}}, \bibinfo {author} {\bibfnamefont {Marios}\ \bibnamefont
  {Galanis}}, \bibinfo {author} {\bibfnamefont {Robert}\ \bibnamefont
  {Lasenby}}, \ and\ \bibinfo {author} {\bibfnamefont {Olivier}\ \bibnamefont
  {Simon}},\ }\bibfield  {title} {\enquote {\bibinfo {title} {{Black hole
  superradiance of self-interacting scalar fields}},}\ }\href {\doibase
  10.1103/PhysRevD.103.095019} {\bibfield  {journal} {\bibinfo  {journal}
  {Phys. Rev. D}\ }\textbf {\bibinfo {volume} {103}},\ \bibinfo {pages}
  {095019} (\bibinfo {year} {2021})},\ \Eprint
  {http://arxiv.org/abs/2011.11646} {arXiv:2011.11646 [hep-ph]} \BibitemShut
  {NoStop}%
\bibitem [{\citenamefont {Hoof}\ \emph {et~al.}(2026)\citenamefont {Hoof},
  \citenamefont {Marsh}, \citenamefont {Sisk-Reyn{\'e}s}, \citenamefont
  {Matthews},\ and\ \citenamefont {Reynolds}}]{Hoof:2024quk}%
  \BibitemOpen
  \bibfield  {author} {\bibinfo {author} {\bibfnamefont {Sebastian}\
  \bibnamefont {Hoof}}, \bibinfo {author} {\bibfnamefont {David J.~E.}\
  \bibnamefont {Marsh}}, \bibinfo {author} {\bibfnamefont {J{\'u}lia}\
  \bibnamefont {Sisk-Reyn{\'e}s}}, \bibinfo {author} {\bibfnamefont {James~H.}\
  \bibnamefont {Matthews}}, \ and\ \bibinfo {author} {\bibfnamefont
  {Christopher}\ \bibnamefont {Reynolds}},\ }\bibfield  {title} {\enquote
  {\bibinfo {title} {{Getting more out of black hole superradiance: a
  statistically rigorous approach to ultralight boson constraints from black
  hole spin measurements}},}\ }\href {\doibase 10.1093/mnras/staf1564}
  {\bibfield  {journal} {\bibinfo  {journal} {Mon. Not. Roy. Astron. Soc.}\
  }\textbf {\bibinfo {volume} {546}},\ \bibinfo {pages} {staf1564} (\bibinfo
  {year} {2026})},\ \Eprint {http://arxiv.org/abs/2406.10337} {arXiv:2406.10337
  [hep-ph]} \BibitemShut {NoStop}%
\bibitem [{\citenamefont {Ning}\ \emph {et~al.}(2026)\citenamefont {Ning},
  \citenamefont {Safdi},\ and\ \citenamefont {Welch}}]{Ning:2026ebu}%
  \BibitemOpen
  \bibfield  {author} {\bibinfo {author} {\bibfnamefont {Orion}\ \bibnamefont
  {Ning}}, \bibinfo {author} {\bibfnamefont {Benjamin~R.}\ \bibnamefont
  {Safdi}}, \ and\ \bibinfo {author} {\bibfnamefont {Catherine}\ \bibnamefont
  {Welch}},\ }\bibfield  {title} {\enquote {\bibinfo {title} {{No Evidence for
  Superradiant Axions in LIGO-Virgo-KAGRA GWTC-5 Binary Black Hole Spins}},}\
  }\href@noop {} {\  (\bibinfo {year} {2026})},\ \Eprint
  {http://arxiv.org/abs/2607.01317} {arXiv:2607.01317 [hep-ph]} \BibitemShut
  {NoStop}%
\bibitem [{\citenamefont {{ITER
  Organization}}(2026{\natexlab{a}})}]{ITER:magnets}%
  \BibitemOpen
  \bibfield  {author} {\bibinfo {author} {\bibnamefont {{ITER Organization}}},\
  }\href@noop {} {\enquote {\bibinfo {title} {{ITER Superconducting
  Magnets}},}\ }\bibinfo {howpublished}
  {\url{https://www.iter.org/machine/magnets}} (\bibinfo {year}
  {2026}{\natexlab{a}}),\ \bibinfo {note} {accessed 9 October 2026}\BibitemShut
  {NoStop}%
\bibitem [{\citenamefont {Bottura}(2000)}]{Bottura:1999vz}%
  \BibitemOpen
  \bibfield  {author} {\bibinfo {author} {\bibfnamefont {L.}~\bibnamefont
  {Bottura}},\ }\bibfield  {title} {\enquote {\bibinfo {title} {{A Practical
  fit for the critical surface of NbTi}},}\ }\href {\doibase 10.1109/77.828413}
  {\bibfield  {journal} {\bibinfo  {journal} {IEEE Trans. Appl. Supercond.}\
  }\textbf {\bibinfo {volume} {10}},\ \bibinfo {pages} {1054--1057} (\bibinfo
  {year} {2000})}\BibitemShut {NoStop}%
\bibitem [{\citenamefont {Aad}\ \emph {et~al.}(2008)\citenamefont {Aad} \emph
  {et~al.}}]{ATLAS:2008xda}%
  \BibitemOpen
  \bibfield  {author} {\bibinfo {author} {\bibfnamefont {G.}~\bibnamefont
  {Aad}} \emph {et~al.} (\bibinfo {collaboration} {ATLAS}),\ }\bibfield
  {title} {\enquote {\bibinfo {title} {{The ATLAS Experiment at the CERN Large
  Hadron Collider}},}\ }\href {\doibase 10.1088/1748-0221/3/08/S08003}
  {\bibfield  {journal} {\bibinfo  {journal} {JINST}\ }\textbf {\bibinfo
  {volume} {3}},\ \bibinfo {pages} {S08003} (\bibinfo {year}
  {2008})}\BibitemShut {NoStop}%
\bibitem [{\citenamefont {Wan}\ \emph {et~al.}(2002)\citenamefont {Wan},
  \citenamefont {Weng}, \citenamefont {Li}, \citenamefont {Yu}, \citenamefont
  {Gao},\ and\ \citenamefont {{HT-7U Team}}}]{Wan:2002fec}%
  \BibitemOpen
  \bibfield  {author} {\bibinfo {author} {\bibfnamefont {Y.~X.}\ \bibnamefont
  {Wan}}, \bibinfo {author} {\bibfnamefont {P.~D.}\ \bibnamefont {Weng}},
  \bibinfo {author} {\bibfnamefont {J.~G.}\ \bibnamefont {Li}}, \bibinfo
  {author} {\bibfnamefont {Q.~Q.}\ \bibnamefont {Yu}}, \bibinfo {author}
  {\bibfnamefont {D.~M.}\ \bibnamefont {Gao}}, \ and\ \bibinfo {author}
  {\bibnamefont {{HT-7U Team}}},\ }\bibfield  {title} {\enquote {\bibinfo
  {title} {{HT-7U Superconducting Tokamak: Physics design, engineering progress
  and schedule}},}\ }in\ \href@noop {} {\emph {\bibinfo {booktitle} {{19th IAEA
  Fusion Energy Conference}}}}\ (\bibinfo {year} {2002})\ \bibinfo {note}
  {paper FT/P2-03,
  \url{https://www-pub.iaea.org/mtcd/publications/pdf/csp_019c/pdf/ftp2_03.pdf}}\BibitemShut
  {NoStop}%
\bibitem [{\citenamefont {{ITER
  Organization}}(2026{\natexlab{b}})}]{ITER:facts}%
  \BibitemOpen
  \bibfield  {author} {\bibinfo {author} {\bibnamefont {{ITER Organization}}},\
  }\href@noop {} {\enquote {\bibinfo {title} {{Facts \& Figures}},}\ }\bibinfo
  {howpublished} {\url{https://www.iter.org/facts-figures}} (\bibinfo {year}
  {2026}{\natexlab{b}}),\ \bibinfo {note} {accessed 9 October 2026}\BibitemShut
  {NoStop}%
\bibitem [{\citenamefont {V{\'e}drine}(2014)}]{Vedrine:2014bdk}%
  \BibitemOpen
  \bibfield  {author} {\bibinfo {author} {\bibfnamefont {P.}~\bibnamefont
  {V{\'e}drine}},\ }\bibfield  {title} {\enquote {\bibinfo {title} {{Large
  Superconducting Magnet Systems}},}\ }in\ \href {\doibase
  10.5170/CERN-2014-005.559} {\emph {\bibinfo {booktitle} {{CERN Accelerator
  School}: {Course on Superconductivity for Accelerators}}}}\ (\bibinfo {year}
  {2014})\ pp.\ \bibinfo {pages} {559--583},\ \Eprint
  {http://arxiv.org/abs/1501.07169} {arXiv:1501.07169 [physics.acc-ph]}
  \BibitemShut {NoStop}%
\bibitem [{\citenamefont {Callen}\ and\ \citenamefont
  {Welton}(1951)}]{Callen:1951vq}%
  \BibitemOpen
  \bibfield  {author} {\bibinfo {author} {\bibfnamefont {Herbert~B.}\
  \bibnamefont {Callen}}\ and\ \bibinfo {author} {\bibfnamefont {Theodore~A.}\
  \bibnamefont {Welton}},\ }\bibfield  {title} {\enquote {\bibinfo {title}
  {{Irreversibility and generalized noise}},}\ }\href {\doibase
  10.1103/PhysRev.83.34} {\bibfield  {journal} {\bibinfo  {journal} {Phys.
  Rev.}\ }\textbf {\bibinfo {volume} {83}},\ \bibinfo {pages} {34--40}
  (\bibinfo {year} {1951})}\BibitemShut {NoStop}%
\bibitem [{\citenamefont {Levin}(1998)}]{Levin:1997kv}%
  \BibitemOpen
  \bibfield  {author} {\bibinfo {author} {\bibfnamefont {Yu.}\ \bibnamefont
  {Levin}},\ }\bibfield  {title} {\enquote {\bibinfo {title} {{Internal thermal
  noise in the LIGO test masses: A Direct approach}},}\ }\href {\doibase
  10.1103/PhysRevD.57.659} {\bibfield  {journal} {\bibinfo  {journal} {Phys.
  Rev. D}\ }\textbf {\bibinfo {volume} {57}},\ \bibinfo {pages} {659--663}
  (\bibinfo {year} {1998})},\ \Eprint {http://arxiv.org/abs/gr-qc/9707013}
  {arXiv:gr-qc/9707013} \BibitemShut {NoStop}%
\bibitem [{\citenamefont {Cremer}\ \emph {et~al.}(2005)\citenamefont {Cremer},
  \citenamefont {Heckl},\ and\ \citenamefont {Petersson}}]{Cremer:2005sbs}%
  \BibitemOpen
  \bibfield  {author} {\bibinfo {author} {\bibfnamefont {L.}~\bibnamefont
  {Cremer}}, \bibinfo {author} {\bibfnamefont {M.}~\bibnamefont {Heckl}}, \
  and\ \bibinfo {author} {\bibfnamefont {B.~A.~T.}\ \bibnamefont {Petersson}},\
  }\href@noop {} {\emph {\bibinfo {title} {{Structure-Borne Sound: Structural
  Vibrations and Sound Radiation at Audio Frequencies}}}},\ \bibinfo {edition}
  {3rd}\ ed.\ (\bibinfo  {publisher} {Springer},\ \bibinfo {year}
  {2005})\BibitemShut {NoStop}%
\bibitem [{\citenamefont {Giaccone}\ \emph {et~al.}(2022)\citenamefont
  {Giaccone} \emph {et~al.}}]{Giaccone:2022pke}%
  \BibitemOpen
  \bibfield  {author} {\bibinfo {author} {\bibfnamefont {B.}~\bibnamefont
  {Giaccone}} \emph {et~al.},\ }\bibfield  {title} {\enquote {\bibinfo {title}
  {{Design of axion and axion dark matter searches based on ultra high Q SRF
  cavities}},}\ }\href@noop {} {\  (\bibinfo {year} {2022})},\ \Eprint
  {http://arxiv.org/abs/2207.11346} {arXiv:2207.11346 [hep-ex]} \BibitemShut
  {NoStop}%
\bibitem [{\citenamefont {Castaneda}(2024)}]{Castaneda:2024shade}%
  \BibitemOpen
  \bibfield  {author} {\bibinfo {author} {\bibfnamefont {F.}~\bibnamefont
  {Castaneda}},\ }\href {\doibase 10.2172/3000318} {\emph {\bibinfo {title}
  {{Preparation for DM searches with high Q SRF cavities}}}},\ \bibinfo {type}
  {Tech. Rep.}\ \bibinfo {number} {FERMILAB-TM-2866-STUDENT}\ (\bibinfo
  {institution} {Fermilab},\ \bibinfo {year} {2024})\BibitemShut {NoStop}%
\bibitem [{\citenamefont {Holmes}\ \emph {et~al.}(2013)\citenamefont {Holmes}
  \emph {et~al.}}]{Holmes:2013vpa}%
  \BibitemOpen
  \bibfield  {author} {\bibinfo {author} {\bibfnamefont {S.~D.}\ \bibnamefont
  {Holmes}} \emph {et~al.},\ }\bibfield  {title} {\enquote {\bibinfo {title}
  {{Project X: Accelerator Reference Design, Physics Opportunities, Broader
  Impacts}},}\ }\href {\doibase 10.2172/1083535} {\  (\bibinfo {year} {2013}),\
  10.2172/1083535},\ \Eprint {http://arxiv.org/abs/1306.5022} {arXiv:1306.5022
  [physics.acc-ph]} \BibitemShut {NoStop}%
\bibitem [{\citenamefont {Thomson}\ \emph {et~al.}(2023)\citenamefont
  {Thomson}, \citenamefont {Goryachev}, \citenamefont {McAllister},
  \citenamefont {Ivanov}, \citenamefont {Altin},\ and\ \citenamefont
  {Tobar}}]{Thomson:2023moc}%
  \BibitemOpen
  \bibfield  {author} {\bibinfo {author} {\bibfnamefont {Catriona~A.}\
  \bibnamefont {Thomson}}, \bibinfo {author} {\bibfnamefont {Maxim}\
  \bibnamefont {Goryachev}}, \bibinfo {author} {\bibfnamefont {Ben~T.}\
  \bibnamefont {McAllister}}, \bibinfo {author} {\bibfnamefont {Eugene~N.}\
  \bibnamefont {Ivanov}}, \bibinfo {author} {\bibfnamefont {Paul}\ \bibnamefont
  {Altin}}, \ and\ \bibinfo {author} {\bibfnamefont {Michael~E.}\ \bibnamefont
  {Tobar}},\ }\bibfield  {title} {\enquote {\bibinfo {title} {{Searching for
  low-mass axions using resonant upconversion}},}\ }\href {\doibase
  10.1103/PhysRevD.107.112003} {\bibfield  {journal} {\bibinfo  {journal}
  {Phys. Rev. D}\ }\textbf {\bibinfo {volume} {107}},\ \bibinfo {pages}
  {112003} (\bibinfo {year} {2023})},\ \Eprint
  {http://arxiv.org/abs/2301.06778} {arXiv:2301.06778 [hep-ex]} \BibitemShut
  {NoStop}%
\bibitem [{\citenamefont {Miyazaki}\ \emph {et~al.}(2023)\citenamefont
  {Miyazaki}, \citenamefont {Duchesne}, \citenamefont {Ledrean}, \citenamefont
  {Longuevergne},\ and\ \citenamefont {Olry}}]{Miyazaki:2023zye}%
  \BibitemOpen
  \bibfield  {author} {\bibinfo {author} {\bibfnamefont {Akira}\ \bibnamefont
  {Miyazaki}}, \bibinfo {author} {\bibfnamefont {Patricia}\ \bibnamefont
  {Duchesne}}, \bibinfo {author} {\bibfnamefont {D.}~\bibnamefont {Ledrean}},
  \bibinfo {author} {\bibfnamefont {David}\ \bibnamefont {Longuevergne}}, \
  and\ \bibinfo {author} {\bibfnamefont {Guillaume}\ \bibnamefont {Olry}},\
  }\bibfield  {title} {\enquote {\bibinfo {title} {{Performance Analysis of
  Spoke Resonators, Statistics from Cavity Fabrication to Cryomodule
  Testing}},}\ }\href {\doibase 10.18429/JACoW-SRF2023-THIAA04} {\bibfield
  {journal} {\bibinfo  {journal} {JACoW}\ }\textbf {\bibinfo {volume}
  {SRF2023}},\ \bibinfo {pages} {THIAA04} (\bibinfo {year} {2023})},\ \Eprint
  {http://arxiv.org/abs/2306.10768} {arXiv:2306.10768 [physics.acc-ph]}
  \BibitemShut {NoStop}%
\bibitem [{\citenamefont {Kolb}\ \emph {et~al.}(2020)\citenamefont {Kolb},
  \citenamefont {Yao}, \citenamefont {Junginger}, \citenamefont {Dury},
  \citenamefont {Fothergill}, \citenamefont {Vanderbanck},\ and\ \citenamefont
  {Laxdal}}]{Kolb:2020jab}%
  \BibitemOpen
  \bibfield  {author} {\bibinfo {author} {\bibfnamefont {P.}~\bibnamefont
  {Kolb}}, \bibinfo {author} {\bibfnamefont {Z.}~\bibnamefont {Yao}}, \bibinfo
  {author} {\bibfnamefont {T.}~\bibnamefont {Junginger}}, \bibinfo {author}
  {\bibfnamefont {B.}~\bibnamefont {Dury}}, \bibinfo {author} {\bibfnamefont
  {A.}~\bibnamefont {Fothergill}}, \bibinfo {author} {\bibfnamefont
  {M.}~\bibnamefont {Vanderbanck}}, \ and\ \bibinfo {author} {\bibfnamefont
  {R.~E.}\ \bibnamefont {Laxdal}},\ }\bibfield  {title} {\enquote {\bibinfo
  {title} {{Coaxial multimode cavities for fundamental superconducting rf
  research in an unprecedented parameter space}},}\ }\href {\doibase
  10.1103/PhysRevAccelBeams.23.122001} {\bibfield  {journal} {\bibinfo
  {journal} {Phys. Rev. Accel. Beams}\ }\textbf {\bibinfo {volume} {23}},\
  \bibinfo {pages} {122001} (\bibinfo {year} {2020})},\ \Eprint
  {http://arxiv.org/abs/2010.02192} {arXiv:2010.02192 [physics.acc-ph]}
  \BibitemShut {NoStop}%
\bibitem [{\citenamefont {Altarelli}\ \emph {et~al.}(2006)\citenamefont
  {Altarelli} \emph {et~al.}}]{Altarelli:2006zza}%
  \BibitemOpen
  \bibinfo {editor} {\bibfnamefont {Massimo}\ \bibnamefont {Altarelli}} \emph
  {et~al.},\ eds.,\ \href {\doibase 10.3204/DESY_06-097} {\emph {\bibinfo
  {title} {{XFEL: The European X-Ray Free-Electron Laser. Technical design
  report}}}}\ (\bibinfo  {publisher} {DESY},\ \bibinfo {address} {Hamburg},\
  \bibinfo {year} {2006})\BibitemShut {NoStop}%
\bibitem [{\citenamefont {Fauve}\ \emph {et~al.}(2024)\citenamefont {Fauve},
  \citenamefont {Pflueckhahn}, \citenamefont {Sevilla}, \citenamefont
  {Ravindranath}, \citenamefont {Chung}, \citenamefont {Hogan}, \citenamefont
  {Bevins},\ and\ \citenamefont {Bhattacharya}}]{Fauve:2024lcls}%
  \BibitemOpen
  \bibfield  {author} {\bibinfo {author} {\bibfnamefont {E.}~\bibnamefont
  {Fauve}}, \bibinfo {author} {\bibfnamefont {D.}~\bibnamefont {Pflueckhahn}},
  \bibinfo {author} {\bibfnamefont {J.}~\bibnamefont {Sevilla}}, \bibinfo
  {author} {\bibfnamefont {V.}~\bibnamefont {Ravindranath}}, \bibinfo {author}
  {\bibfnamefont {C.}~\bibnamefont {Chung}}, \bibinfo {author} {\bibfnamefont
  {J.}~\bibnamefont {Hogan}}, \bibinfo {author} {\bibfnamefont
  {M.}~\bibnamefont {Bevins}}, \ and\ \bibinfo {author} {\bibfnamefont
  {R.}~\bibnamefont {Bhattacharya}},\ }\bibfield  {title} {\enquote {\bibinfo
  {title} {{LCLS-II Helium Refrigeration System: A comprehensive overview of
  the project}},}\ }\href {\doibase 10.1088/1757-899X/1301/1/012117} {\bibfield
   {journal} {\bibinfo  {journal} {IOP Conf. Ser. Mater. Sci. Eng.}\ }\textbf
  {\bibinfo {volume} {1301}},\ \bibinfo {pages} {012117} (\bibinfo {year}
  {2024})}\BibitemShut {NoStop}%
\bibitem [{\citenamefont {Green}(2008)}]{Green:2008refrigerators}%
  \BibitemOpen
  \bibfield  {author} {\bibinfo {author} {\bibfnamefont {M.~A.}\ \bibnamefont
  {Green}},\ }\bibfield  {title} {\enquote {\bibinfo {title} {{The Cost of
  Helium Refrigerators and Coolers for Superconducting Devices as a Function of
  Cooling at 4 K}},}\ }\href {\doibase 10.1063/1.2908683} {\bibfield  {journal}
  {\bibinfo  {journal} {AIP Conf. Proc.}\ }\textbf {\bibinfo {volume} {985}},\
  \bibinfo {pages} {872--878} (\bibinfo {year} {2008})}\BibitemShut {NoStop}%
\bibitem [{\citenamefont {Pflueckhahn}\ \emph {et~al.}(2022)\citenamefont
  {Pflueckhahn} \emph {et~al.}}]{Pflueckhahn:2022zun}%
  \BibitemOpen
  \bibfield  {author} {\bibinfo {author} {\bibfnamefont {D.}~\bibnamefont
  {Pflueckhahn}} \emph {et~al.},\ }\bibfield  {title} {\enquote {\bibinfo
  {title} {{LCLS-II helium cryoplant and cryo distribution system
  installation}},}\ }\href {\doibase 10.1088/1757-899X/1240/1/012113}
  {\bibfield  {journal} {\bibinfo  {journal} {IOP Conf. Ser. Mater. Sci. Eng.}\
  }\textbf {\bibinfo {volume} {1240}},\ \bibinfo {pages} {012113} (\bibinfo
  {year} {2022})}\BibitemShut {NoStop}%
\bibitem [{\citenamefont {Lebrun}(2003)}]{Lebrun:2003zm}%
  \BibitemOpen
  \bibfield  {author} {\bibinfo {author} {\bibfnamefont {P.}~\bibnamefont
  {Lebrun}},\ }\bibfield  {title} {\enquote {\bibinfo {title} {{Large cryogenic
  helium refrigeration system for the LHC}},}\ }in\ \href@noop {} {\emph
  {\bibinfo {booktitle} {{Proc. 5th Int. Conf. on Cryogenics and Refrigeration
  (ICCR 2003), Hangzhou}}}}\ (\bibinfo {year} {2003})\BibitemShut {NoStop}%
\bibitem [{\citenamefont {Benvenuti}\ \emph {et~al.}(1984)\citenamefont
  {Benvenuti}, \citenamefont {Circelli},\ and\ \citenamefont
  {Hauer}}]{Benvenuti:1984er}%
  \BibitemOpen
  \bibfield  {author} {\bibinfo {author} {\bibfnamefont {C.}~\bibnamefont
  {Benvenuti}}, \bibinfo {author} {\bibfnamefont {N.}~\bibnamefont {Circelli}},
  \ and\ \bibinfo {author} {\bibfnamefont {M.}~\bibnamefont {Hauer}},\
  }\bibfield  {title} {\enquote {\bibinfo {title} {{NIOBIUM FILMS FOR
  SUPERCONDUCTING ACCELERATING CAVITIES}},}\ }\href {\doibase 10.1063/1.95289}
  {\bibfield  {journal} {\bibinfo  {journal} {Appl. Phys. Lett.}\ }\textbf
  {\bibinfo {volume} {45}},\ \bibinfo {pages} {583--584} (\bibinfo {year}
  {1984})}\BibitemShut {NoStop}%
\bibitem [{\citenamefont {Miyazaki}\ and\ \citenamefont
  {Venturini~Delsolaro}(2019)}]{Miyazaki:2018pbz}%
  \BibitemOpen
  \bibfield  {author} {\bibinfo {author} {\bibfnamefont {A.}~\bibnamefont
  {Miyazaki}}\ and\ \bibinfo {author} {\bibfnamefont {W.}~\bibnamefont
  {Venturini~Delsolaro}},\ }\bibfield  {title} {\enquote {\bibinfo {title}
  {{Two different origins of the Q-slope problem in superconducting niobium
  film cavities for a heavy ion accelerator at CERN}},}\ }\href {\doibase
  10.1103/PhysRevAccelBeams.22.073101} {\bibfield  {journal} {\bibinfo
  {journal} {Phys. Rev. Accel. Beams}\ }\textbf {\bibinfo {volume} {22}},\
  \bibinfo {pages} {073101} (\bibinfo {year} {2019})},\ \Eprint
  {http://arxiv.org/abs/1812.04658} {arXiv:1812.04658 [physics.acc-ph]}
  \BibitemShut {NoStop}%
\bibitem [{\citenamefont {Slater}(1946)}]{Slater:1946}%
  \BibitemOpen
  \bibfield  {author} {\bibinfo {author} {\bibfnamefont {J.~C.}\ \bibnamefont
  {Slater}},\ }\bibfield  {title} {\enquote {\bibinfo {title} {{Microwave
  Electronics}},}\ }\href {\doibase 10.1103/RevModPhys.18.441} {\bibfield
  {journal} {\bibinfo  {journal} {Rev. Mod. Phys.}\ }\textbf {\bibinfo {volume}
  {18}},\ \bibinfo {pages} {441--512} (\bibinfo {year} {1946})}\BibitemShut
  {NoStop}%
\bibitem [{\citenamefont {Li}\ \emph {et~al.}(2026)\citenamefont {Li},
  \citenamefont {Zhou}, \citenamefont {Oriunno}, \citenamefont {Berlin},
  \citenamefont {Calatroni}, \citenamefont {D'Agnolo}, \citenamefont {Ellis},
  \citenamefont {Schuster}, \citenamefont {Tantawi},\ and\ \citenamefont
  {Toro}}]{Li:2025pyi}%
  \BibitemOpen
  \bibfield  {author} {\bibinfo {author} {\bibfnamefont {Zenghai}\ \bibnamefont
  {Li}}, \bibinfo {author} {\bibfnamefont {Kevin}\ \bibnamefont {Zhou}},
  \bibinfo {author} {\bibfnamefont {Marco}\ \bibnamefont {Oriunno}}, \bibinfo
  {author} {\bibfnamefont {Asher}\ \bibnamefont {Berlin}}, \bibinfo {author}
  {\bibfnamefont {Sergio}\ \bibnamefont {Calatroni}}, \bibinfo {author}
  {\bibfnamefont {Raffaele~Tito}\ \bibnamefont {D'Agnolo}}, \bibinfo {author}
  {\bibfnamefont {Sebastian A.~R.}\ \bibnamefont {Ellis}}, \bibinfo {author}
  {\bibfnamefont {Philip}\ \bibnamefont {Schuster}}, \bibinfo {author}
  {\bibfnamefont {Sami~G.}\ \bibnamefont {Tantawi}}, \ and\ \bibinfo {author}
  {\bibfnamefont {Natalia}\ \bibnamefont {Toro}},\ }\bibfield  {title}
  {\enquote {\bibinfo {title} {{A prototype hybrid mode cavity for heterodyne
  axion detection}},}\ }\href {\doibase 10.1016/j.nima.2026.171474} {\bibfield
  {journal} {\bibinfo  {journal} {Nucl. Instrum. Meth. A}\ }\textbf {\bibinfo
  {volume} {1088}},\ \bibinfo {pages} {171474} (\bibinfo {year} {2026})},\
  \Eprint {http://arxiv.org/abs/2507.07173} {arXiv:2507.07173
  [physics.ins-det]} \BibitemShut {NoStop}%
\bibitem [{\citenamefont {Schiff}(1963)}]{Schiff:1963zz}%
  \BibitemOpen
  \bibfield  {author} {\bibinfo {author} {\bibfnamefont {L.~I.}\ \bibnamefont
  {Schiff}},\ }\bibfield  {title} {\enquote {\bibinfo {title} {{Measurability
  of Nuclear Electric Dipole Moments}},}\ }\href {\doibase
  10.1103/PhysRev.132.2194} {\bibfield  {journal} {\bibinfo  {journal} {Phys.
  Rev.}\ }\textbf {\bibinfo {volume} {132}},\ \bibinfo {pages} {2194--2200}
  (\bibinfo {year} {1963})}\BibitemShut {NoStop}%
\bibitem [{\citenamefont {Pospelov}\ and\ \citenamefont
  {Ritz}(2000)}]{Pospelov:1999mv}%
  \BibitemOpen
  \bibfield  {author} {\bibinfo {author} {\bibfnamefont {Maxim}\ \bibnamefont
  {Pospelov}}\ and\ \bibinfo {author} {\bibfnamefont {Adam}\ \bibnamefont
  {Ritz}},\ }\bibfield  {title} {\enquote {\bibinfo {title} {{Theta vacua, QCD
  sum rules, and the neutron electric dipole moment}},}\ }\href {\doibase
  10.1016/S0550-3213(99)00817-2} {\bibfield  {journal} {\bibinfo  {journal}
  {Nucl. Phys. B}\ }\textbf {\bibinfo {volume} {573}},\ \bibinfo {pages}
  {177--200} (\bibinfo {year} {2000})},\ \Eprint
  {http://arxiv.org/abs/hep-ph/9908508} {arXiv:hep-ph/9908508} \BibitemShut
  {NoStop}%
\bibitem [{\citenamefont {Skripnikov}\ and\ \citenamefont
  {Titov}(2016)}]{Skripnikov:2016pto}%
  \BibitemOpen
  \bibfield  {author} {\bibinfo {author} {\bibfnamefont {L.~V.}\ \bibnamefont
  {Skripnikov}}\ and\ \bibinfo {author} {\bibfnamefont {A.~V.}\ \bibnamefont
  {Titov}},\ }\bibfield  {title} {\enquote {\bibinfo {title} {{LCAO-based
  theoretical study of PbTiO3 crystal to search for parity and time reversal
  violating interaction in solids}},}\ }\href {\doibase 10.1063/1.4959973}
  {\bibfield  {journal} {\bibinfo  {journal} {J. Chem. Phys.}\ }\textbf
  {\bibinfo {volume} {145}},\ \bibinfo {pages} {054115} (\bibinfo {year}
  {2016})}\BibitemShut {NoStop}%
\bibitem [{\citenamefont {Winter}\ \emph {et~al.}(2026)\citenamefont {Winter},
  \citenamefont {{\'A}d{\'a}m}, \citenamefont {Aybas}, \citenamefont {Budker},
  \citenamefont {Kimball}, \citenamefont {Rajendran}, \citenamefont
  {Wickenbrock},\ and\ \citenamefont {Sushkov}}]{Winter:2026nge}%
  \BibitemOpen
  \bibfield  {author} {\bibinfo {author} {\bibfnamefont {Andrew~J.}\
  \bibnamefont {Winter}}, \bibinfo {author} {\bibfnamefont {J{\'a}nos}\
  \bibnamefont {{\'A}d{\'a}m}}, \bibinfo {author} {\bibfnamefont {Deniz}\
  \bibnamefont {Aybas}}, \bibinfo {author} {\bibfnamefont {Dmitry}\
  \bibnamefont {Budker}}, \bibinfo {author} {\bibfnamefont {Derek F.~Jackson}\
  \bibnamefont {Kimball}}, \bibinfo {author} {\bibfnamefont {Surjeet}\
  \bibnamefont {Rajendran}}, \bibinfo {author} {\bibfnamefont {Arne}\
  \bibnamefont {Wickenbrock}}, \ and\ \bibinfo {author} {\bibfnamefont
  {Alexander~O.}\ \bibnamefont {Sushkov}},\ }\bibfield  {title} {\enquote
  {\bibinfo {title} {{Solid-state nuclear magnetic resonance search for
  axion-like dark matter with broadband SQUID magnetometry}},}\ }\href@noop {}
  {\  (\bibinfo {year} {2026})},\ \Eprint {http://arxiv.org/abs/2609.07908}
  {arXiv:2609.07908 [hep-ex]} \BibitemShut {NoStop}%
\bibitem [{\citenamefont {Aybas}\ \emph
  {et~al.}(2021{\natexlab{b}})\citenamefont {Aybas}, \citenamefont {Bekker},
  \citenamefont {Blanchard}, \citenamefont {Budker}, \citenamefont {Centers},
  \citenamefont {Figueroa}, \citenamefont {Gramolin}, \citenamefont {Kimball},
  \citenamefont {Wickenbrock},\ and\ \citenamefont {Sushkov}}]{Aybas:2021cdk}%
  \BibitemOpen
  \bibfield  {author} {\bibinfo {author} {\bibfnamefont {Deniz}\ \bibnamefont
  {Aybas}}, \bibinfo {author} {\bibfnamefont {Hendrik}\ \bibnamefont {Bekker}},
  \bibinfo {author} {\bibfnamefont {John~W.}\ \bibnamefont {Blanchard}},
  \bibinfo {author} {\bibfnamefont {Dmitry}\ \bibnamefont {Budker}}, \bibinfo
  {author} {\bibfnamefont {Gary~P.}\ \bibnamefont {Centers}}, \bibinfo {author}
  {\bibfnamefont {Nataniel~L.}\ \bibnamefont {Figueroa}}, \bibinfo {author}
  {\bibfnamefont {Alexander~V.}\ \bibnamefont {Gramolin}}, \bibinfo {author}
  {\bibfnamefont {Derek F.~Jackson}\ \bibnamefont {Kimball}}, \bibinfo {author}
  {\bibfnamefont {Arne}\ \bibnamefont {Wickenbrock}}, \ and\ \bibinfo {author}
  {\bibfnamefont {Alexander~O.}\ \bibnamefont {Sushkov}},\ }\bibfield  {title}
  {\enquote {\bibinfo {title} {{Quantum sensitivity limits of nuclear magnetic
  resonance experiments searching for new fundamental physics}},}\ }\href
  {\doibase 10.1088/2058-9565/abfbbc} {\bibfield  {journal} {\bibinfo
  {journal} {Quantum Sci. Technol.}\ }\textbf {\bibinfo {volume} {6}},\
  \bibinfo {pages} {034007} (\bibinfo {year} {2021}{\natexlab{b}})},\ \Eprint
  {http://arxiv.org/abs/2103.06284} {arXiv:2103.06284 [quant-ph]} \BibitemShut
  {NoStop}%
\bibitem [{\citenamefont {Dror}\ \emph {et~al.}(2023)\citenamefont {Dror},
  \citenamefont {Gori}, \citenamefont {Leedom},\ and\ \citenamefont
  {Rodd}}]{Dror:2022xpi}%
  \BibitemOpen
  \bibfield  {author} {\bibinfo {author} {\bibfnamefont {Jeff~A.}\ \bibnamefont
  {Dror}}, \bibinfo {author} {\bibfnamefont {Stefania}\ \bibnamefont {Gori}},
  \bibinfo {author} {\bibfnamefont {Jacob~M.}\ \bibnamefont {Leedom}}, \ and\
  \bibinfo {author} {\bibfnamefont {Nicholas~L.}\ \bibnamefont {Rodd}},\
  }\bibfield  {title} {\enquote {\bibinfo {title} {{Sensitivity of
  Spin-Precession Axion Experiments}},}\ }\href {\doibase
  10.1103/PhysRevLett.130.181801} {\bibfield  {journal} {\bibinfo  {journal}
  {Phys. Rev. Lett.}\ }\textbf {\bibinfo {volume} {130}},\ \bibinfo {pages}
  {181801} (\bibinfo {year} {2023})},\ \Eprint
  {http://arxiv.org/abs/2210.06481} {arXiv:2210.06481 [hep-ph]} \BibitemShut
  {NoStop}%
\bibitem [{\citenamefont {Kuenstner}\ \emph {et~al.}(2026)\citenamefont
  {Kuenstner}, \citenamefont {Smith}, \citenamefont {Winter}, \citenamefont
  {Ozdemir}, \citenamefont {Mari{\'c}}, \citenamefont {Matthews},\ and\
  \citenamefont {Sushkov}}]{Kuenstner:2026fuu}%
  \BibitemOpen
  \bibfield  {author} {\bibinfo {author} {\bibfnamefont {Stephen~E.}\
  \bibnamefont {Kuenstner}}, \bibinfo {author} {\bibfnamefont {Declan~W.}\
  \bibnamefont {Smith}}, \bibinfo {author} {\bibfnamefont {Andrew~J.}\
  \bibnamefont {Winter}}, \bibinfo {author} {\bibfnamefont {Eren}\ \bibnamefont
  {Ozdemir}}, \bibinfo {author} {\bibfnamefont {Tanja}\ \bibnamefont
  {Mari{\'c}}}, \bibinfo {author} {\bibfnamefont {Alyssa}\ \bibnamefont
  {Matthews}}, \ and\ \bibinfo {author} {\bibfnamefont {Alexander~O.}\
  \bibnamefont {Sushkov}},\ }\bibfield  {title} {\enquote {\bibinfo {title}
  {{Quantum-limited metrology of macroscopic spin ensembles}},}\ }\href
  {\doibase 10.1038/s41567-026-03187-6} {\bibfield  {journal} {\bibinfo
  {journal} {Nature Phys.}\ }\textbf {\bibinfo {volume} {22}},\ \bibinfo
  {pages} {571--576} (\bibinfo {year} {2026})}\BibitemShut {NoStop}%
\bibitem [{\citenamefont {Bouchard}\ \emph {et~al.}(2008)\citenamefont
  {Bouchard}, \citenamefont {Sushkov}, \citenamefont {Budker}, \citenamefont
  {Ford},\ and\ \citenamefont {Lipton}}]{Bouchard:2007tji}%
  \BibitemOpen
  \bibfield  {author} {\bibinfo {author} {\bibfnamefont {L.~S.}\ \bibnamefont
  {Bouchard}}, \bibinfo {author} {\bibfnamefont {A.~O.}\ \bibnamefont
  {Sushkov}}, \bibinfo {author} {\bibfnamefont {D.}~\bibnamefont {Budker}},
  \bibinfo {author} {\bibfnamefont {J.~J.}\ \bibnamefont {Ford}}, \ and\
  \bibinfo {author} {\bibfnamefont {A.~S.}\ \bibnamefont {Lipton}},\ }\bibfield
   {title} {\enquote {\bibinfo {title} {{Nuclear-spin relaxation of $^{207}$Pb
  in ferroelectric powders}},}\ }\href {\doibase 10.1103/PhysRevA.77.022102}
  {\bibfield  {journal} {\bibinfo  {journal} {Phys. Rev. A}\ }\textbf {\bibinfo
  {volume} {77}},\ \bibinfo {pages} {022102} (\bibinfo {year} {2008})},\
  \Eprint {http://arxiv.org/abs/0711.1392} {arXiv:0711.1392 [cond-mat.other]}
  \BibitemShut {NoStop}%
\bibitem [{\citenamefont {Mukhamedjanov}\ and\ \citenamefont
  {Sushkov}(2005)}]{Mukhamedjanov:2004ki}%
  \BibitemOpen
  \bibfield  {author} {\bibinfo {author} {\bibfnamefont {T.~N.}\ \bibnamefont
  {Mukhamedjanov}}\ and\ \bibinfo {author} {\bibfnamefont {O.~P.}\ \bibnamefont
  {Sushkov}},\ }\bibfield  {title} {\enquote {\bibinfo {title} {{A Suggested
  search for Pb-207 nuclear Schiff moment in PbTiO(3) ferroelectric}},}\ }\href
  {\doibase 10.1103/PhysRevA.72.034501} {\bibfield  {journal} {\bibinfo
  {journal} {Phys. Rev. A}\ }\textbf {\bibinfo {volume} {72}},\ \bibinfo
  {pages} {034501} (\bibinfo {year} {2005})},\ \Eprint
  {http://arxiv.org/abs/physics/0411226} {arXiv:physics/0411226} \BibitemShut
  {NoStop}%
\bibitem [{\citenamefont {Nelmes}\ and\ \citenamefont
  {Kuhs}(1985)}]{Nelmes:1985pto}%
  \BibitemOpen
  \bibfield  {author} {\bibinfo {author} {\bibfnamefont {R.~J.}\ \bibnamefont
  {Nelmes}}\ and\ \bibinfo {author} {\bibfnamefont {W.~F.}\ \bibnamefont
  {Kuhs}},\ }\bibfield  {title} {\enquote {\bibinfo {title} {{The crystal
  structure of tetragonal PbTiO3 at room temperature and at 700 K}},}\ }\href
  {\doibase 10.1016/0038-1098(85)90595-2} {\bibfield  {journal} {\bibinfo
  {journal} {Solid State Commun.}\ }\textbf {\bibinfo {volume} {54}},\ \bibinfo
  {pages} {721--723} (\bibinfo {year} {1985})}\BibitemShut {NoStop}%
\bibitem [{\citenamefont {Wang}\ \emph {et~al.}(2015)\citenamefont {Wang} \emph
  {et~al.}}]{CUPID:2015tlr}%
  \BibitemOpen
  \bibfield  {author} {\bibinfo {author} {\bibfnamefont {G.}~\bibnamefont
  {Wang}} \emph {et~al.} (\bibinfo {collaboration} {CUPID}),\ }\bibfield
  {title} {\enquote {\bibinfo {title} {{R{\&}D towards CUPID (CUORE Upgrade
  with Particle IDentification)}},}\ }\href@noop {} {\  (\bibinfo {year}
  {2015})},\ \Eprint {http://arxiv.org/abs/1504.03612} {arXiv:1504.03612
  [physics.ins-det]} \BibitemShut {NoStop}%
\bibitem [{\citenamefont {{\'A}d{\'a}m}\ \emph {et~al.}(2026)\citenamefont
  {{\'A}d{\'a}m}, \citenamefont {Winter}, \citenamefont {Aybas}, \citenamefont
  {Budker}, \citenamefont {Kimball}, \citenamefont {Wickenbrock},\ and\
  \citenamefont {Sushkov}}]{Adam:2026odx}%
  \BibitemOpen
  \bibfield  {author} {\bibinfo {author} {\bibfnamefont {J{\'a}nos}\
  \bibnamefont {{\'A}d{\'a}m}}, \bibinfo {author} {\bibfnamefont {Andrew~J.}\
  \bibnamefont {Winter}}, \bibinfo {author} {\bibfnamefont {Deniz}\
  \bibnamefont {Aybas}}, \bibinfo {author} {\bibfnamefont {Dmitry}\
  \bibnamefont {Budker}}, \bibinfo {author} {\bibfnamefont {Derek F.~Jackson}\
  \bibnamefont {Kimball}}, \bibinfo {author} {\bibfnamefont {Arne}\
  \bibnamefont {Wickenbrock}}, \ and\ \bibinfo {author} {\bibfnamefont
  {Alexander~O.}\ \bibnamefont {Sushkov}},\ }\bibfield  {title} {\enquote
  {\bibinfo {title} {{Control of relaxation properties of a macroscopic nuclear
  spin ensemble}},}\ }\href@noop {} {\  (\bibinfo {year} {2026})},\ \Eprint
  {http://arxiv.org/abs/2604.25548} {arXiv:2604.25548 [cond-mat.other]}
  \BibitemShut {NoStop}%
\bibitem [{\citenamefont {Ladd}\ \emph {et~al.}(2005)\citenamefont {Ladd},
  \citenamefont {Maryenko}, \citenamefont {Abe}, \citenamefont {Itoh},\ and\
  \citenamefont {Yamamoto}}]{Ladd:2005}%
  \BibitemOpen
  \bibfield  {author} {\bibinfo {author} {\bibfnamefont {T.~D.}\ \bibnamefont
  {Ladd}}, \bibinfo {author} {\bibfnamefont {D.}~\bibnamefont {Maryenko}},
  \bibinfo {author} {\bibfnamefont {E.}~\bibnamefont {Abe}}, \bibinfo {author}
  {\bibfnamefont {K.~M.}\ \bibnamefont {Itoh}}, \ and\ \bibinfo {author}
  {\bibfnamefont {Y.}~\bibnamefont {Yamamoto}},\ }\bibfield  {title} {\enquote
  {\bibinfo {title} {{Coherence Time of a Solid-State Nuclear Qubit}},}\ }\href
  {\doibase 10.1103/PhysRevB.71.014401} {\bibfield  {journal} {\bibinfo
  {journal} {Phys. Rev. B}\ }\textbf {\bibinfo {volume} {71}},\ \bibinfo
  {pages} {014401} (\bibinfo {year} {2005})},\ \Eprint
  {http://arxiv.org/abs/quant-ph/0309164} {arXiv:quant-ph/0309164} \BibitemShut
  {NoStop}%
\bibitem [{\citenamefont {Dong}\ \emph {et~al.}(2008)\citenamefont {Dong},
  \citenamefont {Ramos}, \citenamefont {Li},\ and\ \citenamefont
  {Barrett}}]{Dong:2008}%
  \BibitemOpen
  \bibfield  {author} {\bibinfo {author} {\bibfnamefont {Yanqun}\ \bibnamefont
  {Dong}}, \bibinfo {author} {\bibfnamefont {R.~G.}\ \bibnamefont {Ramos}},
  \bibinfo {author} {\bibfnamefont {Dale}\ \bibnamefont {Li}}, \ and\ \bibinfo
  {author} {\bibfnamefont {S.~E.}\ \bibnamefont {Barrett}},\ }\bibfield
  {title} {\enquote {\bibinfo {title} {{Controlling Coherence Using the
  Internal Structure of Hard $\pi$ Pulses}},}\ }\href {\doibase
  10.1103/PhysRevLett.100.247601} {\bibfield  {journal} {\bibinfo  {journal}
  {Phys. Rev. Lett.}\ }\textbf {\bibinfo {volume} {100}},\ \bibinfo {pages}
  {247601} (\bibinfo {year} {2008})},\ \Eprint {http://arxiv.org/abs/0804.3087}
  {arXiv:0804.3087} \BibitemShut {NoStop}%
\bibitem [{\citenamefont {Geuzaine}\ and\ \citenamefont
  {Remacle}(2009)}]{Geuzaine:2009rot}%
  \BibitemOpen
  \bibfield  {author} {\bibinfo {author} {\bibfnamefont {Christophe}\
  \bibnamefont {Geuzaine}}\ and\ \bibinfo {author} {\bibfnamefont
  {Jean-Fran{\c{c}}ois}\ \bibnamefont {Remacle}},\ }\bibfield  {title}
  {\enquote {\bibinfo {title} {{Gmsh: A 3-D finite element mesh generator with
  built-in pre- and post-processing facilities}},}\ }\href {\doibase
  10.1002/nme.2579} {\bibfield  {journal} {\bibinfo  {journal} {Int. J. Numer.
  Meth. Eng.}\ }\textbf {\bibinfo {volume} {79}},\ \bibinfo {pages}
  {1309--1331} (\bibinfo {year} {2009})}\BibitemShut {NoStop}%
\bibitem [{\citenamefont {Ghysels}\ \emph {et~al.}(2016)\citenamefont
  {Ghysels}, \citenamefont {Li}, \citenamefont {Rouet}, \citenamefont
  {Williams},\ and\ \citenamefont {Napov}}]{Ghysels:2016strumpack}%
  \BibitemOpen
  \bibfield  {author} {\bibinfo {author} {\bibfnamefont {Pieter}\ \bibnamefont
  {Ghysels}}, \bibinfo {author} {\bibfnamefont {Xiaoye~S.}\ \bibnamefont {Li}},
  \bibinfo {author} {\bibfnamefont {Fran{\c{c}}ois-Henry}\ \bibnamefont
  {Rouet}}, \bibinfo {author} {\bibfnamefont {Samuel}\ \bibnamefont
  {Williams}}, \ and\ \bibinfo {author} {\bibfnamefont {Artem}\ \bibnamefont
  {Napov}},\ }\bibfield  {title} {\enquote {\bibinfo {title} {{An Efficient
  Multicore Implementation of a Novel HSS-Structured Multifrontal Solver Using
  Randomized Sampling}},}\ }\href {\doibase 10.1137/15M1010117} {\bibfield
  {journal} {\bibinfo  {journal} {SIAM J. Sci. Comput.}\ }\textbf {\bibinfo
  {volume} {38}},\ \bibinfo {pages} {S358--S384} (\bibinfo {year}
  {2016})}\BibitemShut {NoStop}%
\bibitem [{\citenamefont {N{\'e}d{\'e}lec}(1980)}]{Nedelec:1980mixed}%
  \BibitemOpen
  \bibfield  {author} {\bibinfo {author} {\bibfnamefont {J.~C.}\ \bibnamefont
  {N{\'e}d{\'e}lec}},\ }\bibfield  {title} {\enquote {\bibinfo {title} {{Mixed
  finite elements in $\mathbb{R}^3$}},}\ }\href {\doibase 10.1007/BF01396415}
  {\bibfield  {journal} {\bibinfo  {journal} {Numer. Math.}\ }\textbf {\bibinfo
  {volume} {35}},\ \bibinfo {pages} {315--341} (\bibinfo {year}
  {1980})}\BibitemShut {NoStop}%
\bibitem [{\citenamefont {Hestenes}\ and\ \citenamefont
  {Stiefel}(1952)}]{Hestenes:1952cg}%
  \BibitemOpen
  \bibfield  {author} {\bibinfo {author} {\bibfnamefont {Magnus~R.}\
  \bibnamefont {Hestenes}}\ and\ \bibinfo {author} {\bibfnamefont {Eduard}\
  \bibnamefont {Stiefel}},\ }\bibfield  {title} {\enquote {\bibinfo {title}
  {{Methods of conjugate gradients for solving linear systems}},}\ }\href
  {\doibase 10.6028/jres.049.044} {\bibfield  {journal} {\bibinfo  {journal}
  {J. Res. Natl. Bur. Stand.}\ }\textbf {\bibinfo {volume} {49}},\ \bibinfo
  {pages} {409--436} (\bibinfo {year} {1952})}\BibitemShut {NoStop}%
\bibitem [{\citenamefont {Kolev}\ and\ \citenamefont
  {Vassilevski}(2009)}]{Kolev:2009ams}%
  \BibitemOpen
  \bibfield  {author} {\bibinfo {author} {\bibfnamefont {Tzanio~V.}\
  \bibnamefont {Kolev}}\ and\ \bibinfo {author} {\bibfnamefont {Panayot~S.}\
  \bibnamefont {Vassilevski}},\ }\bibfield  {title} {\enquote {\bibinfo {title}
  {{Parallel auxiliary space AMG for H(curl) problems}},}\ }\href {\doibase
  10.4208/jcm.2009.27.5.013} {\bibfield  {journal} {\bibinfo  {journal} {J.
  Comput. Math.}\ }\textbf {\bibinfo {volume} {27}},\ \bibinfo {pages}
  {604--623} (\bibinfo {year} {2009})}\BibitemShut {NoStop}%
\bibitem [{\citenamefont {Hiptmair}\ and\ \citenamefont
  {Xu}(2007)}]{Hiptmair:2007aux}%
  \BibitemOpen
  \bibfield  {author} {\bibinfo {author} {\bibfnamefont {Ralf}\ \bibnamefont
  {Hiptmair}}\ and\ \bibinfo {author} {\bibfnamefont {Jinchao}\ \bibnamefont
  {Xu}},\ }\bibfield  {title} {\enquote {\bibinfo {title} {{Nodal Auxiliary
  Space Preconditioning in H(curl) and H(div) Spaces}},}\ }\href {\doibase
  10.1137/060660588} {\bibfield  {journal} {\bibinfo  {journal} {SIAM J. Numer.
  Anal.}\ }\textbf {\bibinfo {volume} {45}},\ \bibinfo {pages} {2483--2509}
  (\bibinfo {year} {2007})}\BibitemShut {NoStop}%
\bibitem [{\citenamefont {Falgout}\ and\ \citenamefont
  {Yang}(2002)}]{Falgout:2002hypre}%
  \BibitemOpen
  \bibfield  {author} {\bibinfo {author} {\bibfnamefont {Robert~D.}\
  \bibnamefont {Falgout}}\ and\ \bibinfo {author} {\bibfnamefont
  {Ulrike~Meier}\ \bibnamefont {Yang}},\ }\bibfield  {title} {\enquote
  {\bibinfo {title} {{hypre: A Library of High Performance Preconditioners}},}\
  }in\ \href {\doibase 10.1007/3-540-47789-6_66} {\emph {\bibinfo {booktitle}
  {{Computational Science -- ICCS 2002}}}},\ \bibinfo {series} {Lecture Notes
  in Computer Science}, Vol.\ \bibinfo {volume} {2331}\ (\bibinfo  {publisher}
  {Springer},\ \bibinfo {address} {Berlin, Heidelberg},\ \bibinfo {year}
  {2002})\ pp.\ \bibinfo {pages} {632--641}\BibitemShut {NoStop}%
\bibitem [{\citenamefont {Saad}\ and\ \citenamefont
  {Schultz}(1986)}]{Saad:1986gmres}%
  \BibitemOpen
  \bibfield  {author} {\bibinfo {author} {\bibfnamefont {Youcef}\ \bibnamefont
  {Saad}}\ and\ \bibinfo {author} {\bibfnamefont {Martin~H.}\ \bibnamefont
  {Schultz}},\ }\bibfield  {title} {\enquote {\bibinfo {title} {{GMRES: A
  Generalized Minimal Residual Algorithm for Solving Nonsymmetric Linear
  Systems}},}\ }\href {\doibase 10.1137/0907058} {\bibfield  {journal}
  {\bibinfo  {journal} {SIAM J. Sci. Stat. Comput.}\ }\textbf {\bibinfo
  {volume} {7}},\ \bibinfo {pages} {856--869} (\bibinfo {year}
  {1986})}\BibitemShut {NoStop}%
\bibitem [{\citenamefont {Armijo}(1966)}]{Armijo:1966ls}%
  \BibitemOpen
  \bibfield  {author} {\bibinfo {author} {\bibfnamefont {Larry}\ \bibnamefont
  {Armijo}},\ }\bibfield  {title} {\enquote {\bibinfo {title} {{Minimization of
  functions having Lipschitz continuous first partial derivatives}},}\ }\href
  {\doibase 10.2140/pjm.1966.16.1} {\bibfield  {journal} {\bibinfo  {journal}
  {Pacific J. Math.}\ }\textbf {\bibinfo {volume} {16}},\ \bibinfo {pages}
  {1--3} (\bibinfo {year} {1966})}\BibitemShut {NoStop}%
\bibitem [{\citenamefont {Keller}\ and\ \citenamefont
  {Givoli}(1989)}]{Keller:1989dtn}%
  \BibitemOpen
  \bibfield  {author} {\bibinfo {author} {\bibfnamefont {Joseph~B.}\
  \bibnamefont {Keller}}\ and\ \bibinfo {author} {\bibfnamefont {Dan}\
  \bibnamefont {Givoli}},\ }\bibfield  {title} {\enquote {\bibinfo {title}
  {{Exact non-reflecting boundary conditions}},}\ }\href {\doibase
  10.1016/0021-9991(89)90041-7} {\bibfield  {journal} {\bibinfo  {journal} {J.
  Comput. Phys.}\ }\textbf {\bibinfo {volume} {82}},\ \bibinfo {pages}
  {172--192} (\bibinfo {year} {1989})}\BibitemShut {NoStop}%
\bibitem [{\citenamefont {Ho}\ \emph {et~al.}(1975)\citenamefont {Ho},
  \citenamefont {Ruehli},\ and\ \citenamefont {Brennan}}]{Ho:1975mna}%
  \BibitemOpen
  \bibfield  {author} {\bibinfo {author} {\bibfnamefont {Chung-Wen}\
  \bibnamefont {Ho}}, \bibinfo {author} {\bibfnamefont {Albert~E.}\
  \bibnamefont {Ruehli}}, \ and\ \bibinfo {author} {\bibfnamefont {Pierce~A.}\
  \bibnamefont {Brennan}},\ }\bibfield  {title} {\enquote {\bibinfo {title}
  {{The modified nodal approach to network analysis}},}\ }\href {\doibase
  10.1109/TCS.1975.1084079} {\bibfield  {journal} {\bibinfo  {journal} {IEEE
  Trans. Circuits Syst.}\ }\textbf {\bibinfo {volume} {22}},\ \bibinfo {pages}
  {504--509} (\bibinfo {year} {1975})}\BibitemShut {NoStop}%
\bibitem [{\citenamefont {Kim}\ \emph {et~al.}(2020)\citenamefont {Kim},
  \citenamefont {Jeong}, \citenamefont {Youn}, \citenamefont {Kim},\ and\
  \citenamefont {Semertzidis}}]{Kim:2020kfo}%
  \BibitemOpen
  \bibfield  {author} {\bibinfo {author} {\bibfnamefont {Dongok}\ \bibnamefont
  {Kim}}, \bibinfo {author} {\bibfnamefont {Junu}\ \bibnamefont {Jeong}},
  \bibinfo {author} {\bibfnamefont {SungWoo}\ \bibnamefont {Youn}}, \bibinfo
  {author} {\bibfnamefont {Younggeun}\ \bibnamefont {Kim}}, \ and\ \bibinfo
  {author} {\bibfnamefont {Yannis~K.}\ \bibnamefont {Semertzidis}},\ }\bibfield
   {title} {\enquote {\bibinfo {title} {{Revisiting the detection rate for
  axion haloscopes}},}\ }\href {\doibase 10.1088/1475-7516/2020/03/066}
  {\bibfield  {journal} {\bibinfo  {journal} {JCAP}\ }\textbf {\bibinfo
  {volume} {03}},\ \bibinfo {pages} {066} (\bibinfo {year} {2020})},\ \Eprint
  {http://arxiv.org/abs/2001.05605} {arXiv:2001.05605 [hep-ex]} \BibitemShut
  {NoStop}%
\bibitem [{\citenamefont {Al~Kenany}\ \emph {et~al.}(2017)\citenamefont
  {Al~Kenany} \emph {et~al.}}]{AlKenany:2016trt}%
  \BibitemOpen
  \bibfield  {author} {\bibinfo {author} {\bibfnamefont {S.}~\bibnamefont
  {Al~Kenany}} \emph {et~al.},\ }\bibfield  {title} {\enquote {\bibinfo {title}
  {{Design and operational experience of a microwave cavity axion detector for
  the 20{\textendash}100{\ensuremath{\mu}}eV range}},}\ }\href {\doibase
  10.1016/j.nima.2017.02.012} {\bibfield  {journal} {\bibinfo  {journal} {Nucl.
  Instrum. Meth. A}\ }\textbf {\bibinfo {volume} {854}},\ \bibinfo {pages}
  {11--24} (\bibinfo {year} {2017})},\ \Eprint
  {http://arxiv.org/abs/1611.07123} {arXiv:1611.07123 [physics.ins-det]}
  \BibitemShut {NoStop}%
\bibitem [{\citenamefont {Brubaker}\ \emph {et~al.}(2017)\citenamefont
  {Brubaker}, \citenamefont {Zhong}, \citenamefont {Lamoreaux}, \citenamefont
  {Lehnert},\ and\ \citenamefont {van Bibber}}]{Brubaker:2017rna}%
  \BibitemOpen
  \bibfield  {author} {\bibinfo {author} {\bibfnamefont {B.~M.}\ \bibnamefont
  {Brubaker}}, \bibinfo {author} {\bibfnamefont {L.}~\bibnamefont {Zhong}},
  \bibinfo {author} {\bibfnamefont {S.~K.}\ \bibnamefont {Lamoreaux}}, \bibinfo
  {author} {\bibfnamefont {K.~W.}\ \bibnamefont {Lehnert}}, \ and\ \bibinfo
  {author} {\bibfnamefont {K.~A.}\ \bibnamefont {van Bibber}},\ }\bibfield
  {title} {\enquote {\bibinfo {title} {{HAYSTAC axion search analysis
  procedure}},}\ }\href {\doibase 10.1103/PhysRevD.96.123008} {\bibfield
  {journal} {\bibinfo  {journal} {Phys. Rev. D}\ }\textbf {\bibinfo {volume}
  {96}},\ \bibinfo {pages} {123008} (\bibinfo {year} {2017})},\ \Eprint
  {http://arxiv.org/abs/1706.08388} {arXiv:1706.08388 [astro-ph.IM]}
  \BibitemShut {NoStop}%
\bibitem [{\citenamefont {Ericsson}\ and\ \citenamefont
  {Ruhe}(1980)}]{Ericsson:1980lanczos}%
  \BibitemOpen
  \bibfield  {author} {\bibinfo {author} {\bibfnamefont {Thomas}\ \bibnamefont
  {Ericsson}}\ and\ \bibinfo {author} {\bibfnamefont {Axel}\ \bibnamefont
  {Ruhe}},\ }\bibfield  {title} {\enquote {\bibinfo {title} {{The spectral
  transformation Lanczos method for the numerical solution of large sparse
  generalized symmetric eigenvalue problems}},}\ }\href {\doibase
  10.1090/S0025-5718-1980-0583502-2} {\bibfield  {journal} {\bibinfo  {journal}
  {Math. Comp.}\ }\textbf {\bibinfo {volume} {35}},\ \bibinfo {pages}
  {1251--1268} (\bibinfo {year} {1980})}\BibitemShut {NoStop}%
\bibitem [{\citenamefont {Reuter}\ and\ \citenamefont
  {Sondheimer}(1948)}]{Reuter:1948ase}%
  \BibitemOpen
  \bibfield  {author} {\bibinfo {author} {\bibfnamefont {G.~E.~H.}\
  \bibnamefont {Reuter}}\ and\ \bibinfo {author} {\bibfnamefont {E.~H.}\
  \bibnamefont {Sondheimer}},\ }\bibfield  {title} {\enquote {\bibinfo {title}
  {{The theory of the anomalous skin effect in metals}},}\ }\href {\doibase
  10.1098/rspa.1948.0123} {\bibfield  {journal} {\bibinfo  {journal} {Proc. R.
  Soc. Lond. A}\ }\textbf {\bibinfo {volume} {195}},\ \bibinfo {pages}
  {336--364} (\bibinfo {year} {1948})}\BibitemShut {NoStop}%
\bibitem [{\citenamefont {Mattis}\ and\ \citenamefont
  {Bardeen}(1958)}]{Mattis:1958yid}%
  \BibitemOpen
  \bibfield  {author} {\bibinfo {author} {\bibfnamefont {D.~C.}\ \bibnamefont
  {Mattis}}\ and\ \bibinfo {author} {\bibfnamefont {J.}~\bibnamefont
  {Bardeen}},\ }\bibfield  {title} {\enquote {\bibinfo {title} {{Theory of the
  Anomalous Skin Effect in Normal and Superconducting Metals}},}\ }\href
  {\doibase 10.1103/PhysRev.111.412} {\bibfield  {journal} {\bibinfo  {journal}
  {Phys. Rev.}\ }\textbf {\bibinfo {volume} {111}},\ \bibinfo {pages}
  {412--417} (\bibinfo {year} {1958})}\BibitemShut {NoStop}%
\bibitem [{\citenamefont {Halbritter}(1970)}]{Halbritter:1970rf}%
  \BibitemOpen
  \bibfield  {author} {\bibinfo {author} {\bibfnamefont {J.}~\bibnamefont
  {Halbritter}},\ }\bibfield  {title} {\enquote {\bibinfo {title} {{Comparison
  between measured and calculated RF losses in the superconducting state}},}\
  }\href {\doibase 10.1007/BF01409429} {\bibfield  {journal} {\bibinfo
  {journal} {Z. Phys.}\ }\textbf {\bibinfo {volume} {238}},\ \bibinfo {pages}
  {466--476} (\bibinfo {year} {1970})}\BibitemShut {NoStop}%
\bibitem [{\citenamefont {Braine}(2024)}]{Braine:2024nzi}%
  \BibitemOpen
  \bibfield  {author} {\bibinfo {author} {\bibfnamefont {Thomas}\ \bibnamefont
  {Braine}},\ }\emph {\bibinfo {title} {{Superconducting Resonator Development
  for the Axion Dark Matter eXperiment}}},\ \href@noop {} {Ph.D. thesis},\
  \bibinfo  {school} {Washington U., Seattle} (\bibinfo {year} {2024}),\
  \Eprint {http://arxiv.org/abs/2408.03444} {arXiv:2408.03444 [hep-ex]}
  \BibitemShut {NoStop}%
\bibitem [{\citenamefont {Alesini}\ \emph {et~al.}(2017)\citenamefont
  {Alesini}, \citenamefont {Babusci}, \citenamefont {Di~Gioacchino},
  \citenamefont {Gatti}, \citenamefont {Lamanna},\ and\ \citenamefont
  {Ligi}}]{Alesini:2017ifp}%
  \BibitemOpen
  \bibfield  {author} {\bibinfo {author} {\bibfnamefont {David}\ \bibnamefont
  {Alesini}}, \bibinfo {author} {\bibfnamefont {Danilo}\ \bibnamefont
  {Babusci}}, \bibinfo {author} {\bibfnamefont {Daniele}\ \bibnamefont
  {Di~Gioacchino}}, \bibinfo {author} {\bibfnamefont {Claudio}\ \bibnamefont
  {Gatti}}, \bibinfo {author} {\bibfnamefont {Gianluca}\ \bibnamefont
  {Lamanna}}, \ and\ \bibinfo {author} {\bibfnamefont {Carlo}\ \bibnamefont
  {Ligi}},\ }\bibfield  {title} {\enquote {\bibinfo {title} {{The KLASH
  Proposal}},}\ }\href@noop {} {\  (\bibinfo {year} {2017})},\ \Eprint
  {http://arxiv.org/abs/1707.06010} {arXiv:1707.06010 [physics.ins-det]}
  \BibitemShut {NoStop}%
\bibitem [{\citenamefont {Alesini}\ \emph {et~al.}(2021)\citenamefont {Alesini}
  \emph {et~al.}}]{QUAX:2020uxy}%
  \BibitemOpen
  \bibfield  {author} {\bibinfo {author} {\bibfnamefont {D.}~\bibnamefont
  {Alesini}} \emph {et~al.} (\bibinfo {collaboration} {QUAX}),\ }\bibfield
  {title} {\enquote {\bibinfo {title} {{Realization of a high quality factor
  resonator with hollow dielectric cylinders for axion searches}},}\ }\href
  {\doibase 10.1016/j.nima.2020.164641} {\bibfield  {journal} {\bibinfo
  {journal} {Nucl. Instrum. Meth. A}\ }\textbf {\bibinfo {volume} {985}},\
  \bibinfo {pages} {164641} (\bibinfo {year} {2021})},\ \Eprint
  {http://arxiv.org/abs/2004.02754} {arXiv:2004.02754 [physics.ins-det]}
  \BibitemShut {NoStop}%
\bibitem [{\citenamefont {Di~Vora}\ \emph {et~al.}(2022)\citenamefont {Di~Vora}
  \emph {et~al.}}]{DiVora:2022tro}%
  \BibitemOpen
  \bibfield  {author} {\bibinfo {author} {\bibfnamefont {R.}~\bibnamefont
  {Di~Vora}} \emph {et~al.},\ }\bibfield  {title} {\enquote {\bibinfo {title}
  {{High-Q Microwave Dielectric Resonator for Axion Dark-Matter Haloscopes}},}\
  }\href {\doibase 10.1103/PhysRevApplied.17.054013} {\bibfield  {journal}
  {\bibinfo  {journal} {Phys. Rev. Applied}\ }\textbf {\bibinfo {volume}
  {17}},\ \bibinfo {pages} {054013} (\bibinfo {year} {2022})},\ \Eprint
  {http://arxiv.org/abs/2201.04223} {arXiv:2201.04223 [physics.ins-det]}
  \BibitemShut {NoStop}%
\bibitem [{\citenamefont {Stern}\ \emph {et~al.}(2015)\citenamefont {Stern},
  \citenamefont {Chisholm}, \citenamefont {Hoskins}, \citenamefont {Sikivie},
  \citenamefont {Sullivan}, \citenamefont {Tanner}, \citenamefont {Carosi},\
  and\ \citenamefont {van Bibber}}]{Stern:2015kzo}%
  \BibitemOpen
  \bibfield  {author} {\bibinfo {author} {\bibfnamefont {I.}~\bibnamefont
  {Stern}}, \bibinfo {author} {\bibfnamefont {A.~A.}\ \bibnamefont {Chisholm}},
  \bibinfo {author} {\bibfnamefont {J.}~\bibnamefont {Hoskins}}, \bibinfo
  {author} {\bibfnamefont {P.}~\bibnamefont {Sikivie}}, \bibinfo {author}
  {\bibfnamefont {N.~S.}\ \bibnamefont {Sullivan}}, \bibinfo {author}
  {\bibfnamefont {D.~B.}\ \bibnamefont {Tanner}}, \bibinfo {author}
  {\bibfnamefont {G.}~\bibnamefont {Carosi}}, \ and\ \bibinfo {author}
  {\bibfnamefont {K.}~\bibnamefont {van Bibber}},\ }\bibfield  {title}
  {\enquote {\bibinfo {title} {{Cavity design for high-frequency axion dark
  matter detectors}},}\ }\href {\doibase 10.1063/1.4938164} {\bibfield
  {journal} {\bibinfo  {journal} {Rev. Sci. Instrum.}\ }\textbf {\bibinfo
  {volume} {86}},\ \bibinfo {pages} {123305} (\bibinfo {year} {2015})},\
  \Eprint {http://arxiv.org/abs/1603.06990} {arXiv:1603.06990
  [physics.ins-det]} \BibitemShut {NoStop}%
\bibitem [{\citenamefont {Jeong}\ \emph {et~al.}(2018)\citenamefont {Jeong},
  \citenamefont {Youn}, \citenamefont {Ahn}, \citenamefont {Kim},\ and\
  \citenamefont {Semertzidis}}]{Jeong:2017hqs}%
  \BibitemOpen
  \bibfield  {author} {\bibinfo {author} {\bibfnamefont {Junu}\ \bibnamefont
  {Jeong}}, \bibinfo {author} {\bibfnamefont {SungWoo}\ \bibnamefont {Youn}},
  \bibinfo {author} {\bibfnamefont {Saebyeok}\ \bibnamefont {Ahn}}, \bibinfo
  {author} {\bibfnamefont {Jihn~E.}\ \bibnamefont {Kim}}, \ and\ \bibinfo
  {author} {\bibfnamefont {Yannis~K.}\ \bibnamefont {Semertzidis}},\ }\bibfield
   {title} {\enquote {\bibinfo {title} {{Concept of multiple-cell cavity for
  axion dark matter search}},}\ }\href {\doibase
  10.1016/j.physletb.2017.12.066} {\bibfield  {journal} {\bibinfo  {journal}
  {Phys. Lett. B}\ }\textbf {\bibinfo {volume} {777}},\ \bibinfo {pages}
  {412--419} (\bibinfo {year} {2018})},\ \Eprint
  {http://arxiv.org/abs/1710.06969} {arXiv:1710.06969 [astro-ph.IM]}
  \BibitemShut {NoStop}%
\bibitem [{\citenamefont {Aune}\ \emph {et~al.}(2000)\citenamefont {Aune} \emph
  {et~al.}}]{Aune:2000gb}%
  \BibitemOpen
  \bibfield  {author} {\bibinfo {author} {\bibfnamefont {B.}~\bibnamefont
  {Aune}} \emph {et~al.},\ }\bibfield  {title} {\enquote {\bibinfo {title}
  {{The superconducting TESLA cavities}},}\ }\href {\doibase
  10.1103/PhysRevSTAB.3.092001} {\bibfield  {journal} {\bibinfo  {journal}
  {Phys. Rev. ST Accel. Beams}\ }\textbf {\bibinfo {volume} {3}},\ \bibinfo
  {pages} {092001} (\bibinfo {year} {2000})},\ \Eprint
  {http://arxiv.org/abs/physics/0003011} {arXiv:physics/0003011} \BibitemShut
  {NoStop}%
\bibitem [{\citenamefont {Woltjer}(1958)}]{Woltjer:1958}%
  \BibitemOpen
  \bibfield  {author} {\bibinfo {author} {\bibfnamefont {L.}~\bibnamefont
  {Woltjer}},\ }\bibfield  {title} {\enquote {\bibinfo {title} {{A Theorem on
  Force-Free Magnetic Fields}},}\ }\href {\doibase 10.1073/pnas.44.6.489}
  {\bibfield  {journal} {\bibinfo  {journal} {Proc. Nat. Acad. Sci.}\ }\textbf
  {\bibinfo {volume} {44}},\ \bibinfo {pages} {489--491} (\bibinfo {year}
  {1958})}\BibitemShut {NoStop}%
\bibitem [{\citenamefont {Taylor}(1974)}]{Taylor:1974}%
  \BibitemOpen
  \bibfield  {author} {\bibinfo {author} {\bibfnamefont {J.~B.}\ \bibnamefont
  {Taylor}},\ }\bibfield  {title} {\enquote {\bibinfo {title} {{Relaxation of
  Toroidal Plasma and Generation of Reverse Magnetic Fields}},}\ }\href
  {\doibase 10.1103/PhysRevLett.33.1139} {\bibfield  {journal} {\bibinfo
  {journal} {Phys. Rev. Lett.}\ }\textbf {\bibinfo {volume} {33}},\ \bibinfo
  {pages} {1139--1141} (\bibinfo {year} {1974})}\BibitemShut {NoStop}%
\bibitem [{\citenamefont {Rosenbluth}\ and\ \citenamefont
  {Bussac}(1979)}]{Rosenbluth:1979spheromak}%
  \BibitemOpen
  \bibfield  {author} {\bibinfo {author} {\bibfnamefont {M.~N.}\ \bibnamefont
  {Rosenbluth}}\ and\ \bibinfo {author} {\bibfnamefont {M.~A.}\ \bibnamefont
  {Bussac}},\ }\bibfield  {title} {\enquote {\bibinfo {title} {{MHD stability
  of Spheromak}},}\ }\href {\doibase 10.1088/0029-5515/19/4/007} {\bibfield
  {journal} {\bibinfo  {journal} {Nucl. Fusion}\ }\textbf {\bibinfo {volume}
  {19}},\ \bibinfo {pages} {489--498} (\bibinfo {year} {1979})}\BibitemShut
  {NoStop}%
\bibitem [{\citenamefont {Taylor}(1986)}]{Taylor:1986rmp}%
  \BibitemOpen
  \bibfield  {author} {\bibinfo {author} {\bibfnamefont {J.~B.}\ \bibnamefont
  {Taylor}},\ }\bibfield  {title} {\enquote {\bibinfo {title} {{Relaxation and
  magnetic reconnection in plasmas}},}\ }\href {\doibase
  10.1103/RevModPhys.58.741} {\bibfield  {journal} {\bibinfo  {journal} {Rev.
  Mod. Phys.}\ }\textbf {\bibinfo {volume} {58}},\ \bibinfo {pages} {741--763}
  (\bibinfo {year} {1986})}\BibitemShut {NoStop}%
\bibitem [{\citenamefont {Finn}\ \emph {et~al.}(1981)\citenamefont {Finn},
  \citenamefont {Manheimer},\ and\ \citenamefont {Ott}}]{Finn:1981tilt}%
  \BibitemOpen
  \bibfield  {author} {\bibinfo {author} {\bibfnamefont {J.~M.}\ \bibnamefont
  {Finn}}, \bibinfo {author} {\bibfnamefont {W.~M.}\ \bibnamefont {Manheimer}},
  \ and\ \bibinfo {author} {\bibfnamefont {E.}~\bibnamefont {Ott}},\ }\bibfield
   {title} {\enquote {\bibinfo {title} {{Spheromak tilting instability in
  cylindrical geometry}},}\ }\href {\doibase 10.1063/1.863536} {\bibfield
  {journal} {\bibinfo  {journal} {Phys. Fluids}\ }\textbf {\bibinfo {volume}
  {24}},\ \bibinfo {pages} {1336--1341} (\bibinfo {year} {1981})}\BibitemShut
  {NoStop}%
\bibitem [{\citenamefont {Bondeson}\ \emph {et~al.}(1981)\citenamefont
  {Bondeson}, \citenamefont {Marklin}, \citenamefont {An}, \citenamefont
  {Chen}, \citenamefont {Lee},\ and\ \citenamefont {Liu}}]{Bondeson:1981tilt}%
  \BibitemOpen
  \bibfield  {author} {\bibinfo {author} {\bibfnamefont {A.}~\bibnamefont
  {Bondeson}}, \bibinfo {author} {\bibfnamefont {G.}~\bibnamefont {Marklin}},
  \bibinfo {author} {\bibfnamefont {Z.~G.}\ \bibnamefont {An}}, \bibinfo
  {author} {\bibfnamefont {H.~H.}\ \bibnamefont {Chen}}, \bibinfo {author}
  {\bibfnamefont {Y.~C.}\ \bibnamefont {Lee}}, \ and\ \bibinfo {author}
  {\bibfnamefont {C.~S.}\ \bibnamefont {Liu}},\ }\bibfield  {title} {\enquote
  {\bibinfo {title} {{Tilting instability of a cylindrical spheromak}},}\
  }\href {\doibase 10.1063/1.863579} {\bibfield  {journal} {\bibinfo  {journal}
  {Phys. Fluids}\ }\textbf {\bibinfo {volume} {24}},\ \bibinfo {pages}
  {1682--1688} (\bibinfo {year} {1981})}\BibitemShut {NoStop}%
\end{thebibliography}%

\appendix
\section{The \texttt{axionMultiPhysics} code}\label{app:axionmultiphysics}

\begin{figure*}[t]
    \centering
    \includegraphics[width=\textwidth]{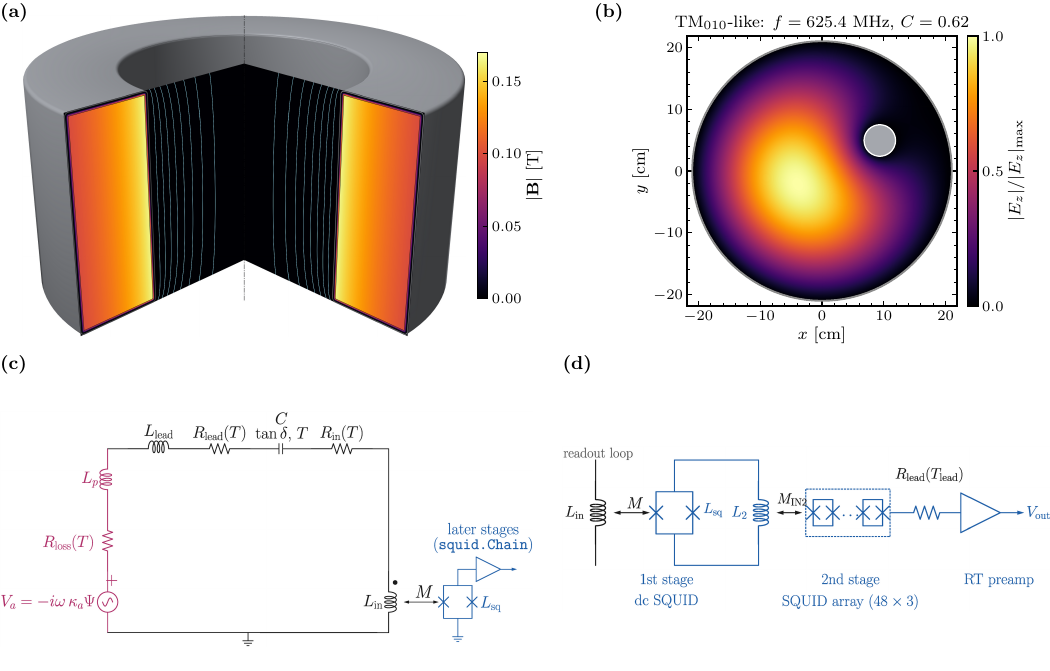}
    \caption{Overview of \texttt{axionMultiPhysics}.
(a)~Axisymmetric toroid (outer radius and height of $2\,\mathrm{m}$), with a $90^\circ$ cutaway using the built-in renderer.
The static magnetic field strength $|\bm{B}|$ is indicated by the colorbar.
Field lines in the central hole trace the pickup field $\bm B_I$, produced by a current driven through the readout port.
(b)~Lowest $\mathrm{TM}_{010}$-like mode of an ADMX-like copper cavity with an off-center tuning rod, computed using the 2D cross-section eigensolver.
The colorbar shows normalized $|E_z|$; $f=625.4\,\mathrm{MHz}$ and $C=0.62$ for uniform axial $B_{\rm lab}$.
(c)~Effective series readout circuit, including pickup loss at temperature $T$, the axion-induced EMF $V_a$ ($|V_a|=m_a\Phi_a$), lead impedances, capacitor loss, input-coil resistance $R_{\rm in}$, and SQUID input coupling, imprecision, backaction, and their correlation.
(d)~The dc SQUID chain, drawn by the \texttt{axionmultiphysics.\allowbreak squid} module: $L_{\rm in}$ coupled via $M$ to the first-stage dc SQUID, $L_2$ coupled via $M_2$ to the second-stage array (48 SQUIDs in series per bank, three parallel banks), lead resistance, and a room-temperature preamplifier with output $V_{\rm out}$~\cite{Ankel:2025nld}.
}
    \label{fig:axionmultiphysics_overview}
\end{figure*}

In this appendix we provide more details on \texttt{axionMultiPhysics}~\href{https://github.com/bsafdi/axionMultiPhysics}{\faGithub}, the finite-element-method (FEM) code used to obtain most of the numerical results in this paper.
The primary purpose of this package is to perform realistic modeling, including electromagnetic, mechanical, and thermal responses, of lumped-element axion detectors, with the goal of computing the sensitivity of the detector in the $(m_a,g_{a\gamma\gamma})$ plane.
Detector geometries are parametrized such that the configuration of minimal cost achieving a given sensitivity can be determined via a scan.
The package can also model resonant-cavity haloscopes (Sec.~\ref{app:axionmultiphysics_cavity}) and superconducting radio-frequency (SRF) cavities (Sec.~\ref{sec:heterodyne}).
Fig.~\ref{fig:axionmultiphysics_overview} gives an overview of key \texttt{axionMultiPhysics} capabilities.
The package also reproduces some capabilities of COMSOL Multiphysics that go beyond calculations relevant for axion detectors, including computational fluid dynamics and chemistry.
We do not discuss these here. The primary modules of \texttt{axionMultiPhysics} are listed in Tab.~\ref{tab:axionmultiphysics_modules}.

The package combines a C++ core built on the \texttt{MFEM} (Modular Finite Element Methods) library~\cite{mfem}, a Python interface, and an interactive viewer.
In addition, \texttt{axionMultiPhysics} interfaces with COMSOL~\cite{multiphysics1998introduction} and includes a Model Context Protocol (MCP) server, through which the detector design and analysis can be performed via natural-language requests to AI assistants.
The \href{https://github.com/bsafdi/axionMultiPhysics}{GitHub repository} and \href{https://bsafdi.github.io/axionMultiPhysics/}{documentation website} provide installation instructions and tutorials.

\subsection{Detector setup}

To compute the static field and pickup response of a given lumped-element detector, the code needs the detector geometry, materials, and currents.
Designs can be supplied in JSON or Python, with templates for the toroid, solenoid, and core detectors discussed in this work.
\texttt{axionMultiPhysics} allows both axisymmetric designs, in which case detector components are regions occupying the meridional plane $(r\geq0,z)$, and 3D designs where the components are defined via solids to which Boolean operations ({\it e.g.}, unions and intersections) and rigid transforms can be applied.
In the latter case, designs can also be imported from a CAD model in STEP format.
Axisymmetric designs can be turned into full 3D models with a single function call.

Built-in materials include vacuum, stranded windings, ideal Meissner superconductors, and normal conductors with radio-frequency (RF) surface resistance.
Soft iron is also included, and used for the core design, with magnetic field
\es{eq:axionmultiphysics_knee}{
H(B)&=\begin{cases}
B/\mu_i, & B\leq B_s,\\
B_s/\mu_i+B-B_s, & B>B_s,
\end{cases}\\
\mu_i&=1000,\qquad B_s=2\,{\rm T},
}
with $\mu_i$ the unsaturated relative permeability.
It is also possible to define a custom material by specifying its electromagnetic properties ($\mu_r$, $\epsilon_r$, dielectric loss tangent, conductivity, tabulated $B$--$H$ curves) and for the thermal and mechanical calculations, its elastic constants, thermal contraction, and temperature-dependent thermal conductivity.

To define currents, the user can set a value for the current in each winding, either manually or through an external circuit.
The code can also check how close the superconductor is to its critical current density $J_c(B,T)$ at the local field and temperature.
For the readout port driving the pickup, we prescribe the current across the port gap; in 3D, the code spreads it over the gap by solving a Laplace problem for a uniform fictitious conductivity, $\bm J=-\sigma\nabla V$.
Lastly, one can also impose a static magnetic field, either uniform or spatially varying according to a tabulation.

\subsection{Electromagnetic solvers}

Given a lumped-element detector setup, \texttt{axionMultiPhysics} computes two magnetic fields which enter in the scan rate.
The first is the static laboratory field $\bm B_{\rm lab}$ produced by the magnet windings, in the presence of any iron and superconductors.
The second is the field $\bm B_I=\nabla\times\bm A_I$ produced when a current $I$ is driven into the pickup through its readout port.
The energy stored in the pickup field determines $L_p$ through~\eqref{eq:Lp}, and the coupling $\Psi$ is its overlap with the laboratory field,~\eqref{eq:flux}.
In the MQS limit, these fields are computed from the vector potential $\bm B=\nabla\times\bm A$, which satisfies
\es{eq:axionmultiphysics_mqs}{
\nabla\times\big(\mu_r^{-1}\nabla\times\bm A\big)=\bm J\,,
}
where $\mu_r$ is the relative permeability (in iron $\mu_r=B/H(B)$ from~\eqref{eq:axionmultiphysics_knee}), and $\bm J$ is the winding or readout current density.

Beyond the MQS limit the static field is unchanged, while the pickup response  at the frequency $\omega$ is determined by the full Maxwell equations, giving the impedance $Z(\omega)$ and coupling $\Psi(\omega)$ seen at the readout port.
The electric field satisfies
\begin{equation}
\begin{gathered}
\nabla\times(\mu_r^{-1}\nabla\times\bm E)
-\omega^2\epsilon_c\bm E=i\omega\bm J,\\
\epsilon_c=\epsilon_r(1+i\tan\delta)+i\sigma/\omega , 
\end{gathered}
\label{eq:axionmultiphysics_fullwave}
\end{equation}
with relative permittivity $\epsilon_r$, dielectric loss tangent $\tan\delta$, and conductivity $\sigma$.
As in the main text, we use natural Heaviside--Lorentz units ($\mu_0=\epsilon_0=c=1$), so materials enter only through $\mu_r$, $\epsilon_r$, $\tan\delta$, and $\sigma$.
\texttt{axionMultiPhysics} solves~\eqref{eq:axionmultiphysics_mqs} and~\eqref{eq:axionmultiphysics_fullwave} variationally with FEM.
In the MQS case, for vacuum and linear magnetic materials, the vector potential that solves~\eqref{eq:axionmultiphysics_mqs} is the one that makes the functional
\es{eq:axionmultiphysics_functional}{
{\mathcal E[\bm A]=\int_\Omega\Big[\frac{1}{2 \mu_r}\,|\nabla\times\bm A|^2+\frac{\epsilon_g}{2}\,|\bm A|^2-\bm J\cdot\bm A\Big]\,dV}
}
stationary.
Here $\epsilon_g$ is a small gauge-fixing regulator added to ensure that $\bm A$ is unique.\footnote{In each element the code sets $\epsilon_g=10^{-12}\,\mu_r^{-1}/h^2$, with $h$ the element size.}
Stationarity of $\mathcal E$ gives
\es{eq:axionmultiphysics_weak_mqs}{
&\int_\Omega\left[\mu_r^{-1}(\nabla\times\bm A)\cdot(\nabla\times\bm A')+\epsilon_g\,\bm A\cdot\bm A'\right]dV=\int_\Omega\bm J\cdot\bm A'\,dV
}
for every test field $\bm A'$.

The variational problem is solved via the Galerkin method: $\bm A$ and $\bm A'$ are expanded in the same set of piecewise-polynomial functions on an unstructured mesh of triangles (tetrahedra) in the axisymmetric  (3D) case, generated with \texttt{gmsh}~\cite{Geuzaine:2009rot}, such that~\eqref{eq:axionmultiphysics_weak_mqs} becomes a sparse linear system for the expansion coefficients.
The mesh sizes are set per region and surface, or chosen automatically by repeatedly refining where the estimated error is largest until $\Psi$, $L_p$, and $W_{\rm mag}$ are converged.
For axisymmetric designs, the unknowns are scalars ($\psi=rA_\phi$, and $rH_\phi$ for currents flowing in the $r$--$z$ plane), expanded in polynomials.
The linear system is small enough to be solved directly, by sparse factorization (optionally with \texttt{STRUMPACK}~\cite{Ghysels:2016strumpack}).
In 3D the unknowns are the line integrals of $\bm A$ along the mesh edges (N\'ed\'elec edge elements~\cite{Nedelec:1980mixed}).
In this case the system is solved iteratively via the conjugate-gradient (CG) method~\cite{Hestenes:1952cg} using the auxiliary-space Maxwell solver (AMS)~\cite{Kolev:2009ams,Hiptmair:2007aux} of \texttt{hypre}~\cite{Falgout:2002hypre} as a preconditioner.

The full-wave equation~\eqref{eq:axionmultiphysics_fullwave} is treated in the same way, with the complex electric field in place of $\bm A$: it makes stationary $\int_\Omega[\tfrac12\mu_r^{-1}(\nabla\times\bm E)^2-\tfrac12\omega^2\epsilon_c\bm E^2-i\omega\bm J\cdot\bm E]\,dV$.
In axisymmetry the problem separates into two independent scalar problems, one for the azimuthal electric field (TE) and one for the azimuthal magnetic field (TM).
In 3D, the frequency-domain system is generally complex and is not positive definite.
We therefore use GMRES~\cite{Saad:1986gmres} in place of the CG method, again with AMS preconditioning.

Note that for designs containing iron, such as the core detector, the field-dependent permeability makes the MQS equation nonlinear.
In this case the quadratic magnetic energy density in~\eqref{eq:axionmultiphysics_functional} is replaced by $w(B)=\int_0^B H(b)\,db$, where $H(B)$ is given by~\eqref{eq:axionmultiphysics_knee}.
We find the stationary vector potential using Newton's method with an Armijo line search~\cite{Armijo:1966ls}, and compute the stored magnetic energy as $W_{\rm mag}=\int w(B)\,dV$.
For the pickup calculation, we approximate the iron as a linear material, fixing its permeability at each position to $\mu=B/H$ evaluated from the static solution.

Lastly, let us discuss boundary conditions for the variational problem.
In the MQS regime, ideal superconductors exclude magnetic flux from their interiors.
We impose this condition through constraints on the vector potential, together with fluxoid constraints for multiply connected superconducting bodies.
In the full-wave calculations, conducting surfaces are treated as perfect electric conductors (PEC), with \(\hat{\bm n}\times\bm E=0\); finite surface resistance is included perturbatively when calculating losses.
In 3D full-wave calculations, \texttt{axionMultiPhysics} can also solve on a reduced domain bounded by symmetry planes.
On a symmetry plane that the magnetic field crosses at right angles, leaving the boundary values free, the stationarity of the full-wave functional imposes $\hat{\bm n}\times\bm H=0$.
This is the condition of a perfect magnetic conductor (PMC).
On a plane that the field runs along, we instead impose the PEC condition.
For full-wave calculations in unshielded detectors, the mesh ends on a sphere which surrounds the detector, and we match the field there to a multipole expansion of the vacuum outside, through a Dirichlet-to-Neumann map~\cite{Keller:1989dtn} truncated at finite multipole order.

\begin{table*}[t]
\caption{Principal \texttt{axionMultiPhysics} modules, the physical quantities each computes, and the numerical method used.
JSON keys refer to the design file; Python names to the \texttt{axionmultiphysics} package.}
\label{tab:axionmultiphysics_modules}
\footnotesize
\begin{ruledtabular}
\begin{tabular}{@{}llll@{}}
\parbox[t]{2.6cm}{\raggedright Module\strut} & \parbox[t]{4.3cm}{\raggedright Physics\strut} & \parbox[t]{4.7cm}{\raggedright Method\strut} & \parbox[t]{3.4cm}{\raggedright Entry point\strut} \\
\hline
\parbox[t]{2.6cm}{\raggedright Axisymmetric MQS\strut} & \parbox[t]{4.3cm}{\raggedright Static magnetic field, stored energy $W_{\rm mag}$, pickup inductance $L_p$ and coupling $\Psi$ for rotationally symmetric detectors, including superconductors and saturating iron\strut} & \parbox[t]{4.7cm}{\raggedright 2D FEM for the azimuthal vector potential and the azimuthal magnetic field. 
\strut} & \parbox[t]{3.4cm}{\raggedright design file (default)\strut} \\
\hline
\parbox[t]{2.6cm}{\raggedright 3D MQS\strut} & \parbox[t]{4.3cm}{\raggedright Same, for geometries without rotational symmetry (discrete coils, ports, joints)\strut} & \parbox[t]{4.7cm}{\raggedright 3D FEM for the vector potential; iterative solver
\strut} & \parbox[t]{3.4cm}{\raggedright \texttt{"symmetry": "3d"}\strut} \\
\hline
\parbox[t]{2.6cm}{\raggedright Full-wave\strut} & \parbox[t]{4.3cm}{\raggedright Frequency-dependent impedance $Z(\omega)$ and coupling $\Psi(\omega)$ beyond the MQS limit; self-resonances, radiation, eigenmodes\strut} & \parbox[t]{4.7cm}{\raggedright Full Maxwell equations in the frequency domain (axisymmetric or 3D), with absorbing and lossy-wall boundaries\strut} & \parbox[t]{3.4cm}{\raggedright \texttt{fullwave}\strut} \\
\hline
\parbox[t]{2.6cm}{\raggedright Circuits, multiport\strut} & \parbox[t]{4.3cm}{\raggedright Signal and noise at the amplifier, with the pickup connected to tuning capacitors, transformers, and leads\strut} & \parbox[t]{4.7cm}{\raggedright Standard circuit-network solution (modified nodal analysis)\strut} & \parbox[t]{3.4cm}{\raggedright \texttt{axionmultiphysics.\allowbreak circuit}, \texttt{axionmultiphysics.\allowbreak multiport}\strut} \\
\hline
\parbox[t]{2.6cm}{\raggedright SQUID chains\strut} & \parbox[t]{4.3cm}{\raggedright Noise and input coupling of the dc SQUID amplifier chain\strut} & \parbox[t]{4.7cm}{\raggedright Model built from the two-stage chain proposed in Ref.~\cite{Ankel:2025nld}, based on measured device parameters\strut} & \parbox[t]{3.4cm}{\raggedright \texttt{axionmultiphysics.\allowbreak squid}\strut} \\
\hline
\parbox[t]{2.6cm}{\raggedright Scan\strut} & \parbox[t]{4.3cm}{\raggedright Scan rate and live time needed to reach a target coupling; the required figure of merit $\mathcal F_{\rm req}$,~\eqref{eq:Freq_req}\strut} & \parbox[t]{4.7cm}{\raggedright SNR integrated over the axion line and the detector bandwidth\strut} & \parbox[t]{3.4cm}{\raggedright \texttt{scan}, \texttt{axionmultiphysics.\allowbreak requirement}\strut} \\
\hline
\parbox[t]{2.6cm}{\raggedright Multi-resonance scan\strut} & \parbox[t]{4.3cm}{\raggedright Allocation of scan time across several tunable bands, including sensitivity away from resonance\strut} & \parbox[t]{4.7cm}{\raggedright Uses the full-wave impedance and coupling tabulated as a function of frequency\strut} & \parbox[t]{3.4cm}{\raggedright \texttt{multipole}\strut} \\
\hline
\parbox[t]{2.6cm}{\raggedright Mechanics, thermal\strut} & \parbox[t]{4.3cm}{\raggedright Magnetic forces and stresses, cooldown contraction, steady heat flow with temperature-dependent conductivity $k(T)$\strut} & \parbox[t]{4.7cm}{\raggedright FEM elasticity and heat conduction
\strut} & \parbox[t]{3.4cm}{\raggedright \texttt{mechanics}, \texttt{thermal}\strut} \\
\hline
\parbox[t]{2.6cm}{\raggedright Electrostatics\strut} & \parbox[t]{4.3cm}{\raggedright Capacitances between conductors, including the pickup gap\strut} & \parbox[t]{4.7cm}{\raggedright FEM electrostatics with open boundaries\strut} & \parbox[t]{3.4cm}{\raggedright \texttt{electrostatics}\strut} \\
\hline
\parbox[t]{2.6cm}{\raggedright High-temperature-superconductor (HTS) critical state\strut} & \parbox[t]{4.3cm}{\raggedright Screening currents, trapped flux, and losses in HTS windings\strut} & \parbox[t]{4.7cm}{\raggedright Time-dependent superconducting screening with a power-law \(E\)--\(J\) relation\strut} & \parbox[t]{3.4cm}{\raggedright \texttt{hts}\strut} \\
\hline
\parbox[t]{2.6cm}{\raggedright Shape optimization\strut} & \parbox[t]{4.3cm}{\raggedright Change of the figure of merit with each geometric parameter, for design optimization\strut} & \parbox[t]{4.7cm}{\raggedright Adjoint method: derivatives with respect to all parameters from one additional solve per output\strut} & \parbox[t]{3.4cm}{\raggedright \texttt{Results.sensitivities}\strut} \\
\hline
\parbox[t]{2.6cm}{\raggedright Cavity haloscopes\strut} & \parbox[t]{4.3cm}{\raggedright Cavity modes, form factor $C$, intrinsic quality factor $Q_0$, mode overlaps, and haloscope sensitivity\strut} & \parbox[t]{4.7cm}{\raggedright 2D and 3D eigenmode solvers; wall losses for normal metals (including the anomalous skin effect) and superconductors\strut} & \parbox[t]{3.4cm}{\raggedright \texttt{cavity}, \texttt{axionmultiphysics.\allowbreak haloscope}\strut} \\
\hline
\parbox[t]{2.6cm}{\raggedright SRF heterodyne\strut} & \parbox[t]{4.3cm}{\raggedright Pump and signal modes, their mixing by wall vibrations, and SRF-cavity sensitivity \strut} & \parbox[t]{4.7cm}{\raggedright Axisymmetric eigenmodes tuned to degeneracy; Slater perturbation theory for deformed cells\strut} & \parbox[t]{3.4cm}{\raggedright \texttt{axionmultiphysics.\allowbreak srf}, \texttt{axionmultiphysics.\allowbreak heterodyne}\strut} \\
\hline
\parbox[t]{2.6cm}{\raggedright COMSOL interface\strut} & \parbox[t]{4.3cm}{\raggedright Exchange of geometry, meshes, materials, and physics settings to or from COMSOL\strut} & \parbox[t]{4.7cm}{\raggedright Reads and writes COMSOL model and mesh files\strut} & \parbox[t]{3.4cm}{\raggedright \texttt{axmp export-comsol}, \texttt{axmp import-comsol}\strut} \\
\hline
\parbox[t]{2.6cm}{\raggedright MCP server\strut} & \parbox[t]{4.3cm}{\raggedright Natural-language control of all modules by AI assistants\strut} & \parbox[t]{4.7cm}{\raggedright Model Context Protocol (MCP) server\strut} & \parbox[t]{3.4cm}{\raggedright \texttt{axionmultiphysics-\allowbreak mcp}\strut}
\end{tabular}
\end{ruledtabular}
\end{table*}

\subsection{Readout and auxiliary modules}

To obtain a detector's sensitivity, we insert the simulated pickup, described by its impedance $Z(\omega)$ and coupling $\Psi$, into a readout-circuit model and compute the signal and noise at the amplifier input.
The \texttt{circuit} and \texttt{multiport} modules achieve this by combining the pickup impedance with capacitors, inductors, resistors, and transformers using modified nodal analysis~\cite{Ho:1975mna} (as in the effective circuit of Fig.~\ref{fig:axionmultiphysics_overview}).
They propagate the axion signal and thermal noise through the circuit, including correlations between SQUID imprecision and backaction noise.
\texttt{axionMultiPhysics} also includes a module, \texttt{axionmultiphysics.\allowbreak squid}, for the dc SQUID amplifier chain, built from measured device parameters~\cite{Ankel:2025nld}; it calculates the frequency-dependent noise and determines the input coupling that minimizes it.
From the resulting signal and noise, the \texttt{scan} module determines the integration time needed at each tuning step to reach a target sensitivity.
Requiring the complete scan to fit within the available live time sets the minimum figure of merit $\mathcal F_{\rm req}$, which can then be compared with the detector's figure $\mathcal F=\Psi^4/L_p^2$  (see~\eqref{eq:scanrate} and~\eqref{eq:Freq_req}).
For calculations beyond the MQS limit, the \texttt{multipole} module can be used to determine the optimal allocation of scan time across disjoint tunable bands, accounting for sensitivity away from resonance~\cite{Benabou:2022qpv}.

\texttt{axionMultiPhysics} also models mechanical and thermal responses.
The \texttt{mechanics} module solves the linear-elastic problem for the magnet structure, with the Lorentz force on the windings and the differential thermal contraction on cooldown as loads, and returns the displacements and stresses.
The \texttt{thermal} module solves for the steady temperature across the cold structure, given the heat that leaks in (for example through mechanical supports) and the temperatures held fixed by the cryogenic system, using temperature-dependent thermal conductivities.

We validate the electromagnetic, thermal, and mechanical solvers against analytic solutions and, where available, published numerical results.
The full-wave solver reproduces the COMSOL impedance and open-circuit voltage calculations of Ref.~\cite{Benabou:2022qpv} beyond the MQS regime.
Further details on validation are given in the code documentation.

\subsection{Resonant-cavity haloscopes and SRF cavities}\label{app:axionmultiphysics_cavity}

For a resonant cavity, sensitivity to $g_{a\gamma\gamma}$ depends on the overlap of the mode's electric field with the applied magnetic field, and on the mode's losses.
The axion-induced power at $f\simeq m_a/2\pi$ is~\cite{Kim:2020kfo}
\es{eq:axionmultiphysics_cavity_power}{
P_{\rm sig}=\gagg^2\,\frac{\rho_{\rm DM}}{m_a}\,B_{\rm lab}^2\,V\,C\,\frac{\beta}{1+\beta}\,\frac{Q_LQ_{\rm axion}}{Q_L+Q_{\rm axion}}\,,
}
where $V$ is the volume, $B_{\rm lab}^2\equiv V^{-1}\int_V|\bm B_{\rm lab}|^2\,dV$, $C$ the form factor, $\beta$ the receiver coupling, and $Q_L=Q_0/(1+\beta)$ the loaded quality factor.
For system noise temperature $T_{\rm sys}$, target signal-to-noise ratio (SNR), and $Q_L\ll Q_{\rm axion}$, the scan rate is~\cite{AlKenany:2016trt,Brubaker:2017rna,Kim:2020kfo}
\es{eq:axionmultiphysics_cavity_scan}{
\frac{df}{dt}\propto\frac{\gagg^4\,\rho_{\rm DM}^2}{{\rm SNR}^2\,m_a^2}\,\frac{B_{\rm lab}^4\,V^2\,C^2\,Q_L\,Q_{\rm axion}}{T_{\rm sys}^2}\left(\frac{\beta}{1+\beta}\right)^2\,,
}
maximized at $\beta=2$ for fixed $Q_0$.
Thus $B_{\rm lab}^4V^2C^2Q_0$ corresponds to the lumped-element figure of merit $\mathcal F Q$.

For a cavity in a given tuning configuration and for a given applied magnetic field $\bm B_{\rm lab}$, the \texttt{cavity} module of \texttt{axionMultiPhysics} computes the resonant frequencies and the electric and magnetic field patterns of the modes.
From each mode it computes the form factor, the normalized overlap of the mode's electric field with $\bm B_{\rm lab}$,
\es{eq:axionmultiphysics_formfactor}{
C=\frac{\left|\int \bm E\cdot\bm B_{\rm lab}\,dV\right|^2}{\int\epsilon_r|\bm E|^2\,dV\,\int|\bm B_{\rm lab}|^2\,dV}\,,
}
and the intrinsic quality factor $Q_0=\omega_0W/P$, the ratio of the stored energy $W$ to the power $P$ lost in the walls and dielectrics.

The cavity modes are the source-free, lossless solutions of~\eqref{eq:axionmultiphysics_fullwave}, with vanishing tangential electric field on the conducting walls, $\hat{\bm n}\times\bm E=0$.
\texttt{axionMultiPhysics} solves this via FEM in the cavity interior, as a matrix eigenvalue problem for $\omega^2$, for the modes nearest a chosen frequency~\cite{Ericsson:1980lanczos}.
Losses are then computed from these mode fields: the power lost in the walls, $P=\tfrac12\oint R_s|\bm H_t|^2\,dS$, depends on the magnetic field along the wall and on the surface resistance $R_s$ of a normal metal (including the anomalous skin effect~\cite{Reuter:1948ase}) or a superconductor~\cite{Mattis:1958yid,Halbritter:1970rf}; dielectric loss enters through $\tan\delta$.
\texttt{axionmultiphysics.\allowbreak haloscope} converts $f,C,Q_0,\beta$ into the signal power~\eqref{eq:axionmultiphysics_cavity_power}, noise (including squeezed-state receivers), scan rate~\eqref{eq:axionmultiphysics_cavity_scan}, and the sensitivity.
Note that we validate the mode frequency, form-factor, and quality factor calculations against published values for ADMX~\cite{Braine:2024nzi}, KLASH~\cite{Alesini:2017ifp}, QUAX~\cite{QUAX:2020uxy,DiVora:2022tro}, multi-post~\cite{Stern:2015kzo}, and CAPP multi-cell~\cite{Jeong:2017hqs}, finding agreement at the percent level in most cases.

In the SRF heterodyne scheme (Sec.~\ref{sec:heterodyne}) the axion transfers photons from a strongly driven pump mode into an empty signal mode whose frequency differs by only $m_a\ll\omega_0$, such that the superconducting cell must have two nearly degenerate modes.
The scan rate~\eqref{eq:SRF_scan} then depends on the cell's shape through the overlap $\eta_{10}$, the quality factor, and the peak fields on the wall, which limit how hard the cell can be pumped.
To find such a pair of modes, the \texttt{axionmultiphysics.\allowbreak srf} module stretches an elliptical or cylindrical superconducting cell along its axis until the pump (${\rm TM}_{020}$, whose frequency is set mainly by the radius) and signal (${\rm TE}_{011}$, which also depends on the length) modes have the same frequency.
It returns the frequency $\omega_0=\xi V^{-1/3}$, with $\xi$ the dimensionless frequency constant of Sec.~\ref{sec:heterodyne}, the pump--signal overlap $\eta_{10}$, the geometry factor $G=QR_s$, the peak wall fields, and the wall area.
\texttt{axionmultiphysics.\allowbreak heterodyne} then evaluates the scan rate~\eqref{eq:SRF_scan}, the volume required to reach a given sensitivity, and the cost~\eqref{eq:cost_SRF}.
We validate \texttt{axionmultiphysics.\allowbreak srf} against published results for accelerator cavities and for heterodyne mode pairs~\cite{Aune:2000gb,Lasenby:2019prg}, finding agreement at the percent level.

Mechanical vibrations of the cavity walls mix the pump and signal modes, leaking pump photons into the signal mode and producing a background.
The mixing vanishes for an axisymmetric cell.
For a 3D cell with a small manufacturing deviation from roundness, \texttt{axionMultiPhysics} computes the mixing from the two mode fields at the wall using Slater perturbation theory~\cite{Slater:1946}.

\subsection{Running a design and examining results}
The example below builds an axisymmetric toroidal lumped-element detector, computes its fields, prints a summary, and saves an interactive visualization:
\begin{verbatim}
# pip install "axionmultiphysics[mesh]"
import axionmultiphysics as axmp
d = axmp.template("toroid", R=2., H=2.)
d.save("design.json")
res = d.solve("run")
print(res.summary())
res.export_html("toroid.html")
\end{verbatim}
This is the toroid shown in Fig.~\ref{fig:axionmultiphysics_overview}(a), with a radius and height of $2\,\mathrm{m}$.
The printed summary includes $\Psi$, $L_p$, $W_{\rm mag}$, and peak fields; for each sensitivity benchmark, the magnet strength, stored energy, and cost needed to reach the target coupling, and, for designs without iron, the size that minimizes the cost at fixed shape; and warnings, for example when a nonlinear solve does not converge or the estimated self-resonance of the readout lies below the top of the scan.

For a given detector topology, the shape minimizing the cost can be determined by varying the design's named parameters.
The optima of Sec.~\ref{sec:numerics} were found by a global search over these parameters (differential evolution), with the field amplitude at each geometry set so that $\mathcal F=\mathcal F_{\rm req}$.
\texttt{axionMultiPhysics} also provides a local gradient-based optimizer that accepts constraints such as field limits.
For the latter it computes the derivatives of $\Psi$, $L_p$, $W_{\rm mag}$, $\mathcal F$, and the cost with respect to every design parameter.
One extra linear solve (the adjoint) gives the derivatives with respect to all parameters at once, with the mesh deformed smoothly as a parameter changes.
With nonlinear iron, or when a parameter changes the mesh structure, each derivative is instead found by re-solving at slightly shifted parameter values.

Lastly, \texttt{axionMultiPhysics} exports geometry, meshes with named selections, and supported 3D full-wave port setups to COMSOL~\cite{multiphysics1998introduction}, as a COMSOL Java model, an \texttt{.mphtxt} mesh, or a STEP file.
It can also import COMSOL models saved in \texttt{.mph} format, either keeping the mesh stored in the file or, less reliably, rebuilding the geometry as an editable \texttt{axionMultiPhysics} design.

\section{Optimality of the toroid and core designs}
\label{sec:optimality}

In Sec.~\ref{sec:duality} we show that, for the geometry shared by the toroid and core (an annular superconducting box around a central hole), the idealized toroid and core designs are optimal.
Here we relax this geometry and prove that among all long detectors whose laboratory and pickup fields do not overlap (and whose fields do not vary along the vertical axis), no configuration does better than the toroid and core.
We also show that letting the fields overlap would lower the cost somewhat but would require placing the pickup in the magnet's static field, which would, {\it e.g.}, increase the loss in the pickup and the magnet noise.

\subsection{Geometric figure of merit}

Recall that in Sec.~\ref{sec:long} we identified the figure of merit $\mathcal F =\Psi^4/L_p^2$, which enters in the scan rate. Achieving a target sensitivity fixes $\mathcal F\ge\mathcal F_{\rm req}$, such that the detector needs stored energy at least
\begin{equation}
W_{\rm mag}=\frac{\sqrt{\mathcal F_{\rm req}}}{\Gamma}\,,\qquad \Gamma\equiv\frac{\Psi^2}{L_pW_{\rm mag}}\,.
\end{equation}
Thus larger $\Gamma$ corresponds to less stored energy, and hence a lower cost~\eqref{eq:cost_HB}, needed to reach the target sensitivity. For a detector of outer radius $R$, $\Gamma=gR^2$, where the dimensionless factor $g$ of~\eqref{eq:g_def} depends only on the shape of the design (including its aspect ratio $H/R$), not on its overall size or strength. We therefore maximize $g$: for designs in the same enclosure, optimizing the overall size at fixed shape gives a minimum cost $\mathcal C_{\min}\propto g^{-28/75}$. 

Given the pickup field $\bm{B}_I=\nabla \times \bm{A}_I$ and the laboratory field $\bm{B}_{\rm lab}$, let us first define the pickup and laboratory field norms, 
\begin{equation}
N_I \equiv \int d^3 x\,\left|\bm{B}_I\right|^2, \qquad N_{\rm lab} \equiv \int d^3 x\,\left|\bm{B}_{\rm lab}\right|^2,
\label{eq:norms}
\end{equation}
and the linkage\footnote{The linkage $M$ does not depend on the choice of $\bm A_I$, provided no laboratory field crosses the outer wall.}
\begin{equation}
M \equiv \int d^3 x\, \bm{A}_I \cdot \bm{B}_{\rm lab},
\end{equation}
such that $\Psi=M/I$ (as in~\eqref{eq:flux}), the pickup inductance is $L_p=N_I/I^2$, and $W_{\rm mag}=N_{\rm lab}/2$.
This allows us to express
\begin{equation}
\Gamma=\frac{2 M^2}{N_I N_{\rm lab}} \,,
\label{eq:app_opt_gamma}
\end{equation}
which will be useful in the following.
Below, we study idealized geometries, allowing for ideal sheet currents and arbitrarily large fields; we do not impose any critical field limits and compute the magnetic energy in vacuum (therefore, our reasoning does not apply to, {\it e.g.}, the core design with iron).

\subsection{Non-overlapping laboratory and pickup fields}

Let us first consider detectors in which the pickup is shielded from the magnet ({\it i.e.}, for which the field $\bm{B}_I$ produced by the pickup current vanishes wherever the laboratory field $\bm{B}_{\rm lab}$ is nonzero, and vice versa). The ideal toroid and ideal core geometries belong to this class (the solenoid does not). The detector therefore has superconducting sheets (such as the hermetic sheath of the toroid) dividing the cross-section of radius $R$ into regions holding either $\bm{B}_I$ or $\bm{B}_{\rm lab}$. The regions are otherwise arbitrary: they may have any shape and need not be axisymmetric; there may be any number of them, and each field may have axial and transverse components. To begin, let us consider long detectors, of height $H$ much larger than their radius $R$, so that edge effects can be neglected. We further assume that the fields are independent of $z$, and vanish for $r > R$. Under these assumptions, we show that
\begin{equation}
    g \le \frac{1}{2e}\,,
    \label{eq:gbound}
\end{equation}
with equality realized only by the toroid and core.

\subsubsection{The toroid and core}
Before considering the general case, let us explain the argument summarized in Sec.~\ref{sec:duality} that the bound~\eqref{eq:gbound} holds for the toroid and core geometries specifically. 

Let $V$ be the region holding the laboratory field: the box for the toroid, and, for the core, the hole together with the return region outside the box. Let $\Phi_{\rm lab}$ be the laboratory flux through a cut $S$ of $V$ ({\it e.g.}, a vertical slice $\phi={\rm const}$ through the toroid's box, or a horizontal disk of the hole for the core). Let $\bm h=\nabla\chi$ be the current-free field in $V$ of a unit current linked by $V$ (flowing axially through the hole for the toroid, and azimuthally around the hole, on the box, for the core), with $\chi$ jumping by one across $S$ and $\int_V|\bm h|^2=L_{\rm static}$. Since $\nabla\cdot\bm B_{\rm lab}=0$ and no laboratory field crosses the boundary of $V$, we have $\int_V\bm B_{\rm lab}\cdot\bm h=\Phi_{\rm lab}$. The Cauchy--Schwarz (CS) inequality then gives
\begin{equation}
\Phi_{\rm lab}^2\le\int_V|\bm B_{\rm lab}|^2\int_V|\bm h|^2=2W_{\rm mag}L_{\rm static},
\end{equation}
with equality if and only if $\bm B_{\rm lab}\propto\bm h$, {\it i.e.}, if the laboratory field is current-free inside $V$. With $\Psi=L_p\Phi_{\rm lab}$, this gives $\Gamma\le2L_pL_{\rm static}=2L_{\rm box}L_{\rm out}$. 

For both the toroid and core, let us parametrize the hole radius by $yR$, with the box spanning $yR<r<R$ as in Sec.~\ref{sec:long_toroid}. As the flux $\Phi=L_{\rm out}I$ crosses every horizontal disk of the hole, the CS inequality on each disk gives $\int|\bm B|^2\ge H\Phi^2/(\pi y^2R^2)$ and hence $L_{\rm out}\le\pi y^2R^2/H$, with equality for $H\gg R$. Using $L_{\rm box}=(H/2\pi)\log(1/y)$, we obtain
\begin{equation}
\Gamma\le y^2\log(1/y)\,R^2\le\frac{R^2}{2e},
\end{equation}
maximized at $y=e^{-1/2}$. 

\subsubsection{General case}

We now show that the same bound holds beyond the toroid and core geometries. Let us abbreviate $\bm B_1=\bm B_I$ and $\bm B_2=\bm B_{\rm lab}$. In cylindrical coordinates $(r,\phi,z)$ let us take these fields to be independent of $z$, supported on a cylinder of radius $R$ and height $H \gg R$, and with no normal component on the sheets. By assumption, $|\bm{B}_1(r,\phi)|\,|\bm{B}_2(r,\phi)|=0$ for almost every $(r,\phi)$ (we allow for tangential jumps, corresponding to sheet currents). We first decompose each field into axial and transverse components,
\begin{equation}
\begin{aligned}
\bm{B}_j & =b_j \hat{\bm{z}}+\nabla_{\perp} \psi_j \times \hat{\bm{z}} \\
& =\frac{\partial_\phi \psi_j}{r} \hat{\bm{r}}-\partial_r \psi_j \hat{\bm{\phi}}+b_j \hat{\bm{z}}, \quad j=1,2 ,
\label{eq:Bj}
\end{aligned}
\end{equation}
with $\psi_j(R,\phi)=0$, and we define $u_j=-\partial_r \psi_j$, so that $\psi_j(r,\phi)=\int_r^R ds\, u_j(s,\phi)$. A vector potential is then given by $\bm{A}_j=\alpha_j\hat{\bm{\phi}}+\psi_j\hat{\bm{z}}$, with $\alpha_j(r,\phi)=r^{-1}\int_0^r ds\, s\, b_j(s,\phi)$. The linkage is then
\begin{equation}
M = H\int d^2 x\, \bm{A}_1 \cdot \bm{B}_2=H\int_0^{2 \pi} d \phi \int_0^R r\, d r\left(\psi_1 b_2+\psi_2 b_1\right) .
\label{eq:mlink}
\end{equation}

On a ray at fixed $\phi$, let $E$ be the region where $b_2 \neq 0$ and $F$ where $u_1 \neq 0$, such that $E$ and $F$ are disjoint sets by our assumption that $\bm{B}_1$ and $\bm{B}_2$ do not overlap. Defining the radial $L^2$ norms on the ray, $\|f\|^2\equiv\int_0^R r\, dr\, f(r,\phi)^2$ (we will include a subscript on the norm to indicate restriction to a given set), the CS inequality gives
\begin{equation}
\left\|\psi_1\right\|_E \leq \frac{R}{2 \sqrt{e}}\left\|u_1\right\| ,
\end{equation}
and likewise for the second term in~\eqref{eq:mlink}.\footnote{To see this, note that $u_1$ vanishes outside $F$, so $\psi_1(r)=\int_{F\cap(r,R)}ds\,u_1(s)$, and the CS inequality gives
\begin{equation}
\psi_1(r)^2\le\int_Fs\,ds\,u_1^2\int_{(r,R)\setminus E}\frac{ds}{s}=\|u_1\|^2\int_{(r,R)\setminus E}\frac{ds}{s},
\end{equation}
where the last integral avoids $E$ because $E$ and $F$ are disjoint. Integrating over $E$ and changing variables to $\xi=r^2/R^2$ and $\eta=s^2/R^2$,
\begin{equation}
\|\psi_1\|_E^2\le\frac{R^2}{4}\,K\,\|u_1\|^2,\qquad K\equiv\int_{E'}d\xi\int_{(\xi,1)\setminus E'}\frac{d\eta}{\eta},
\end{equation}
where $E'$ is the image of $E$. To bound $K$, let $G(\eta)=|E'\cap(0,\eta)|$ and $a=G(1)$, so that $G'$ is the indicator function of $E'$ and $K=\int_0^1 d\eta\,(1-G')\,G/\eta$. Integrating 
by parts
gives $K=\int_0^1 d\eta\,(t-t^2/2)-a^2/2$, where $t\equiv G/\eta$. 
This gives $K\le a\log(1/a)\le1/e$, with equality only for $E'=(0,1/e)$.} In addition, since $b_2$ vanishes outside $E$, $\int_0^Rr\,dr\,\psi_1b_2\le\|\psi_1\|_E\|b_2\|$.

Lastly, given~\eqref{eq:Bj}, the field norms in~\eqref{eq:norms} obey \begin{equation}
N_j\ge H\int_0^{2\pi}d\phi\,(\|b_j\|^2+\|u_j\|^2)\,. 
\end{equation}
Combining the previous inequalities, and applying the CS inequality over the two terms in~\eqref{eq:mlink} and over $\phi$ then gives
\begin{equation}
\begin{aligned}
|M| & \leq \frac{HR}{2 \sqrt{e}} \int_0^{2 \pi} d \phi\left(\left\|b_1\right\|\left\|u_2\right\|+\left\|u_1\right\|\left\|b_2\right\|\right) \\
& \leq \frac{R}{2 \sqrt{e}} \sqrt{N_1 N_2}\,,
\end{aligned}
\end{equation}
so that~\eqref{eq:app_opt_gamma} gives $g\leq 1/(2e)$.
Equality requires one field to be axial and uniform for $r < R/\sqrt{e}$ and the other azimuthal, proportional to $1/r$, outside. That is, only the toroid and core saturate the bound within this class of geometries.

\subsection{Overlapping fields}

\begin{figure*}[!t]
\centering
\includegraphics[width=\textwidth]{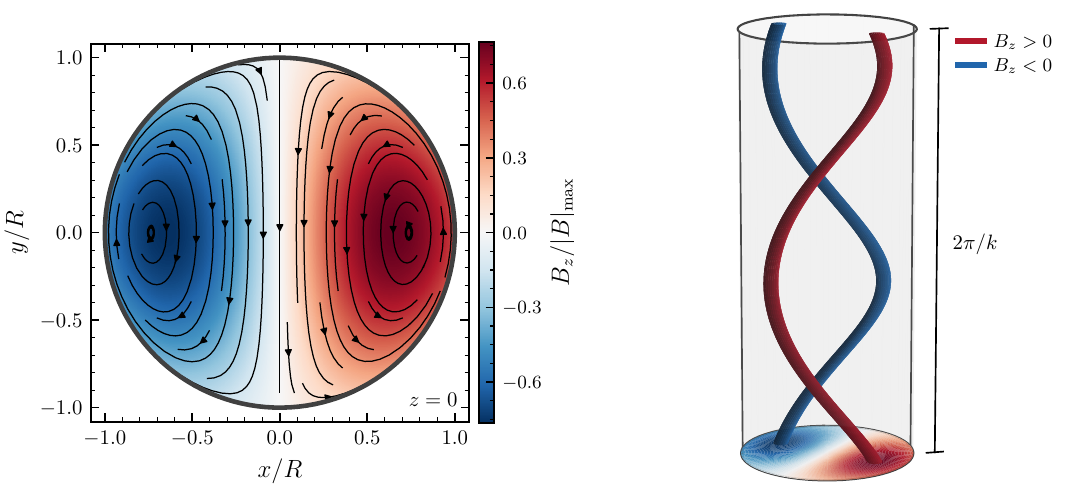}
\caption{The magnetic field corresponding to the optimal geometry with overlapping laboratory and pickup magnetic fields, in the limit of a superconducting shield of
radius $R$ and infinite height. The field is the force-free field $\nabla\times\bm B=\mu\bm B$ solving~\eqref{eq:Br0} with
$qR=2.86$, $kR=1.23$ and $\mu R=3.11$. Both $\bm B_{\rm lab}$
and $\bm B_I$ are proportional to this field.  \textit{Left:} Cross-section
at $z=0$: $B_z$ (color) and field lines of the transverse field.  \textit{Right:} One axial
period $2\pi/k$: two tubes tracing the maxima of $|B_z|$ (red, $B_z>0$; blue, $B_z<0$), which wind around each other as a
double helix; the bottom face shows $B_z$ as in the left panel.}
\label{fig:app_opt_helix}
\end{figure*}

Let us now consider the case where the pickup and static laboratory fields overlap. We will also relax the assumption that the fields are independent of $z$, requiring only that they are confined by a superconducting boundary. Note that without overlap,
$\bm A_I$ is curl-free wherever $\bm B_{\rm lab}$ is nonzero, so it is fixed by the
pickup flux that the laboratory field lines encircle; with overlap, $\bm A_I$ can instead
be aligned with $\bm B_{\rm lab}$ throughout the volume, which allows exceeding the upper bound on $g$ of the previous subsection.
In fact, allowing for overlap, we find that $\Gamma$ is maximized by making $\bm B_{\rm lab}$ and $\bm B_{I}$ proportional to the same field. 

To see this, let us write
$M=\int d^3x\,(\mathcal K\bm{B}_I)\cdot\bm{B}_{\rm lab}$, where the inverse curl $\mathcal K$ is defined via $\bm{A} = \mathcal K \bm{B}$.\footnote{We choose $\bm A$ divergence free and tangent to the wall, so that $\mathcal K$ maps
such fields to themselves and has eigenfields.} Viewing fields as vectors with inner product $\langle\bm B,\bm C\rangle=\int d^3x\,\bm B\cdot\bm C$, $\mathcal K$ is symmetric: $\langle\mathcal K\bm B,\bm C\rangle=\langle\bm B,\mathcal K\bm C\rangle$. Indeed, we have $\int\bm A_I\cdot\bm B_{\rm lab}=\int\bm A_{\rm lab}\cdot\bm B_I$. By the spectral theorem for the compact self-adjoint operator $\mathcal{K}$, $M/\sqrt{N_IN_{\rm lab}}$ is then maximized when both $\bm B_I$ and $\bm B_{\rm lab}$ are proportional to the eigenfield of $\mathcal K$ with the largest eigenvalue in magnitude, $|\lambda|$. Taking the curl of $\mathcal{K} \bm{B}=\lambda \bm{B}$ gives 
\begin{equation}
\nabla\times\bm{B}=\mu\bm{B}\,,
\end{equation}
with $\mu=1/\lambda$. This corresponds to a force-free field~\cite{Woltjer:1958,Taylor:1974}. Thus, $\Gamma \le 2/\mu_{\min}^2$, where $\mu_{\min}$ is the smallest $|\mu|$ allowed by the boundary.

Let us now compute the optimal $\Gamma$ in an infinitely long superconducting shield, closed at its far ends (the same kind of enclosure our shielded toroid sits in), and compare it with a shielded toroid, which does not have overlapping fields. In the shield, the admissible force-free fields are of the form
\begin{equation}
\bm{B}=\mathcal{T}[\psi]+\mu^{-1}\nabla \times \mathcal{T}[\psi]\,,
\end{equation}
with $\mathcal{T}[\psi]=\nabla \times(\psi \hat{\bm{z}})$, $\psi=J_m(qr)e^{i(m\phi+kz)}$ and $\mu^2=q^2+k^2$, where $J_m$ is the Bessel function of the first kind, $k$ the axial wavenumber, $q$ the radial wavenumber, and $m$ the azimuthal mode number. The shield imposes that $B_r(R)=0$:
\begin{equation}
m\, \mu R\, J_m(q R)+k R\, q R\, J_m^{\prime}(q R)=0 \,.
\label{eq:Br0}
\end{equation}
Since the shield is closed, the net axial flux through every cross-section must also vanish. The axisymmetric $m=k=0$ fields satisfy~\eqref{eq:Br0} for any $\mu$, but the flux condition requires $J_1(qR)=0$, so they have $\mu R\ge j_{1,1}=3.83$, where $j_{1,1}$ is the first zero of $J_1$.
The field with the smallest $\mu$ is realized instead for $m=1$, $qR=2.86$, $kR=1.23$, and $\mu R=3.11$. This field, shown in Fig.~\ref{fig:app_opt_helix}, is helical and gives $g=2/(\mu R)^2 \approx 0.206$. 

For comparison, a toroid with outer radius $uR$ and, as in Sec.~\ref{sec:long_toroid}, inner-to-outer radius ratio $y$ in the same shield has
\begin{equation}
g=y^2\log(1/y)\,\frac{u^2(1-u^2)}{1-u^2+y^2u^2}\,,
\end{equation}
which is maximized at $g=0.0772$ (for $y=0.478$ and $u=0.823$), {\it i.e.}, 37\% of the optimum. Using the cost-optimal scaling $\mathcal{C}_{\min} \propto g^{-28/75}$, this implies that the shielded toroid costs only $1.44$ times as much as the optimal overlapping design in a long closed shield.

Let us now relax the infinite-height idealization, and take the fields to be confined in a cylindrical shield of radius $R$ and finite height $H$. In this case, we minimize $\mu$ numerically with a variational Rayleigh--Ritz method: writing $\bm B=\nabla\times\bm A$ with the tangential part of $\bm A$ vanishing on the walls,\footnote{By gauge invariance, $\int\bm A\cdot\bm B$ does not depend on this choice.} which guarantees $\bm B\cdot\bm n=0$ there, we have $1/\mu_{\min}=\max\int\bm A\cdot\bm B\big/\int|\bm B|^2$ over all such $\bm A$. 
Because the cylinder is axisymmetric, fields with different azimuthal dependence $e^{im\phi}$ decouple, and for each $m$ we maximize this ratio over a polynomial basis for $\bm A$ in $r^2$ and $z$. 

We find that for $H/R<1.67$ the optimum is the axisymmetric $m=0$ field solving~\eqref{eq:Br0}, with $qR=j_{1,1}$ and $k=\pi/H$ set by the end caps, so that $(\mu_{\min}R)^2=j_{1,1}^2+(\pi R/H)^2$.\footnote{This solution is known as the spheromak~\cite{Rosenbluth:1979spheromak,Taylor:1986rmp}.} For $H/R>1.67$, the optimal solution is instead a twisted $m=1$ field (in agreement with Refs.~\cite{Finn:1981tilt,Bondeson:1981tilt}), which tends towards the double helix solution discussed previously in the limit $H/R\to \infty$. We may compare these optima with the optimized shielded toroids (see Tab.~\ref{tab:numerics}), which were bounded by cylinders with $H/R\approx1.4$ and reached $g\approx0.048$ for both benchmarks. The optimal overlapping geometry at this aspect ratio instead has $g=2/(\mu_{\min}R)^2\approx0.101$, so the shielded toroid reaches 47\% of the optimum and costs about 1.3 times as much. 

In summary, any possible cost savings beyond the shielded toroid design are only at the level of a few tens of percent. In addition, we emphasize that for any $H/R$ the optimal overlapping geometry is not realistic: its pickup and magnet fields have the same shape, so the pickup is fully coupled to the magnet's own current noise and sits in the full static field.

\end{document}